\documentclass[12pt,a4paper]{article}
\usepackage[latin1]{inputenc}
\usepackage{amssymb,amsmath,mathrsfs}
\usepackage{hyperref}
\usepackage{comment}

 \usepackage{mdframed}
 \usepackage{graphicx}
 \usepackage{setspace}
 \usepackage{a4wide}

\usepackage{cite}

\usepackage{tikz}
\usepackage{color}

\makeatletter 
\@addtoreset{equation}{section}% \@addtoreset{equation}{subsection} 
\makeatother

\hypersetup{
  linktocpage,
  colorlinks  = true, %Colours links instead of ugly boxes
  urlcolor    = black, %Colour for external hyperlinks
  linkcolor   = black, %Colour of internal links
  citecolor   = black %Colour of citations
}

\newcommand{\ltr}{k}
\newcommand{\ltro}{k}

\newcommand{\cO}{{\cal O}}

\def \be  {\begin{equation}}
\def \ee  {\end{equation}}
\def \ba  {\begin{eqnarray}}
\def \ea  {\end{eqnarray}}

\def \cO{\mathcal{O}}

\def\x{x}

\usepackage{titlesec}

\titleformat*{\section}{\large\bfseries}
\begin{document}

\usetikzlibrary{arrows}
% Front page here
\thispagestyle{empty}

~\\[-2.25cm]

%\null\vskip-12pt \hfill  \\
%\null\vskip-12pt \hfill   \\

%\vskip2.2truecm
\begin{center}
%\vskip 0.2truecm 
{\Large\bf
%\titleline
%
{\Large The Virasoro-Shapiro amplitude in AdS$_5\times$S$^5$ and\\[.25cm] 
		level splitting of 10d conformal symmetry }
}\\
\vskip 1.25truecm
	{\bf F.~Aprile${}^{1}$, J.~M. Drummond${}^{2}$, H.~Paul${}^{2,3}$, M.~Santagata${}^{2}$ \\
	}
	
	\vskip 0.4truecm
	
	{\it
		${}^{1}$ Dipartimento di Fisica, Universit\`a di Milano-Bicocca \& INFN, 
		Sezione di Milano-Bicocca, I-20126 Milano,\\
		\vskip .2truecm }
	\vskip .2truecm
	{\it
		${}^{2}$ School of Physics and Astronomy and STAG Research Centre, \\
		University of Southampton,
		Highfield,  SO17 1BJ,\\
		\vskip .2truecm                        }
	\vskip .2truecm
	{\it
		${}^{3}$ Institut de Physique  Th\'eorique, CEA Saclay, CNRS, Orme des Merisiers, 91191 Gif-sur-Yvette, France         }	

\end{center}
\vskip 1.25truecm %\Large

\centerline{\bf Abstract} 
\vskip .4truecm
The genus zero contribution to the four-point correlator $\langle {\cal O}_{p_1}{\cal O}_{p_2}{\cal O}_{p_3}{\cal O}_{p_4}\rangle$  of half-BPS 
single-particle operators ${\cal O}_p$ in $\mathcal{N}=4$ super Yang-Mills, at strong coupling,
computes the Virasoro-Shapiro amplitude of closed superstrings in $AdS_5\times S^5$. 
Combining Mellin space techniques, the large $p$ limit, and data about the spectrum of 
two-particle operators at tree level in supergravity, we design a bootstrap algorithm which 
heavily constrains its $\alpha'$ expansion. We use crossing symmetry, polynomiality in the 
Mellin variables and the large $p$ limit to stratify the Virasoro-Shapiro amplitude away from the 
ten-dimensional flat space limit. Then we analyse the spectrum of exchanged two-particle operators 
at fixed order in the $\alpha'$ expansion. We impose that the ten-dimensional spin of the spectrum visible 
at that order is bounded above in the same way as in the flat space amplitude. This constraint determines the 
Virasoro-Shapiro amplitude in $AdS_5\times S^5$ up to a small number of ambiguities at each order. 
We compute it explicitly for $(\alpha')^{5,6,7,8,9}$. As the order of $\alpha'$ grows, the ten dimensional 
spin grows, and the set of visible two-particle operators opens up. Operators illuminated for the first time 
receive a string correction to their anomalous dimensions which is uniquely determined and 
lifts the residual degeneracy of tree level supergravity, due to ten-dimensional conformal symmetry.
We encode the lifting of the residual degeneracy in a characteristic polynomial. This object carries information 
about all orders in $\alpha'$. It is analytic in the quantum numbers, symmetric under an $AdS_5 \leftrightarrow S^5$ 
exchange, and it enjoys intriguing properties, which we explain and detail in various cases.

\noindent
%\vskip 0.25truecm

\vskip 1truecm 
%\centerline{\bf Abstract}\normalsize

%\medskip

\newpage
\tableofcontents

\newpage

%%==============================================================================
%%==============================================================================

\section{Introduction and summary of results}\label{intro_sec}

%%==============================================================================
%%==============================================================================
                                                                                                                                
\setcounter{page}{1}\setcounter{footnote}{0}

If we were to solve a four-dimensional quantum field theory, analytically, and in the near future, 
that would certainly be $\mathcal{N}=4$ super Yang-Mills in four dimensions with $SU(N)$ gauge group. 
Being the most symmetric field theory, it has long been appreciated that the interplay of supersymmetry,
conformal symmetry and integrability would manifest in many non trivial ways, ultimately
leading to remarkable simplicity at the quantum level \cite{Beisert:2010jr}.  
More of this beauty has been recently uncovered at large $N$ and strong 't Hooft 
coupling $\lambda$, where the field theory provides the completion of quantum
gravity in $AdS_5\times S^5$. A number of results at tree level and one loop, 
which otherwise would have been out of reach, are now available and systematised thanks to
 the bootstrap program in $\mathcal{N}=4$ SYM
 \cite{Rastelli:2016nze,Rastelli:2017udc,Alday:2017xua,Aprile:2017bgs,Aprile:2017xsp,Alday:2017vkk,Aprile:2017qoy,Aprile:2018efk,
 Caron-Huot:2018kta,Aprile:2019rep,Alday:2018kkw,Alday:2019nin,Drummond:2019hel,Goncalves:2019znr,Bissi:2020wtv,Drummond:2020uni}.

In this paper we will consider genus zero string corrections to
four-point correlators of single-particle operators ${\cal O}_p$ in $AdS_5\times S^5$,
following up on previous work \cite{Goncalves:2014ffa,Alday:2018pdi,Drummond:2019odu}  
and especially \cite{Drummond:2020dwr},  where a formula for the $(\alpha')^5$ amplitude was 
obtained for arbitrary charges $p_{i=1,2,3,4}$.  Our approach here will have two important upgrades described below.

The first improvement is to manifest crossing symmetry and the 10d flat space limit\footnote{For the flat space amplitude 
we use conventions as in \cite{Green:2008uj}.} of the correlators, by using the $AdS_5\times S^5$ Mellin transform of 
\cite{Aprile:2020luw} as reviewed in Sect. \ref{VS_section}. The Mellin 
amplitude we will work with then takes the form
\begin{align}\label{ini_M}
\mathcal{M}&= \frac{1}{(1+{\bf s})(1+{\bf t})(1+{\bf u})}  + \sum_{n=0}^{\infty} \left(\frac{\alpha'}{4}\right)^{\!\!n+3}\,
\mathcal{V}_n({\bf s},{\bf t},\tilde{\bf s},\tilde{\bf t},p_1,p_2,p_3,p_4)
\end{align}
where generically $\mathcal{M}$ depends on two Mellin variables $s,t$, conjugate to spacetime 
cross-ratios $U,V$ at the boundary of $AdS_5$, two Mellin variables $\tilde s,\tilde t$, conjugate to internal space 
cross-ratios $\tilde U,\tilde V$ on the $S^5$, and four external charges $p_{i=1,2,3,4}$. 
Instead, we will use a more convenient set of Mellin variables,  
denoted by the bold font letters ${\bf s,t},\tilde{\bf s},\tilde{\bf t}$, 
such that the physical picture that accompanies the Mellin amplitude is clear, %which are constructed accordingly with 
and moreover these new variables transform in a simple way under crossing. 
The first ones are
\begin{align}
&
\rule{1cm}{0pt}
\begin{array}{ccccc}
{\bf s}=s+\tilde s\qquad&;&\qquad {\bf t}=t+\tilde t\qquad&;&\qquad {\bf s}+{\bf t}+{\bf u}=-4   %\\[.3cm]
\end{array}% \\[.3cm]
\end{align}
and were defined in \cite{Aprile:2020luw}. 
There it was shown that in the limit in which the charges $p_{i=1,2,3,4}$ are large, %say of order $p\gg 1$,
 the correlator localises on a classical saddle point, 
$s\rightarrow  s_{cl}$, $\tilde s\rightarrow  \tilde s_{cl}$, $t\rightarrow  t_{cl}$, $\tilde t\rightarrow  \tilde t_{cl}$, 
 where the Mellin variables are also large, %of order $p$, 
 and quite remarkably the combinations 
 ${\bf s}_{cl}$, ${\bf t}_{cl}$ and ${\bf u}_{cl}=-{\bf s}_{cl}-{\bf t}_{cl}$ become (proportional to)
the 10d flat space Mandelstam invariants of a scattering process 
which is focussed on a bulk point of $AdS_5\times S^5$. 
The amplitude for this process is simply the ten-dimensional 
flat space Virasoro-Shapiro (VS) amplitude,
\begin{align}
\overline{\mathcal{M}}^{flat}= \frac{1}{ {\bf s}_{cl} {\bf t}_{cl} {\bf u}_{cl}}  
\exp\left[\, 2\sum_{n\ge 1} \frac{\zeta_{2n+1}}{ 2n+1} (\Sigma\alpha')^{2n+1}\Big({\bf s}_{cl}^{2n+1} +{\bf t}_{cl} ^{2n+1}+{\bf u}_{cl}^{2n+1} \Big) \,\right]
\end{align}
where $\Sigma=\frac{1}{2}(p_1+p_2+p_3+p_4)$.
It follows that
\begin{align}
\lim_{p\rightarrow \infty}\mathcal{V}_n({\bf s}_{cl},{\bf t}_{cl},\tilde{\bf s}_{cl},\tilde{\bf t}_{cl},p_1,p_2,p_3,p_4)=  
\overline{\mathcal{M}}^{flat}\Big|_{(\alpha')^{n+3}}
\label{boldflatlimit}
\end{align} 
and that the full VS amplitude in $AdS_5\times S^5$ acquires a novel expansion, away from the 10d flat space limit, 
which is particularly meaningful since %, as discussed in Sect. \ref{VS_section}, 
$\mathcal{V}_n$ will be polynomial in the Mellin variables. We shall call this expansion ``the large $p$ limit".\footnote{This terminology is meant to abbreviate
the limit $p_{i}=\epsilon \tilde p_i$ for $i=1,2,3,4$, with $\epsilon\rightarrow \infty$ and $\tilde p_i$ fixed.}

The $AdS_5\times S^5$ VS amplitude starts with tree level supergravity, i.e.~the first term in \eqref{ini_M}.
This is special, it depends just on ${\bf s}$, ${\bf t}$ and ${\bf u}$, as a consequence of a surprising 
accidental ten-dimensional symmetry, which at tree level is conformal \cite{Caron-Huot:2018kta}. This symmetry 
is recovered in the large $p$ limit, at all orders in $\alpha'$, but 
away from the large $p$ limit the amplitude is expected to depend on all variables, 
thus implementing curvature effects of the $AdS_5\times S^5$ background.
Nevertheless, the large $p$ limit implies a certain stratification of the amplitudes ${\cal V}_n$, 
which we discuss in section \ref{VS_section}, and goes as follows,
\begin{align}
&
\mathcal{V}_{n} = 
\sum_{\ell=0}^{n-1} (\Sigma-1)_{\ell+3}\ \mathcal{M}_{n,\ell}({\bf s},{\bf t},\tilde{\bf s},\tilde{\bf t},p_1,p_2,p_3,p_4)+
 (\Sigma-1)_{n+3}\, \mathcal{M}_{n,n}^{flat} ({\bf s},{\bf t},{\bf u})
 \label{stratification}
\end{align}
where schematically\footnote{
This expansion will include similar terms where $\tilde{\bf s}$ is replaced by combinations of the charges. }
\begin{align}
\lim_{p\rightarrow \infty} 
\mathcal{M}_{n,\ell}
\ \ \sim\ \ (p\, {\bf s})^{\ell}\!\times\!\Big( ({p}\, \tilde{\bf s})^{(n-\ell)} + ({p}\, \tilde{\bf s})^{(n-1-\ell)}+ \ldots \Big)
\end{align}%

In (\ref{stratification}) $\mathcal{M}_{n,n}^{flat}({\bf s,t,u})$ is a homogeneous symmetric function 
obtained from the flat space amplitude chosen so that (\ref{boldflatlimit}) is satisfied. Its completion
in $AdS_5\times S^5$ will \emph{not} be homogeneous as in flat space.
We will have additional strata $\mathcal{M}_{n,\ell}$ for $\ell \leq n-1$ built out of all possible crossing 
invariant polynomials in ten (constrained) variables, of fixed degree $\ell$ in ${\bf s}$, ${\bf t}$ and ${\bf u}$, 
and maximal degree $n$ in the large $p$ limit.  

From the Operator Product Expansion (OPE) point of view, the tree level amplitude at strong 't Hooft coupling is quite simple. Its (single) logarithmic discontinuity is 
determined only by the exchange of two-particle operators \cite{Aprile:2017xsp}, and thus is directly related to $\mathcal{M}$ in Mellin space. 
Then, the full tree level amplitude is immediately reconstructed. This would be the case
if we only knew in advance, as function of $\alpha'$, both the planar three-point couplings of the two-particle operators with the external operators $\mathcal{O}_{p_i}\mathcal{O}_{p_j}$, 
and their anomalous dimension. Unfortunately we don't have this data, and instead we will try to constraint $\mathcal{V}_n$, % , in the $\alpha'$ expansion, 
beyond $\mathcal{M}_{n,n}^{flat}({\bf s,t,u})$, by using bootstrap techniques. 

At given order in the $\alpha'$ expansion, we understand that $\mathcal{V}_n$ is a polynomial, 
bounded by $\mathcal{M}_{n,n}^{flat}({\bf s,t,u})$. Therefore, if we write $\mathcal{V}_n$ as a conformal block decomposition in $AdS_5\times S^5$,
in the flat space limit  we have to recover the partial wave expansion of the flat VS amplitude, by construction.\footnote{In Mellin space the conformal block expansion 
is an expansion in Mack's polynomial. In the flat space limit these precisely reduce to partial waves. See for example the discussion in \cite{Penedones:2019tng} section 4.2.}  
This implies that the spin of the two-particle operators exchanged in $\mathcal{V}_n$ is bounded by 
greatest spin contribution in $\mathcal{M}_{n,n}^{flat}({\bf s,t,u})$ %and this bound is
\begin{align}
l_{10}\leq n\quad,\quad n\in\mathbb{N}\ {\rm even} \qquad\quad \longleftrightarrow \qquad\quad  {\bf s}^n+ {\bf t}^n + {\bf u}^n
\end{align}
Since $\mathcal{M}_{n,n}^{flat}$ is a 10d object, the bound on the spin applies not just to a particular two-particle operator exchanged, 
but to a family of $AdS_5\times S^5$ operators that at tree level in supergravity have the same $l_{10}$ as defined below. 
%
%spin is actually a 10d spin decomposed in \ldots. 
%This is an important point on which we will come back in section .
%
We will then constrain the lower strata $\mathcal{M}_{n,l}$ in (\ref{stratification}) by assuming that the inequality $l_{10}\leq n$
holds not just in the flat space limit, but in the full $AdS_5\times S^5$.\footnote{This assumption turns out to be equivalent 
to the assumptions made in previous studies of genus zero corrections in \cite{Drummond:2019odu,Drummond:2020dwr}, 
whose results we will reproduce.}  

%where $l_{10}$ is suitably expressed in terms of the labels of the (long) two-particle 
%operators that participate in the OPE at tree level in supergravity.

\begin{figure}
\begin{center}
\begin{tikzpicture}[scale=.54]
%
%\draw[step=2cm,gray,very thin] (-4,2) grid (8,8);
%
\def\prop{.5}
\def\shifthor{\prop*2}
\def\ptuno{(\prop*2-\shifthor,\prop*8)}
\def\ptdue{(\prop*5-\shifthor,\prop*5)}
\def\pttree{(\prop*9-\shifthor,\prop*15)}
\def\ptquattro{(\prop*12-\shifthor,\prop*12)}

\draw[red,dashed,thick] (\prop*9-2*\shifthor,\prop*16) --(\prop*9-2*\shifthor,\prop*2);
\filldraw[ gray!50] (\prop*5+\prop*3-\shifthor, \prop*6+\prop*5-\prop*3)   -- \ptquattro -- \pttree -- (\prop*8+\prop*0-\shifthor, \prop*6+\prop*8-\prop*0) -- cycle;

\node[scale=1] (legend) at (10.5,5) {$\eta_{(m)}\Big|_{(\alpha')^{n+3}}=0\ \ \forall m>m^*$};
\node[scale=1] (legend) at (10.5,3) {$\langle \mathcal{O}_{p_i}\mathcal{O}_{p_j}|\mathcal{K}_{(m)}\rangle\Big|_{(\alpha')^{n+3}}=0\ \ \forall m> m^*$};
\node[scale=1] (legend) at (3.3,1) {$m^*$};

%
%axis horizontal
\draw[-latex, line width=.6pt]		(\prop*1   -\shifthor-4,         \prop*14          -0.5*\shifthor)    --  (\prop*1  -\shifthor-2.5  ,   \prop*14-      0.5*\shifthor) ;
\node[scale=.8] (oxxy) at 			(\prop*1   -\shifthor-2.5,  \prop*16.5     -0.5*\shifthor)  {};
\node[scale=.9] [below of=oxxy] {$p$};
%
%axis vertical
\draw[-latex, line width=.6pt] 		(\prop*1   -\shifthor-4,     \prop*14       -0.5*\shifthor)     --  (\prop*1   -\shifthor-4,        \prop*17-      0.5*\shifthor);
\node[scale=.8] (oxyy) at 			(\prop*1   -\shifthor-2,   \prop*16.8   -0.5*\shifthor) {};
\node[scale=.9] [left of= oxyy] {$q$};
%
%rectangle
\draw[] 								\ptuno -- \ptdue;
\draw[black]							\ptuno --\pttree;
\draw[black]							\ptdue --\ptquattro;
\draw[]								\pttree--\ptquattro;
%
%		
%dots
%
\foreach \indeyc in {0,1,2,3}
\foreach \indexc  in {2,...,9}
\filldraw   					 (\prop*\indexc+\prop*\indeyc-\shifthor, \prop*6+\prop*\indexc-\prop*\indeyc)   	circle (.07);
%
%letters
%
\node[scale=.8] (puntouno) at (\prop*4-\shifthor,\prop*8) {};
\node[scale=.8]  [left of=puntouno] {$A$};   
\node[scale=.8] (puntodue) at (\prop*5-\shifthor,\prop*6+.5) {};
\node[scale=.8] [below of=puntodue]  {$B$}; 
\node[scale=.8] (puntoquattro) at (\prop*13-\shifthor,\prop*15) {};
\node[scale=.8] [below of=puntoquattro] {$C$};
\node[scale=.8] (puntotre) at (\prop*9-\shifthor,\prop*13) {};
\node[scale=.8] [above of=puntotre] {$D$}; 

\node at  (8.5,\prop*13) {\phantom{space}};
													
\end{tikzpicture}
\caption{\small The rectangle $R_{\vec{\tau}}$ of operators $\mathcal{K}_{pq}$ which are degenerate at leading order. 
The lifting in supergravity is only partial with the anomalous dimension depending only on the column. At order $(\alpha')^{n+3}$ 
we impose that the CFT data of the operators in the grey area are uncorrected.}
\label{procedure_intro}
\end{center}
\end{figure}
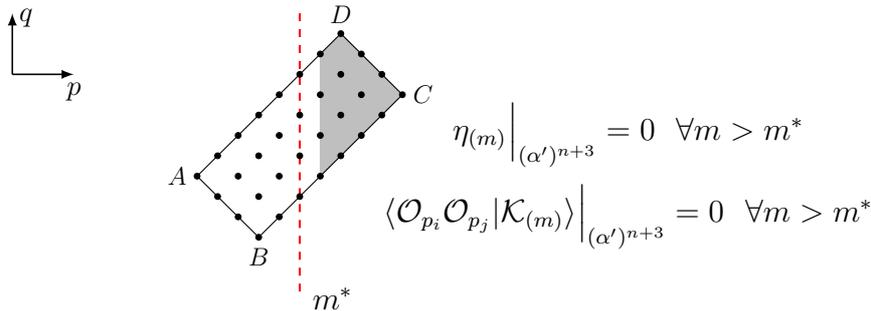
%

%As described in \cite{Aprile:2017xsp}, 
For fixed free theory quantum numbers\footnote{The 
twist $\tau_{}=\Delta_{free}-l$ in free theory, four dimensional spin $l$ and $su(4)$ rep $[aba]$.}
$\vec{\tau}=(\tau_{},l,[aba])$ there are as many two-particle operators ${\cal K}_{(pq),\vec{\tau}}$ as pairs $(pq)$ in 
a certain rectangle $R_{\vec{\tau}}$ in the $(p,q)$-plane \cite{Aprile:2018efk}. This is exemplified in Fig. \ref{procedure_intro}.
As shown in \cite{Aprile:2017xsp,Aprile:2018efk,Caron-Huot:2018kta}, 
the operators ${\cal K}_{(pq),\vec{\tau}}$ are defined by unmixing a basis of degenerate two-particle 
operators in free theory $\mathcal{O}_{(pq),\vec{\tau}} = \mathcal{O}_p \partial^l \Box^{\frac{1}{2}(\tau-p-q)} \mathcal{O}_q$. 
This yields the rotation matrix ${\cal O}_{(pq)}\rightarrow {\cal K}_{(pq)}$ and the leading correction to the 
dimensions $\Delta = \Delta_{free} + \tfrac{2}{N^2} \eta^{(0)}_{(pq),\vec{\tau}}$. 
The result of \cite{Aprile:2018efk} for the anomalous dimension in supergravity is 
\begin{align}\label{intro_ell10d}
	\eta^{(0)}_{(pq),\vec{\tau}} = -\frac{ 2\delta^{(8)}_{\tau,l,[aba]}}{ (l_{10}+1 )_6}\qquad;\ \qquad
	%\begin{array}{rl} 
	%\displaystyle 
	%l_{10} &= l + a +2\, m_{\mathcal{K}_{(pq)}}- \tfrac{3+(-1)^{a+l}}{2}\,.
	l_{10} &= l + a +2\,m_{\mathcal{K}_{(pq)}}
- \tfrac{1+(-1)^{a+l}}{2}-1\,
	%\\[.2cm]  \displaystyle %m_{\mathcal{K}_{(pq)}} &\equiv p-a-1 
	%\end{array}
\end{align}
where the normalisation $\delta^{(8)}_{\tau,l,[aba]}$ is given in \eqref{delta8_factor} and depends 
only on the quantum numbers $\vec{\tau}$ and not on the rectangle $R_{\vec{\tau}}$. 
Instead $m_{\mathcal{K}_{(pq)}}=1,2,\ldots$ counts the number of columns in the rectangle $R_{\vec{\tau}}$, it is defined by  $m_{\mathcal{K}_{(pq)}}\equiv p-a-1$, and will be 
called  the \emph{level-splitting label} of the operator $\mathcal{K}_{(pq),\vec{\tau}}$.

Remarkably, the tree level anomalous dimension $\eta^{(0)}_{(pq)}$ do not distinguish operators $\mathcal{K}_{(pq)}$
%only depends on $p$ (or equivalently $m$) and not on 
with different value of $q$, since they only depend on the column index, $m_{\mathcal{K}_{(pq)}}$. 
The unmixing in supergravity is thus not complete and a partial degeneracy remains. 
The partial degeneracy is then explained by the an accidental 10d \emph{conformal} symmetry of $AdS_5\times S^5$ supergravity \cite{Caron-Huot:2018kta}. 
The number of degenerate operators with a given $\eta^{(0)}_{(pq)}$ is counted by the number of points in a column $m=const.$ in $R_{\vec{\tau}}$.

Under our assumption, $l_{10}\le n$, and the relation
\begin{align}\label{master_relation_intro}
m_{\mathcal{K}_{pq}}\leq m^*_{}\qquad;\qquad
m^*(a,l,n)=\tfrac{1}{2} \bigl(n-(a+l)-\tfrac{1}{2}(1-(-1)^{a+l}) \bigr)+1
\end{align}
follows. We conclude that the VS amplitude at order $(\alpha')^{n+3}$ contributes 
to the CFT data of all operators $\mathcal{K}_{(pq),\vec{\tau}}\in R_{\vec{\tau}}$ 
with level-splitting label $m\leq m^*$. Note that if we pick a rectangle $R_{\vec{\tau}}$ and a level splitting label $m$, and we scan over the $\alpha'$ expansion,
operators with level splitting label $m=m^*(a,l,n)$ become visible \emph{for the first time} 
precisely at order $(\alpha')^{n+3}$.
We expect that $\mathcal{V}_n(\alpha')^{n+3}$ is responsible for lifting their residual degeneracy, thus exhibiting the breaking 
of the 10d conformal symmetry by $\alpha'$ corrections.

On the other hand, the amplitude $\mathcal{V}_n(\alpha')^{n+3}$ is   blind to operators $\mathcal{K}_{(pq),\vec{\tau}}$ 
with $m> m^*$ (depicted in the grey-shaded area in Fig. \ref{procedure_intro}). This means that their dimensions and three-point 
functions are uncorrected at that order.  Consequently, it will follow %from the above constraints 
that the rank of the matrix $\mathbf{N}_{\vec{\tau}}$, from which we compute the anomalous dimensions of the two-particle 
operators with $m\leq m^*$, is simply given by the number of pairs $(pq)$ in the unshaded area, as discussed in section \ref{spectum_sect}.
Thus,
our second main improvement over the methodology of \cite{Drummond:2020dwr} is simply to impose the expected 
value for the rank on the CFT data obtained from our ansatz for $\mathcal{V}_n$. This information goes beyond $\mathcal{M}_{n,n}^{flat}$ 
and thus the rank constraints become constraints on the strata $\mathcal{M}_{n,\ell<n}$. 

The rank constraints fix the Mellin amplitudes up to a handful 
of ambiguities, at least at low orders in $\alpha'$. For low orders, results obtained from localisation can be used to further 
constrain the result \cite{Binder:2019jwn,Chester:2019pvm}. The results we obtain exhibit some further remarkable simplicity, and 
motivate an integral representation which generalises the one originally proposed by Penedones in relation to the flat space limit \cite{Penedones:2010ue}. 
In turn, the integral representation hints at a possible $AdS_5 \leftrightarrow S^5$ symmetry which we also see exhibited at the level of the spectrum.

Quite remarkably, the CFT data living on the edge at  $m=m^*$ are \emph{fully determined} 
by imposing the rank constraints. The CFT data for operators with $m<m^*$ instead is affected by ambiguities 
which cannot be resolved by the tree level bootstrap. However, the ambiguities are totally irrelevant for the level 
splitting problem at  $m=m^*$ since they have to do with shifts of $\mathcal{V}_n$ by terms with $l_{10}<n$.
We tested this statement for $(\alpha')^{5,7,9}$ and in section \ref{tailoring_sec} we will given an argument 
to explain the general mechanism behind it.

A new feature of the level splitting problem is the following: If we keep $m^*$ fixed,  since $m^*\sim n-l$,  we find that as the 
spin $l$ increases, so must $n$. In other words, we have 
to increase the order of the $\alpha'$ expansion in order to study operators with large spin.
This new feature is quite intriguing because, in combination with analyticity in $\vec{\tau}$ of the 
characteristic polynomial of the mixing matrix, it allows us to extract information about 
the spectrum of two-particle operators at \emph{all orders} in $\alpha'$.
We give a glimpse from section \ref{sec_mstar2} of the simplicity of the level splitting problem by looking 
at the characteristic polynomial for $m^*=2$ and $a+l$ even,\footnote{$T$ and $B$ are simply related 
to the twist $\tau$ and $b$, respectively. $T\equiv\tfrac{1}{4}\tau(\tau+2l+4)$, $B\equiv \tfrac{1}{4}b(b+2a+4)$.}
\begin{gather}\label{sample_intro_ch_poly}
(\tilde{\eta}+r )^2- (\tilde{\eta}+r )\frac{(n+2)(n+3)}{2n+5}\gamma(T,B)_{} + \frac{(n+2)^2(n+3)^2}{2n+5}BT =0 \\[.2cm]
r={ (T-B)^2}{} +  (2+l)B +(2+a)T \quad;\quad
\gamma_{}=(2l+5)B + (2a+5)T-  (a+2)(l+2) \notag
\end{gather}
The roots in $\tilde{\eta}$ of this polynomial will give the split anomalous dimensions of the corresponding operators ${\cal K}$, as function of $T$ and $B$.
Our result here is valid at all orders in $(\alpha')^{n+3}$ with $n$ even! It generalises the result 
of \cite{Drummond:2020dwr} where the first splitting of two residually degenerate states with $a=l=0$ 
was observed at order $(\alpha')^5$. Note the presence of an $AdS_5 \leftrightarrow S^5$ symmetry under 
the exchange $T \leftrightarrow B$ and $a \leftrightarrow l$.

Following the outline of this introduction, the bulk of the paper is divided mainly into two parts.
In section \ref{VS_section} we discuss the $AdS_5\times S^5$ Mellin representation of the VS amplitude, 
in the $\alpha'$ expansion, and we show explicitly our solutions for $(\alpha')^{5,6,7}$, and comment 
on $(\alpha')^{8,9}$ (attached in an ancillary file). In section \ref{spectum_sect}, we describe the details 
and the ideas that led us to our algorithm. This amounts to explain the spectral properties of the VS amplitude, 
and the level splitting problem, encoded in the  characteristic polynomial.
More features of the characteristic polynomial for general $m^*$ will be explained in section \ref{sec_mstar3}.\\

{\bf Note added:} While our work was being completed, 
we were informed by the authors of \cite{NEWPAUL} about their beautiful results, 
which nicely complement ours. We thank them for coordinating the release on the arXiv.

%%==============================================================================
%%==============================================================================

\section{Virasoro-Shapiro amplitude in AdS$_5\times$S$^5$}\label{VS_section}

%%==============================================================================
%%==============================================================================

We split the correlators of single particle operators\footnote{These single-particle half-BPS 
operators ${\cal O}_p$ are those identified in \cite{Aprile:2018efk} and studied in \cite{Aprile:2020uxk}. 
For $\alpha'$ corrections at tree level the difference with single-trace half-BPS operators is not important.} $\langle \cO_{p_1} \cO_{p_2} \cO_{p_3} \cO_{p_4}\rangle$
into free and dynamical part, according to the partial non renormalisation theorem of \cite{Eden:2000bk}. Then we define the amplitude of the correlator, 
in position space, by stripping off a {\tt kinematic} factor from the dynamical contribution. The Mellin transform of this amplitude is performed 
by using the $AdS_5\times S^5$  kernel of $\Gamma$ functions denoted by $\Gamma_{\otimes}$.  
Details are reviewed in appendix \ref{CONVENTIONS}. Analyticity of the Mellin amplitude will be manifest in our conventions. 
The Virasoro-Shapiro amplitude in Mellin space will be the focus of this section.

Unlike the supergravity contribution in (\ref{ini_M}), which has simple poles to cancel unwanted 
string states from recombined connected free theory contributions, we expect the $\alpha'$ corrections to the Mellin amplitude 
to be polynomial at each order. This is because all possible poles are already contained in $\Gamma_{\otimes}$ in the Mellin integral.
Thus on very general grounds we expect to accommodate  the $AdS_5\times S^5$ version of the VS amplitude 
in the polynomial ansatz, 
\be
\mathcal{V}_{n} = 
\sum_{\ell=0}^{n-1} (\Sigma-1)_{\ell+3} 
\left( \sum_{  0 \leq d_1+ d_2\leq \ell }^{} C^{(n)}_{\ell;\, d_1d_2}(\tilde s,\tilde t,\vec{p}) \, {\bf s}^{d_1}{\bf t}^{d_2} \right) +
 (\Sigma-1)_{n+3}\, \mathcal{M}_{n,n}^{flat} ({\bf s},{\bf t},{\bf u})
\label{VS_ansatz_1}
\ee
where $\vec{p}$ will abbreviate $\vec{p}=p_1,p_2,p_3,p_4$ and $C^{(n)}_{\ell;\, d_1d_2}$ are polynomial coefficient functions  determined as we explain in this section.

The first piece of data in the VS amplitude is $\mathcal{M}_{n,n}^{flat}$ and is obtained by covariantising the 
flat space VS amplitude, as shown in \cite{Aprile:2020luw}. To do so we start from the flat VS amplitude 
where it will be important to keep $u$ as an independent variable w.r.t.~$s,t$. 
Therefore define $\mathcal{M}_{n,n}^{flat}$ by the relation
\begin{align}
\rule{.5cm}{0pt}
\mathcal{M}^{flat}_{n,n}( s,t,u)\Bigg|_{u=-s-t}= \left[ \frac{1}{ s t u }\, 
e^{ \,2\sum_{k\ge 1} \frac{\zeta_{2k+1}}{ 2k+1} (\alpha')^{2k+1}\left({s}_{}^{2k+1} +{t}_{} ^{2k+1}+{u}_{}^{2k+1}\right)}  \right]\Bigg|^{(\alpha')^{n+3}}_{u=-s-t}\,.
\end{align}
This $\mathcal{M}_{n,n}^{flat}$ 
is a symmetric homogenous degree $n$ polynomial in the variables, $s,t,u$.
Then, $\mathcal{M}^{flat}_{n,n}$ so defined enters $\mathcal{V}_n$ evaluated as $\mathcal{M}^{flat}_{n,n}( {\bf s},{\bf t},{\bf u})$ where now
\begin{align}\label{var_bolds}
{\bf s}=s+\tilde s\qquad;\qquad {\bf t}=t+\tilde t\qquad;\qquad {\bf s}+{\bf t}+{\bf u}=-4\,.
\end{align}

In the large $p$ limit both Mellin variables and charges scale in the same way, let's say with $p$. Thus the 
large $p$ limit of $\mathcal{V}_n$ is $\Sigma^{n+3}\mathcal{M}^{flat}_{n,n}$ 
by construction, and enjoys a 10d symmetry. The completion 
of it in $AdS_5\times S^5$ has more structures, which we parametrise in strata $\mathcal{M}_{n,\ell}$
w.r.t.~$\ell=n$, with the following definition
\begin{mdframed}
\begin{center}
$\mathcal{M}_{n,\ell}({\bf s},{\bf t}, {\bf u},\tilde s,\tilde t, \vec{p})$ 
is a crossing symmetric polynomial
in all its variables, of maximum degree $n$, such that only monomials of 
degree $\ell$ in ${\bf s}$, ${\bf t}$ and ${\bf u}$ appear.
\end{center}
\end{mdframed}
Therefore $\mathcal{M}_{n,\ell}$ is not an homogeneous polynomial, 
but of course can be written recursively,
\begin{align}
span\big(\,\mathcal{M}_{n,\ell}\,\big)= span\big(\,\mathcal{H}_{n,(\ell,n-\ell)}\,,\, \mathcal{M}_{n-1,\ell}\big)
\end{align}
by isolating each time a new homogeneous polynomial. 
In fact, 
\begin{mdframed}
\begin{center}
$\mathcal{H}_{n, (\ell,n-\ell)}({\bf s},{\bf t},{\bf u},\tilde s,\tilde t, \vec{p})$ is  a crossing symmetric polynomial
in all its variables, of fixed degree $n$, such that only monomials of degree $\ell$ in ${\bf s}$, ${\bf t}$ and ${\bf u}$ appear.
\end{center}
\end{mdframed}
Consequently, $\mathcal{H}_{n, (\ell,n-\ell)}$ has degree $n-\ell$ in all other variables ${\tilde s},\tilde t,$ and $p_1p_2p_3p_4$.

The next task is to read off an ansatz for $\mathcal{M}_{n,\ell}$ out of the most general ansatz we wrote above, 
\begin{align}
\sum_{0 \leq d_1+ d_2\leq \ell }^{} C^{(n)}_{\ell;\, d_1d_2}(\tilde s,\tilde t, \vec{p}) \, {\bf s}^{d_1}{\bf t}^{d_2} \qquad\rightarrow\qquad \mathcal{M}_{n,\ell}\qquad
\end{align}
Notice that the starting point on the l.h.s.~is a sum over monomials ${\bf s}^{d_1}{\bf t}^{d_2}$, 
but rather than $d_1+ d_2=\ell$, we do need to include all lower powers $d_1+ d_2\leq \ell$. 
The reason is that a crossing symmetric polynomial will depend on ${\bf s,t},$ and ${\bf u}=-{\bf s}-{\bf t}-4$, 
thus any power of ${\bf u}$ brings down lower powers of ${\bf s}$ and ${\bf t}$ in the stratum.
To restrict the ansatz we will need crucially crossing symmetry.

In order to proceed we first have to understand
the polynomial $C^{(n)}_{\ell;\, d_1d_2}(\tilde s,\tilde t , \vec{p})$,
and to do so we will now argue what is the expected scaling with $p$. 
Observe that the leading term in $\mathcal{V}_{n}\sim \Sigma^{n+3} \mathcal{M}_{n,n}^{flat}$ 
scales like $p^{2n+3}$, thus the next-to-leading term should scale at most as 
$p^{2n+3-1}$ in order not to conflict with the large $p$ limit.
But as far as powers of ${\bf s}$ and ${\bf t}$ are concerned, $\mathcal{M}_{n,n-1}$
is analogous to the top term of $\mathcal{V}_{n-1}$, and would be equivalent if $C^{(n)}_{\ell}$ 
was just a constant. If this is non constant then we should find that $\mathcal{M}_{n-1,n-1}$ scales 
at least with one more power than $\Sigma^{n+2}\mathcal{M}_{n,n-1}$,
in order for it to contribute with new terms. That's indeed the case $p^{[2(n-1)+3]+1}=p^{2n+3-1}$. 
If we apply recursively this argument for each strata we fill  a table as follows,
 \be
 \begin{array}{c|ccccc}
 \ell & 0 & 1 & 2 & 3 & \ldots \\[.2cm]
 \hline\\[-.2cm]
 \mathcal{V}_0 & p^3 \\[.2cm]
 \mathcal{V}_1 & p^4 & p^5 \\[.2cm]
 \mathcal{V}_2 & p^5 & p^6  & p^7 \\[.2cm]
 \mathcal{V}_3 & p^6 & p^7  & p^8 & p^9  \\
 \vdots &  
 \end{array}
 \ee
Extracting the various pochhammers $(\Sigma-1)_{\ell+3}$ we simply find $p^n$, and we deduce that 
\begin{align}
\lim_{p\rightarrow \infty} C^{(n)}_{\ell;\, d_1d_2}(\tilde s,\tilde t, \vec{p})\ \sim\ p^{(n-\ell)}.
\end{align}
In formulas, 
\begin{align}
C^{(n)}_{\ell;\, d_1d_2}=\sum_{ 0 \leq \delta_1+\delta_2\leq (n-\ell) } c^{(n)}_{\ell;\, d_1d_2, \delta_1\delta_2}( \vec{p})\, \tilde{s}^{\delta_1} \tilde{t}^{\delta_2}
\end{align}
where finally $c^{(n)}_{\ell;\, d_1d_2, \delta_1\delta_2}$ is a polynomial in 
$p_1,p_2,p_3,p_4$ of max degree $(n-\ell)-\delta_1-\delta_2$.

With the information about $C^{(n)}_{\ell}$ at hand, we can impose crossing symmetry. 
This is a statement about the  full correlator  and in particular about the equality 
\begin{align}
\langle {\cal O}_{p_1}({\tt x }_{\sigma_1}) {\cal O}_{p_2}({\tt x }_{\sigma_2}) {\cal O}_{p_3}({\tt x }_{\sigma_3}) {\cal O}_{p_4}({\tt x }_{\sigma_4}) \rangle=
\langle {\cal O}_{p_{\sigma_1}}({\tt x }_{1}) {\cal O}_{p_{\sigma_2}}({\tt x }_{2}) {\cal O}_{p_{\sigma_3}}({\tt x }_{3}) {\cal O}_{p_{\sigma_4}}({\tt x }_{4}) \rangle
\end{align}
The possible permutations $\sigma$ are six. 
Considering then the Mellin transform of the correlator,
as defined in appendix \ref{CONVENTIONS},
we deduce what relations the Mellin amplitude satisfies
\be
\begin{array}{rcl}
\mathcal{M}(s,u,\tilde s,\tilde u; p_2,p_1,p_3,p_4)&=&\mathcal{M}(s,t,\tilde s,\tilde t; p_1,p_2,p_3,p_4) \\[.2cm]
\mathcal{M}(t,s,\tilde t,\tilde s; p_1,p_4,p_3,p_2)&=&\mathcal{M}(s,t,\tilde s,\tilde t; p_1,p_2,p_3,p_4) \\[.2cm]
\mathcal{M}(u,t,\tilde u,\tilde t; p_4,p_2,p_3,p_1)&=&\mathcal{M}(s,t,\tilde s,\tilde t; p_1,p_2,p_3,p_4) \\[.2cm]
\mathcal{M}(s,u-c_u,\tilde s,\tilde u+c_u; p_1,p_2,p_4,p_3)&=&\mathcal{M}(s,t,\tilde s,\tilde t; p_1,p_2,p_3,p_4) \\[.2cm]
\mathcal{M}(t-c_t,s-c_s,\tilde t+c_t,\tilde s+c_s; p_3,p_2,p_1,p_4)&=&\mathcal{M}(s,t,\tilde s,\tilde t; p_1,p_2,p_3,p_4) \\[.2cm]
\mathcal{M}(u-c_u,t,\tilde u+c_u,t; p_1,p_3,p_2,p_4)&=&\mathcal{M}(s,t,\tilde s,\tilde t; p_1,p_2,p_3,p_4) \\[.2cm]
\end{array}
\rule{1.2cm}{0pt}
\ee

The best we can do at this point is to make crossing symmetry 
manifest by identifying variables such that the transformations above
act in a `block diagonal' form. The large $p$ limit suggests first to pick ${\bf s,t,u}$ 
and we will accompany this with another set. In total
\begin{gather}
%\begin{array}{c}
{\bf s}=s+\tilde s\qquad;\qquad {\bf t}=t+\tilde t\qquad\qquad {\bf s}+{\bf t}+{\bf u}=-4   \notag \\[.3cm]
\,\tilde{\bf s}=c_s+2\tilde s\qquad;\qquad \,\tilde{\bf t}=c_t+2\tilde t\qquad;\qquad \tilde{\bf s}+\tilde{\bf t}+\tilde{\bf u}=\Sigma-4 \\[.3cm]
\begin{array}{c}
c_s=\tfrac{ p_1+p_2-p_3-p_4}{2}\quad;\quad 
c_t=\tfrac{p_1+p_4-p_2-p_3}{2} \quad;\quad 
c_u=\tfrac{ p_2+p_4-p_3-p_1}{2}\quad;\quad  
\Sigma=\tfrac{p_1+p_2+p_3+p_4}{2} 
\end{array}\notag
\end{gather}
In these variables, crossing becomes
\begin{align}
\begin{array}{rcl}
\mathcal{M}({\bf s},{\bf u},{\bf t},\tilde{\bf s},\tilde{\bf u},\tilde{\bf t},+c_s,+c_u,+c_t,\Sigma)&=&\mathcal{M}({\bf s},{\bf t},{\bf u},\tilde{\bf s},\tilde{\bf t},\tilde{\bf u},c_s,c_t,c_u,\Sigma)\\[.2cm]
\mathcal{M}({\bf t},{\bf s},{\bf u},\tilde{\bf t},\tilde{\bf s},\tilde{\bf u},+c_t,+c_s,+c_u,\Sigma)&=&\mathcal{M}({\bf s},{\bf t},{\bf u},\tilde{\bf s},\tilde{\bf t},\tilde{\bf u},c_s,c_t,c_u,\Sigma)\\[.2cm]
\mathcal{M}({\bf u},{\bf t},{\bf s},\tilde{\bf u},\tilde{\bf t},\tilde{\bf s},+c_u,+c_t,+c_s,\Sigma)&=&\mathcal{M}({\bf s},{\bf t},{\bf u},\tilde{\bf s},\tilde{\bf t},\tilde{\bf u},c_s,c_t,c_u,\Sigma)\\[.2cm]
\mathcal{M}({\bf s},{\bf u},{\bf t},\tilde{\bf s},\tilde{\bf u},\tilde{\bf t},+c_s,-c_u,-c_t,\Sigma)&=&\mathcal{M}({\bf s},{\bf t},{\bf u},\tilde{\bf s},\tilde{\bf t},\tilde{\bf u},c_s,c_t,c_u,\Sigma)\\[.2cm]
\mathcal{M}({\bf t},{\bf s},{\bf u},\tilde{\bf t},\tilde{\bf s},\tilde{\bf u},-c_t,-c_s,+c_u,\Sigma)&=&\mathcal{M}({\bf s},{\bf t},{\bf u},\tilde{\bf s},\tilde{\bf t},\tilde{\bf u},c_s,c_t,c_u,\Sigma)\\[.2cm]
\mathcal{M}({\bf u},{\bf t},{\bf s},\tilde{\bf u},\tilde{\bf t},\tilde{\bf s},-c_u,+c_t,-c_s,\Sigma)&=&\mathcal{M}({\bf s},{\bf t},{\bf u},\tilde{\bf s},\tilde{\bf t},\tilde{\bf u},c_s,c_t,c_u,\Sigma)
\end{array}
\end{align}
Each set of three transforms in the same way, modulo $\pm1$ signs. $\Sigma$ is obviously singlet.

%%==============================================================================
%%=================================================

\subsection{Genus zero amplitudes from the bootstrap} %towards VS

%%=================================================
%%==============================================================================

The combination of crossing symmetry and large $p$ stratification, 
discussed in the previous section, provides us with the initial ansatz for 
the VS amplitude in $AdS_5\times S^5$.  The results are summarised by the formula
\begin{align}
&
\mathcal{V}_{n} = 
\sum_{\ell=0}^{n-1} (\Sigma-1)_{\ell+3}\ \mathcal{M}_{n,\ell}({\bf s},{\bf t},{\bf u},\tilde{\bf s},\tilde{\bf t},\tilde{\bf u},c_s,c_t,c_u,\Sigma)+
 (\Sigma-1)_{n+3}\, \mathcal{M}_{n,n}^{flat} ({\bf s},{\bf t},{\bf u})
\end{align}
where recursively we get
\begin{align}\label{recursion_M_main}
span\big(\,\mathcal{M}_{n,\ell}\,\big)= span\big(\,\mathcal{H}_{n,(\ell,n-\ell)}\,,\, \mathcal{M}_{n-1,\ell}\big)
\end{align}
by constructing $\mathcal{H}_{n, (\ell,n-\ell)}({\bf s},{\bf t},{\bf u},\tilde{\bf s},\tilde{\bf t},\tilde{\bf u}, c_s,c_t,c_u,\Sigma)$, which is
a homogeneous polynomial of degree $n$ such that only monomials of degree $\ell$ in ${\bf s}$, ${\bf t}$ and ${\bf u}$ appear.

The $(\alpha')^3$ amplitude is just a constant, since it would correspond to $\mathcal{H}_{0,(0,0)}$. 
The $(\alpha')^4$ amplitude vanishes in flat space, but in $AdS_5\times S^5$ can in principle get a 
contribution from $\mathcal{H}_{1,(0,1)}$ and $\mathcal{H}_{1,(1,0)}$. Notice that only $\mathcal{H}_{1,(0,1)}$ exists, 
and it is spanned just by $\Sigma$. Indeed, the constraints on ${\bf s}+{\bf t} +{\bf u}$, $\tilde{\bf s}+\tilde{\bf t} +\tilde{\bf u}$ 
prevent other terms to be present at this order. 
The construction of all possible 
terms which can contribute to the amplitude is therefore quite interesting.  In appendix \ref{app_ansatz} we describe the method we used in this paper. 
A counting of initial parameters is given in the table below.  
The notation $|\mathcal{H}|$ denotes the number of crossing invariant terms,

\be
\begin{tikzpicture}[scale=1]
\node[scale=.8] at (0,0)  {$\begin{array}{|c|c|c|c|c|c|c|}
  		(\alpha')^3 & (\alpha')^4 & (\alpha')^5 & (\alpha')^6 & (\alpha')^7 & (\alpha')^8  & (\alpha')^9  \\[.2cm]
		 \hline
	\rule{0pt}{.6cm}  |\mathcal{H}_{0,(0,0)}|=1& |\mathcal{H}_{1,(0,1)}|=1& |\mathcal{H}_{2,(0,2)}|=3&  |\mathcal{H}_{3,(0,3)}|= 6 &  |\mathcal{H}_{4,(0,4)}|= 11  &  |\mathcal{H}_{5,(0,5 )}|= 18 &  |\mathcal{H}_{6,(0,6 )}|= 32 \\[.2cm]
		 \hline
		\rule{0pt}{.6cm} & | \mathcal{H}_{1,(1,0)}|=0 & |\mathcal{H}_{2,(1,1)}|=1 &  |\mathcal{H}_{3,(1,2)}|=3  & |\mathcal{H}_{4,(1,3)}|=  6  &  |\mathcal{H}_{5,(1,4)}|= 14 &  |\mathcal{H}_{6,(1,5)}|= 26  \\[.2cm]
		 \hline
		\rule{0pt}{.6cm}  & & |\mathcal{H}_{2,(2,0)}|=1	&  |\mathcal{H}_{3,(2,1)}|=2  & |\mathcal{H}_{4,(2,2)}|=  6  &  |\mathcal{H}_{5,(2,3)}|=  12 &  |\mathcal{H}_{6,(2,4)}|=25 \\[.2cm]
		\hline
		\rule{0pt}{.6cm} 			&	&	&  |\mathcal{H}_{3,(3,0)}|=1  &  |\mathcal{H}_{4,(3,1)}|= 2  &  |\mathcal{H}_{5,(3,2)}|= 6\,  &  |\mathcal{H}_{6,(3,3)}|=14 \\[.2cm]
		\hline
		\rule{0pt}{.6cm} 			&	&	&  					     &    |\mathcal{H}_{4,(4,0)}|=1 &  |\mathcal{H}_{5,(4,1)}|= 3\, &  |\mathcal{H}_{6,(4,2)}|= 9 \\[.2cm]	
		\hline 	
		\rule{0pt}{.6cm} 			&	&	&  					     &   					   & |\mathcal{H}_{5,(5,0)}|= 1    &  |\mathcal{H}_{6,(5,1)}|=3 \\[.2cm]	
		\hline 	
		\rule{0pt}{.6cm} 			&	&	&  					     &   					   & 					       &  |\mathcal{H}_{6,(6,0)}|=2 \\[.2cm]										
 \end{array}$};
\end{tikzpicture}
\label{tabella_H}
\ee
The ansatz for $\mathcal{M}_{n,\ell}$ is obtained from the recursion in \eqref{recursion_M_main},
 so that the total number of terms is given by summing along the rows, from right to left.

We point out that inside a given $\mathcal{H}_{n,(\ell,n-\ell)}$ we can add another level, which is the 
one given by terms of the form $(\Sigma^\#\,\times$ crossing invariants$)$, where usually the latter already appeared at previous orders. 
For example, 
\begin{align}
span( \, \mathcal{H}_{3,(2,1)} \, )=\big\{\,
{\bf s}^2 \tilde{\bf s} +{\bf t}^2 \tilde{\bf t} + {\bf u}^2\tilde{\bf u}\ ,\ \Sigma\times\left({\bf s}^2+ {\bf t}^2+{\bf u}^2\right)  \big\}
\end{align}
Notice that the terms of the form $(\Sigma^\#\,\times$ crossing invariants$)$ are the 
first instance of the more general class of terms of the form $($crossing invariant$)\times($crossing invariant$)$. 
In the case above one of the two is simply a power of $\Sigma$.

The next step is to impose constraints on the free parameters in our initial ansatz, 
at each order in the $\alpha'$ expansion.  The idea, already sketched in 
section \ref{intro_sec}, is to solve  for the VS amplitude in perturbation theory 
by using polynomiality in Mellin space and imposing a bound on the spectrum of 
two-particle operators visible by ${\cal V}_n$. This statement translates into a statement on the rank 
of a certain matrix of CFT data. Let us point out that there will be an infinite number of constraints, but finitely 
many parameters in our ansatz. The outcome will be our proposal for the VS amplitude in 
$AdS_5\times S^5$ up to certain ambiguities, at its first stage. Indeed, we know from the very beginning that we will not be able 
to fix the ambiguity of adding previous amplitudes $\mathcal{V}_{k\leq n-1}$ to our result for $\mathcal{V}_{n}$, 
within the bootstrap. 
Nevertheless, the problem of finding the CFT data at the edge at
$m=m^*$ \emph{is fully determined} at each order in $\alpha'$, therefore for each new amplitude that we bootstrap, 
we can extract novel CFT data out of it, and feed this new data into 
the OPE relations governing the amplitudes at higher orders,
thus reducing the number of free parameters at the first stage.

%=====================================================================================
%================================================================

\subsection{Explicit results and remarkable simplifications}\label{sec_solutions}

%================================================================
%=====================================================================================

Our results and main observations about the amplitudes up to order $(\alpha')^{9}$ are presented in this section. 
We begin by revisiting the amplitude at $(\alpha')^5$ first found in \cite{Drummond:2020dwr}.  
This case is simple enough to see how much the rank constraints fix the amplitude, leaving nevertheless 
some ambiguities. Then we show explicitly the new results at $(\alpha')^{6}$ and before moving on to 
$(\alpha')^{7,8}$ we observe additional simplicity in the structure of the results. This simplicity will be nicely 
packaged into an integral transform which allows us to rewrite the initial results in a remarkably compact form.

To fully fix the VS amplitude, i.e.~the ambiguities, we will need additional input. One source of such information is 
the relation between the integrated correlators and derivatives of the partition function w.r.t.~deformations 
of $\mathcal{N}=4$ SYM on the sphere, computed by supersymmetric localisation \cite{Binder:2019jwn,Chester:2019pvm,Chester:2019,SHAI}. 
Some ambiguities can be fixed with the currently available data, and our formalism will make more 
transparent how these  contribute. For example, at $(\alpha')^4$ the flat space contribution vanishes
but we do find a non zero ansatz in $AdS_5\times S^5$, consisting of $k_1 + \Sigma\times k_2$. 
These constants are set to zero by localisation \cite{Binder:2019jwn}. 
Independently, the rank constraints will also set to zero the term $\Sigma\times k_2$.
For the amplitude at $(\alpha')^5$, localisation results were already used in \cite{Drummond:2020dwr}.

%======================================================================================
\subsubsection*{Warming up with $(\alpha')^{5,6}$}%~\\[-.2cm]
%======================================================================================

The parametrisation of the VS amplitude at $(\alpha')^5$ is
\begin{equation}
\mathcal{V}_2=\ (\Sigma-1)_3 \mathcal{M}_{2,0}+(\Sigma-1)_{4} \mathcal{M}_{2,1}+(\Sigma-1)_{5}\times( {\bf s}^2 + {\bf t}^2+{\bf u}^2 )
\label{V2}
\end{equation}
with the strata given by
\begin{align}
\mathcal{M}_{2,0}=	&\ \ltr_{3,1}^{} \Sigma^2+ \ltr_{3,2}^{}(c_s^2+c_t^2+c_u^2) + \ltr_{3,3}^{}\left( \tilde{\bf s}^2+\tilde{\bf t}^2+{\bf \tilde{u}}^2 \right)+ \ltr_{3,4}^{} \Sigma  + \ltr_{3,5}^{}\\[.2cm]
\mathcal{M}_{2,1}=	&\ 
				\ltr_{4,1}^{} \left( {\bf s}\tilde{\bf s} + {\bf t}\tilde{\bf t}+ {\bf u}\tilde{\bf u} \right).\notag
\end{align}
This form was also presented in \cite{Aprile:2020luw}. The rank constraints impose
\be\label{rank_constraints_V2}
\ltr_{4,1}^{}=-5\qquad;\qquad \ltr_{3,3}^{}=5\qquad;\qquad \ltr_{3,2}^{}-\ltr_{3,1}^{}=11\qquad;\qquad \ltr_{3,4}=0
\ee
and there are two free parameters, $k_{3,5}$ is a constant, 
as the amplitude at $(\alpha')^3$,
the other, say $\ltr_{3,1}$, goes with the combination $\Sigma^2+c_s^2+c_t^2+c_u^2$.
Localisation results then imply \cite{Drummond:2020dwr},
\be
\ltr_{3,1}=-\tfrac{27}{2}\qquad;\qquad \ltr_{3,5}=\tfrac{33}{2}\,.
\label{locV2}
\ee
In section \ref{tailoring_sec} we will also point out that the CFT data at the edge 
is uniquely determined by \eqref{rank_constraints_V2}
regardless of the value of $\ltr_{3,1}^{}$ and $\ltr_{3,5}^{}$.\\%[.15cm]

%======================================================================================
%\paragraph{Amplitude at $(\alpha')^6$.}~\\[-.2cm]
%======================================================================================

The parametrisation of the VS amplitude at $(\alpha')^6$ is
\begin{align}\label{VS_3}
%\!\!\!\!\!\!\!
\mathcal{V}_3%\left[ \substack{ \displaystyle {\bf s,t, u} \\ \displaystyle \tilde s, \tilde t, \tilde u };c_s,c_t,c_u,\Sigma\right]
=(\Sigma-1)_3 \mathcal{M}_{3,0}+(\Sigma-1)_4\mathcal{M}_{3,1}+(\Sigma-1)_5\mathcal{M}_{3,2}+(\Sigma-1)_6\times  \tfrac{2}{3}\big({\bf s}^3+{\bf t}^3+{\bf u}^3 \big)
\end{align}
with the covariantised flat space amplitude $\mathcal{M}^{flat}_{3,3}$, 
and the lower strata given by
\begin{align}
& 
\!\!\!\!\!\mathcal{M}_{3,2}=\ltro_{5,1} \left({\bf s}^2\tilde{\bf s} + {\bf t}^2\tilde{\bf t} +{\bf u}^2\tilde{\bf u}  \right)+ \left(  {\bf s}^2+ {\bf t}^2+{\bf u}^2\right)\left( \Sigma  \ltro_{5,2} + \ltro_{5,3} \right) \\[.25cm]
& 
\!\!\!\!\!\mathcal{M}_{3,1}=
\ltro_{4,1}\left( {\bf s}\tilde{\bf s}^2 + {\bf t} \tilde{\bf t}^2 +  {\bf u} \tilde{\bf u}^2 \right) +
\ltro_{4,2} \left( {\bf s}  c_s^2+{\bf t} c_t^2+ {\bf u} c_u^2\right)
+(\Sigma \ltro_{4,3}+\ltro_{4,4}) \left( {\bf s}   \tilde{\bf s} +{\bf t} \tilde{\bf t}+  {\bf u}  \tilde{\bf u} \right)\notag\\[.25cm]
&
\!\!\!\!\!\mathcal{M}_{3,0} =
				 \ltro_{3,1} ( \tilde{\bf s}^3 + \tilde{\bf t}^3 +\tilde{\bf u}^3) 
				 + \ltro_{3,2} \left(c_s^2 \tilde{\bf s} + c_t^2 \tilde{\bf t} + c_u^2 \tilde{\bf u}\right)+  \ltro_{3,3} \Sigma^3  + \ltro_{3,4} (c_s^2 + c_t^2 + c_u^2) \Sigma  \notag \\
&\rule{1cm}{0pt} + \ltro_{3,6}^{(3)}c_s\,c_t\,c_u+\left(\ltro_{3,5} \Sigma +\ltro_{3,9}\right) (\tilde{\bf s}^2  + \tilde{\bf t}^2 + \tilde{\bf u}^2)   %+ \notag\\
			%&
			+ \ltro_{3,7} \Sigma^2 +  \ltro_{3,8} (c_s^2 + c_t^2 + c_u^2)  + \Sigma \ltro_{3,10}+ \ltro_{3,11}  \notag 
\end{align}
From left to right we  first wrote the terms corresponding 
to the homogeneous polynomial $\mathcal{H}_{3,(\ell,3-\ell)}$ and 
then the terms coming from previous orders, which in this case are simple to recognise. 
Notice that $\mathcal{M}^{flat}_{3,3}$ has 10d spin equal to two, 
even though the degree of the polynomial is three.\footnote{It is useful to have in mind  the 
flat space relation $s t u =\frac{1}{3} (s^3+t^3+u^3)$ with $u=-s-t$.}
It goes down to two because of the constraint on $\bf{u}$.

The rank constraints impose
\be
\!\!\!\!\!\!\!\begin{array}{lll}
 \ltro_{5,1}^{}=-6&;\quad &\ltr_{5,2}=4 \\[.3cm]
 \ltro_{4,1}^{}=+15&;\quad & \ltro_{4,2}^{} =-\frac{7}{3}-\frac{1}{32}\ltro_{3,10}\quad;\quad \ltro_{4,3}^{} =-\frac{58}{3}+\frac{1}{16} \ltro_{3,10}  \quad;\quad \ltro_{4,4}^{}= -\frac{4}{3}-5\ltro_{5,3}^{}-\frac{1}{8}\ltro_{3,10}\\[.3cm]
 \ltro_{3,1}^{}= -10&;\quad& \ltro_{3,2}^{}= \frac{14}{3}+\frac{1}{16} \ltro_{3,10}\quad;\quad \ltro_{3,3}^{}= -32+\frac{1}{8}\ltro_{3,10}\quad;\quad \ltro_{3,4}^{}= -\frac{7}{3}-\frac{1}{32}\ltro_{3,10} \\[.3cm]
 				& &    \ltro_{3,5}=\frac{55}{3}-\frac{5}{32}\ltro_{3,10}\quad;\quad \ltro_{3,7}=-\frac{22}{3}-\frac{11}{16} \ltro_{3,10}+\ltro_{3,8}-11\ltro_{5,3}  \\[.3cm]
				&  &    \ltro_{3,9}=\frac{10}{3}+\frac{5}{16}\ltro_{3,10}+5\ltro_{5,3}\quad;\quad \ltro_{3,6}=0
\end{array}
\ee
The four free parameters are, $k_{3,11}$ and $k_{3,8}$, i.e.~the ambiguities we also found at order 
$(\alpha')^5$, then $k_{5,3}$, i.e.~the ambiguity corresponding to a shift by the same amplitude 
as $\mathcal{V}_2$, and finally $\ltro_{3,10}^{}$. 
At this point we can use the OPE once more by considering what information at $(\alpha')^6$
comes from the amplitude at $(\alpha')^5$, in particular from the solution of the partial 
degeneracy of operators at $m^*=2$. We explain the details of this procedure in section 
\ref{tailoring_sec}. Remarkably, the new constraints we obtain in this way are automatically satisfied,
therefore we are left with four genuine bootstrap ambiguities. 

By imposing on our bootstrapped amplitude consistency with the results from supersymmetric 
localisation for $\langle \cO_{2}\cO_{2}\cO_{p}\cO_{p} \rangle$ 
and $\langle \cO_{2}\cO_{2}\cO_{2}\cO_{2} \rangle$,
see~\cite{Binder:2019jwn} and~\cite{Chester:2019}, respectively,
we obtain three additional equations~\footnote{We thank Shai Chester for sharing these results at orders 
$(\alpha')^6$ \cite{SHAI}, obtained using the methods described in~\cite{Binder:2019jwn}.} 
\begin{align}
	\ltro_{3,8}=\tfrac{4}{3}-\tfrac{1}{16}\ltro_{3,10} ,\qquad \ltro_{3,11}=0,\qquad \ltro_{5,3}=-2,
	\label{alpha6loc}
\end{align}
leaving us finally with only one free parameter, $\ltro_{3,10}$. In contrast to order $(\alpha')^5$, 
here localisation is not yet sufficient to fix the full amplitude.

%======================================================================================
\paragraph{Remarkable simplicity and a possible symmetry.}~\\[-.2cm]
%======================================================================================

The results for $(\alpha')^{7,8,9}$ can be presented as above, and we do so in an ancillary file. 
Here, we will observe a further structure in the pattern of the coefficients leading to remarkable simplicity. 
For example, returning to ${\cal V}_2$ given in \eqref{V2}-\eqref{locV2},
if we expand in terms of  the original $AdS_5 \times S^5$ Mellin variables $s,t,u$ and 
$\tilde{s},\tilde{t},\tilde{u}$ we observe the terms of the form $s^l \tilde{s}^a$ with $a+l=2$ have coefficients 
\be
(\Sigma-1-a)_{a+l+3} \frac{(a+l)!}{a!l!}s^l \tilde{s}^a\,. \label{binomials}
\ee
Note that these coefficients arise from different strata in $\mathcal{V}_2$ thus they are non trivial.  
A similar pattern is observed at higher orders for the terms with $a+l=n$.\footnote{From the $rank=1$ 
solution in section \ref{sec_rank_equal1_anadim} we know this will work similarly at all orders in $\alpha'$.} 
This observation suggests a rescaling of the variables according to an integral transform which generalises 
the one used by Penedones in \cite{Penedones:2010ue}.
The integral transform we have in mind is
\be\label{integral_tr_James}
\mathcal{V}_n = \frac{i}{2 \pi} \int_0^\infty  d\alpha \int_\mathcal{C} d\beta\ e^{-\alpha-\beta} \alpha^{1+\Sigma}  (-\beta)^{1-\Sigma} \, \tilde{\mathcal{V}}_n(\alpha,\beta)
\ee
where $\mathcal{C}$ is the Hankel contour. Here $\mathcal{V}_n$ is given by our bootstrap results, 
and $\tilde{\cal V}_n$ is a simplified amplitude, defined in terms of the following variables,
\be
S = \alpha \hat{s} - \beta \check{s}\quad;\quad\tilde{S} = \alpha \hat{s} + \beta \check{s}\,\qquad;\qquad \left\{\begin{array}{l}  \hat{s} = s - \tfrac{1}{2} c_s  + 1\,, \\[.2cm]
 																		\check{s} = \tilde{s} + \tfrac{1}{2} c_s  + 1\,,  \end{array}\right.
\ee
and similarly for $t$-type and $u$-type variables. The integral transform \eqref{integral_tr_James} 
provides $\Gamma$ functions, direct and inverse, and produces the Pochhammer in eq. (\ref{binomials}) for the relevant terms.

Quite remarkably all terms  $s^l \tilde{s}^a$ in $\tilde{\cal V}$ such that $a+l=n$ then recombine into the 
binomial expansion of powers of the combinations $S,T,U$, while the combinations $\tilde{S},\tilde{T},\tilde{U}$ only arise from terms with $a+l<n$. 

Let us quote the results for the $\tilde{\mathcal{V}}_n$. For completeness, $\tilde{\mathcal{V}}_0 = \zeta_3$ and $\tilde{\mathcal{V}}_4=0$. 
Then, we will split the amplitude as a \emph{particular} contributions plus a choice of ambiguities. At order $(\alpha')^5$ we have 
\be
\tilde{\mathcal{V}}_2 = \zeta_5\bigl[\tilde{\mathcal{V}}_2^{\rm ptic} + \tilde{\mathcal{V}}_2^{\rm amb} \bigr]
\ee
with the remaining free parameters (after the rank constraints have been imposed) in the second term. The two terms are given explicitly by
\begin{align}
\tilde{\mathcal{V}}^{\rm ptic}_2 = S^2+T^2+U^2 + 3 \Sigma^2  \qquad;\qquad \tilde{\mathcal{V}}^{\rm amb}_2 = b_1 I_2 + b_2\ \ \ 
\label{Vt2}
\end{align}
where 
$
I_2 \equiv c_s^2+c_t^2+c_u^2 + \Sigma^2 = {\textstyle \sum}_i p_i^2
$.
Localisation fixes $b_1=-\tfrac{5}{2}$ and $b_2 = \tfrac{41}{2}$. As we mentioned above, constraints from localisation at this order fully fix the Virasoro-Shapiro amplitude. 

At order $(\alpha')^6$ we have
\be
\tilde{\mathcal{V}}_3 = \zeta_3^2\bigl[\tilde{\mathcal{V}}_3^{\rm ptic} + \tilde{\mathcal{V}}_3^{\rm amb} \bigr] 
%\ee
%where
%\ee
\qquad;\qquad\left\{\begin{array}{ll} 
\tilde{\mathcal{V}}^{\rm ptic}_3 &= \tfrac{2}{3} \bigl(S^3 + T^3 +U^3 - 2 \Sigma(\Sigma^2-4)\bigr) \\[.3cm]
\tilde{\mathcal{V}}^{\rm amb}_3& = b_1 \tilde{\mathcal{V}}_3^{{\rm amb},1}  +b_2 \tilde{\mathcal{V}}^{\rm ptic}_2 +b_3 I_2 + b_4  
\end{array}\right.
\label{Vt3}
\ee
where the new ambiguous contribution, compared to the three known at $(\alpha')^5$, is
\be
\tilde{\mathcal{V}}_3^{{\rm amb},1} = S (2\tilde{S} +c_s^2) + T (2\tilde{T} + c_t^2) + U (2 \tilde{U} +c_u^2) + \Sigma(12-c_s^2-c_t^2-c_u^2)\,.
\ee
In this case the constraints from localisation quoted in (\ref{alpha6loc}) become
\be
b_1 = -3 - \overline{k}\,, \quad  b_2 = 2  \overline{k} \,, \quad  b_3 = -2  \overline{k} \,, \quad  b_4 = 8  \overline{k}\,,
\label{alpha6finalparam}
\ee
for some free parameter $ \overline{k}$.

At order $(\alpha')^7$ we have
$
\tilde{\mathcal{V}}_4 = \zeta_7\bigl[\tilde{\mathcal{V}}_4^{\rm ptic} + \tilde{\mathcal{V}}_4^{\rm amb} \bigr]
$
with
\ba
\tilde{\mathcal{V}}_4^{\rm ptic} &= &
S^4+T^4+U^4 +8 (S^2+T^2+U^2)\Sigma^2 + 9 (S\tilde{S} + T \tilde{T} + U \tilde{U})\Sigma \notag \\
& &
- \tfrac{1}{2}(\tilde{S} c_s^2 + \tilde{T} c_t^2 + \tilde{U} c_u^2) -\tfrac{1}{4} \Sigma \bigl[\Sigma (I_2 - 16)-6c_s c_t c_u - 56 \Sigma^3\bigr]  \,.
\label{Vt4fixed}
\ea
and ten ambiguities in total,
\ba
\tilde{\mathcal{V}}_4^{\rm amb} &=  &
b_1 \tilde{\mathcal{V}}_4^{{\rm amb},1} + b_2 \tilde{\mathcal{V}}_4^{{\rm amb},2}  +  b_3 \tilde{\mathcal{V}}_4^{{\rm amb},3} + b_4 I_2 \tilde{\mathcal{V}}_2^{\rm ptic} +b_5 (I_2)^2 \notag \\
& &
+\ b_6 \tilde{\mathcal{V}}_3^{\rm ptic} +b_7 \tilde{\mathcal{V}}_3^{{\rm amb},1} +b_8 \tilde{\mathcal{V}}_2^{\rm ptic}  + b_9I_2  +b_{10}\,.
 \label{alpha7ambigs}
\ea
Those in the first line above are either products of terms from previous orders or given by
\begin{align}
\tilde{\mathcal{V}}_4^{{\rm amb},1} &= S^2 (2\tilde{S} +c_s^2+\Sigma^2) + T^2 (2\tilde{T} + c_t^2+\Sigma^2) + U^2 (2 \tilde{U} +c_u^2+\Sigma^2) \notag \\
&\qquad \,\,\, - \Sigma\bigl(2(Sc_s^2+Tc_t^2+Uc_u^2)+3c_sc_tc_u\bigr)+\Sigma^2\bigl(\tfrac{5}{2}I_2 -2\Sigma^2 + 8\bigr) \,, \notag \\[.2cm]
\tilde{\mathcal{V}}_4^{{\rm amb},2} &= c_s^4 + c_t^4 + c_u^4 + 12 c_s c_t c_u \Sigma + \Sigma^4 \,, \notag \\[.2cm]
\tilde{\mathcal{V}}_4^{{\rm amb},3} &= (c_s^2 +2 \tilde{S})^2 + (c_t^2 + 2 \tilde{T})^2 +  (c_u^2 + 2\tilde{U})^2 + 28 \Sigma^2+2(c_s^2+c_t^2+c_u^2)\Sigma^2-\Sigma^4  \,.
\end{align}

At order $(\alpha')^8$ we just quote the full result in the form, 
\begin{align}
\tilde{\mathcal{V}}_5 = \tfrac{4}{5}\zeta_3 \zeta_5 \Bigl[ &\bigl[S^5+ 15 S^3 \Sigma^2 + 25 S^2\tilde{S} \Sigma - \tfrac{5}{2}S \tilde{S} (c_s^2+\Sigma^2) -\tfrac{5}{8} S c_s^4  -\tfrac{5}{4}S c_s^2 \Sigma^2 +\tfrac{5}{2} \tilde{S}c_s^2 \Sigma +\tfrac{5}{4} c_s^4 \Sigma \notag \\
&+\tfrac{5}{4} c_s^2 \Sigma^3 - 5 c_s^2 \Sigma + (\text{$t$-type }) + (\text{$u$-type}) \bigr] - \Sigma(32\Sigma^4-135\Sigma^2+88) \notag \\
&+ 16 \text{ ambiguities}\Bigr] \,.
\label{Vt4}
\end{align}
The 16 ambiguities for this case can be found in the ancillary file.

The ambiguities at $(\alpha')^{7,8}$ can be further constrained by using the OPE and the data extracted from the amplitude at $(\alpha')^5$, 
as we tried to do with $(\alpha')^6$ where it turned out the extra constraints were automatically satisfied. In section \ref{tailoring_sec} we will find
a new constraint at $(\alpha')^7$ and two new constraints at $(\alpha')^8$.

The simplicity of the rescaled amplitudes is quite remarkable, with $\tilde{\mathcal{V}}_n$ simply given by the corresponding 
term in the Virasoro-Shapiro amplitude in terms of $S,T,U$ plus terms of lower order in $S,T,U,\tilde{S},\tilde{T},\tilde{U}$. Note that 
this continues to hold even at order $(\alpha')^9$ where there are two distinct contributions coming with different combinations of zeta values,
\be 
\tilde{\mathcal{V}}_6 = \zeta_9(S^6 +T^6 + U^6) - \tfrac{1}{27}(7\zeta_9 - 4\zeta_3^3)(S^3+T^3+U^3)^2 + \ldots
\ee
These relations are strongly suggestive of an even more restrictive relation of the  Mellin amplitudes to the flat space amplitudes, 
enhancing that of \cite{Aprile:2020luw} which itself enhanced that of \cite{Penedones:2010ue} in the case of $AdS_5 \times S^5$.

Furthermore we observe that the rescaled amplitudes exhibit properties under an interesting 
$Z_2$ transformation which exchanges $AdS_5$ and $S^5$ quantities, 
\be
\{S,T,U\} \leftrightarrow \{-S,-T,-U\}\,, \qquad \{\tilde{S},\tilde{T},\tilde{U}\} \leftrightarrow \{\tilde{S},\tilde{T},\tilde{U}\}\,, \qquad p_i \leftrightarrow -p_i\,.
\ee
At each order the term $\tilde{\mathcal{V}}_n^{\rm ptic}$ is even/odd under the transformation 
depending on whether $n$ is even or odd. Each ambiguity also has a definite parity under the transformation.
We will see in the next section that this behaviour is reflected in the double-trace spectrum. 
%for which the particular contribution for even $n$ play the central role. 
If one insists that at each order ambiguities of the opposite parity 
compared to $\tilde{\mathcal{V}}_n^{\rm ptic}$ are ruled out, then we find that the remaining parameter $\overline{k}$ in eq. (\ref{alpha6finalparam}) 
vanishes and that imposing the possible symmetry is simultaneously consistent with the three conditions from localisation. 
A similar statement also holds at order $(\alpha')^4$ where the constant contribution removed by localisation is also 
of odd parity.\footnote{We are very grateful for discussions regarding this possible $AdS \leftrightarrow S$ symmetry 
with the authors of \cite{NEWPAUL} who reached a similar conclusion.} In a similar way the symmetry would also imply $b_6=b_7=0$ in (\ref{alpha7ambigs}).

%%==============================================================================
%%==============================================================================

\section{The spectrum of two-particle operators at genus zero}\label{spectum_sect}

%%==============================================================================
%%==============================================================================

Our bootstrap algorithm will use in a crucial way the spectral properties of the VS amplitude in $AdS_5\times S^5$.
To explain what we mean by this,  we first need to review in some more details results 
in \cite{Aprile:2017xsp,Aprile:2018efk}, reloading quickly the introduction given in section \ref{intro_sec}.
At that point, we will be able to formulate the STRINGY eigenvalue problem which on one side
defines the level splitting problem, and on the other identifies the `rank constraints'.

%%==============================================================================
%%=================================================

\subsection{SUGRA eigenvalue problem: a review}\label{rew_sec_SUGRA}

%%=================================================
%======================================================================

At leading order in the large $N$ expansion only long two-particle multiplets receive an anomalous 
dimension in the interacting theory.  The corresponding primaries in the free theory have the 
schematic form\footnote{This formula here will also count protected operators when $\tau=p+q$}
\be
\mathcal{O}_{pq} = \mathcal{O}_p \partial^l \Box^{\frac{1}{2}(\tau - p - q)} \mathcal{O}_q\,, \qquad ( p < q )\
\label{dbltrace}
\ee
For given quantum numbers $\vec{\tau}=(\tau_{free},l,[aba])$,
many operators are degenerate. We count them by the number of 
pairs $(pq)$ filling in a rectangle \cite{Aprile:2018efk}
\begin{align}\label{ir}
%	(p,q)\in
 {R}_{\vec{\tau}}:=\left\{(p,q):\begin{array}{l}
	p=i+a+2+r\\q=i+a+2+b-r\end{array}\text{ for } \begin{array}{l}
	i=\,0,\ldots,(t-2)\\ r=0,\ldots,(\mu -1)\end{array}
	\right\}
\end{align}
where 
\be\label{multiplicity}
t\equiv \frac{(\tau-b)}{2}-a\qquad;\qquad 
\mu \equiv   \left\{\begin{array}{ll}
\bigl\lfloor{\frac{b+2}2}\bigr\rfloor \quad &a+l \text{ even,}\\[.2cm]
\bigl\lfloor{\frac{b+1}2}\bigr\rfloor \quad &a+l \text{ odd.}
\end{array}\right.
\ee
This rectangle $ {R}_{\vec{\tau}}$ consists of $d=\mu(t-1)$ allowed lattice points, as the figure below shows

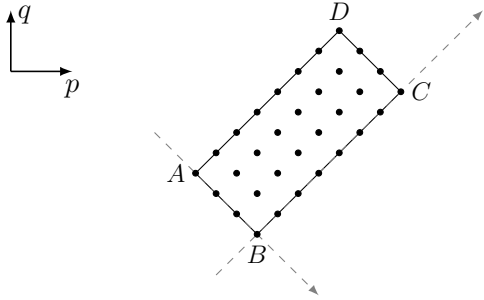
\be
\rule{.7cm}{0pt}
\begin{tikzpicture}[scale=.54]
%
%\draw[step=2cm,gray,very thin] (-4,2) grid (8,8);
%
\def\prop{.5}
\def\shifthor{\prop*2}
\def\ptuno{(\prop*2-\shifthor,\prop*8)}
\def\ptdue{(\prop*5-\shifthor,\prop*5)}
\def\pttree{(\prop*9-\shifthor,\prop*15)}
\def\ptquattro{(\prop*12-\shifthor,\prop*12)}
%
%axis horizontal
\draw[-latex, line width=.6pt]		(\prop*1   -\shifthor-4,         \prop*14          -0.5*\shifthor)    --  (\prop*1  -\shifthor-2.5  ,   \prop*14-      0.5*\shifthor) ;
\node[scale=.8] (oxxy) at 			(\prop*1   -\shifthor-2.5,  \prop*16.5     -0.5*\shifthor)  {};
\node[scale=.9] [below of=oxxy] {$p$};
%
%axis vertical
\draw[-latex, line width=.6pt] 		(\prop*1   -\shifthor-4,     \prop*14       -0.5*\shifthor)     --  (\prop*1   -\shifthor-4,        \prop*17-      0.5*\shifthor);
\node[scale=.8] (oxyy) at 			(\prop*1   -\shifthor-2,   \prop*16.8   -0.5*\shifthor) {};
\node[scale=.9] [left of= oxyy] {$q$};
%
%rectangle
\draw[] 								\ptuno -- \ptdue;
\draw[black]							\ptuno --\pttree;
%\draw[green!50!black,thin]						(\prop*3-\shifthor,\prop*7) --(\prop*10-\shifthor,\prop*14);
%\draw[orange!70!white,thin]						(\prop*4-\shifthor,\prop*6) --(\prop*11-\shifthor,\prop*13);
\draw[black]							\ptdue --\ptquattro;
\draw[]								\pttree--\ptquattro;
\draw[-latex,gray, dashed]					(\prop*0-\shifthor,\prop*10) --(\prop*8-\shifthor,\prop*2);
\draw[-latex,gray, dashed]					(\prop*3-\shifthor,\prop*3) --(\prop*16-\shifthor,\prop*16);
%		
%dots
%
\foreach \indeyc in {0,1,2,3}
\foreach \indexc  in {2,...,9}
\filldraw   					 (\prop*\indexc+\prop*\indeyc-\shifthor, \prop*6+\prop*\indexc-\prop*\indeyc)   	circle (.07);
%
%letters
%
\node[scale=.8] (puntouno) at (\prop*4-\shifthor,\prop*8) {};
\node[scale=.8]  [left of=puntouno] {$A$};   
\node[scale=.8] (puntodue) at (\prop*5-\shifthor,\prop*6+.5) {};
\node[scale=.8] [below of=puntodue]  {$B$}; 
\node[scale=.8] (puntoquattro) at (\prop*13-\shifthor,\prop*15) {};
\node[scale=.8] [below of=puntoquattro] {$C$};
\node[scale=.8] (puntotre) at (\prop*9-\shifthor,\prop*13) {};
\node[scale=.8] [above of=puntotre] {$D$}; 
%
%
%legend
\node[scale=.84] (legend) at (16,5) {$\begin{array}{l}  
													\displaystyle A=(a+2,a+b+2); \\[.1cm]
													\displaystyle B=(a+1+\mu,a+b+3-\mu); \\[.1cm]
													\displaystyle C=(a+\mu+t-1,a+b+1+t-\mu); \\[.1cm]
													\displaystyle D=(a+t,a+b+t); \\[.1 cm] \end{array}$  };
\end{tikzpicture}
\ee

Some degenerations of the $R_{\vec{\tau}}$ will have a meaning later on. 
The first one corresponds to $\mu=1$, i.e.~the rectangle collapses to a line with $+45^{\circ}$ orientation.
The second one corresponds to first available twist in $[aba]$, i.e. $\tau=2a+b+4$, for $\mu>1$.
Also in this case the rectangle collapses to a line, this time with $-45^{\circ}$ orientation.
Then, as the twist increases the rectangle opens up in the plane. 

Free theory long operators $\mathcal{O}_{pq}$ will mix when interactions are turned on and 
we will denote the true two-particle operators in the interacting theory, i.e.~the eigenstates 
with well-defined scaling dimensions, by $\mathcal{K}_{pq}$.
To address the mixing problem arrange first a $(d \times d)$ matrix of correlators 
$\langle \cO_{p_1}\cO_{p_2}|$ and $|\cO_{p_3}\cO_{p_4}\rangle$  with both $(p_1p_2)$ 
and $(p_3p_4)$ ranging over the same ${R}_{\vec{\tau}}$. Define the matrices 
$\mathbf{L}_{\vec{\tau}}$ from the long sector of disconnected free theory 
and $\mathbf{M}_{\vec{\tau}}(\alpha')$ from the leading $log$ discontinuity at tree level 
(including all $\alpha'$ corrections), 
\begin{align}
\begin{array}{rll}
O( N^0): &\qquad &  
\langle \mathcal{O}_{p_1} \mathcal{O}_{p_2} \mathcal{O}_{p_3} \mathcal{O}_{p_4} \rangle\Big|_{free\ long} =\  {\textstyle \sum_{\vec{\tau}}}\  \mathbf{L}_{\vec{\tau}}\, \mathbb{L}_{\vec{\tau}}\,, \\[.3cm]
O(N^{-2}): &\qquad& 
\langle \mathcal{O}_{p_1} \mathcal{O}_{p_2} \mathcal{O}_{p_3} \mathcal{O}_{p_4} \rangle\Big|_{\log u} = \ {\textstyle \sum_{\vec{\tau}}}\  \mathbf{M}_{\vec{\tau}}(\alpha')  \mathbb{L}_{\vec{\tau}}\,
\end{array}
\end{align}
Here $\mathbb{L}_{\vec{\tau}}$ is the superblock for long multiplets \cite{Doobary:2015gia} (see 
also appendix \ref{superb_deco_app}, where we wrote it in terms of $\tau,l$ and $[aba]$).
From the OPE it follows that 
\begin{align}
\label{unmixing}
\mathbf{C}_{\vec{\tau}}(\alpha')  \mathbf{C}_{\vec{\tau}} ^T(\alpha') = \mathbf{L}_{\vec{\tau}}\qquad;\qquad 
\mathbf{C}_{\vec{\tau}}(\alpha')  \pmb{\eta}_{\vec{\tau}}(\alpha')  \mathbf{C}_{\vec{\tau}}^T(\alpha') = \mathbf{M}_{\vec{\tau}}(\alpha') \,.
\end{align}
where $\mathbf{C}_{(pq),(\tilde{p}\tilde{q})}$ is a $(d\times d)$ matrix of three-point functions 
$\langle \mathcal{O}_p \mathcal{O}_q \mathcal{K}_{\tilde{p}\tilde{q}} \rangle$ and $\pmb{\eta}$ is a diagonal 
matrix encoding the anomalous dimensions of the eigenstates $\mathcal{K}_{pq}$,
\be
\Delta_{pq} = \tau-l + \frac{2}{N^2} \eta_{pq}(\alpha') + O\left(\frac{1}{N^4}\right)\,.
\ee
Our notation for the $\alpha'$ expansion will be
\be\label{def_three_pt}
\ \ \pmb{\eta}_{} = \pmb{\eta}^{(0)}_{} + \alpha'^{3}\pmb{ \eta}^{(3)}_{} \ + \alpha'^{5} \pmb{\eta}^{(5)}_{} + \ldots\, \quad\ ;\quad\ 
\mathbf{C} = \mathbf{C}^{(0)} + \alpha'^{3} \mathbf{C}^{(3)} + \alpha'^{5} \mathbf{C}^{(5)} + \ldots\,.
\ee

The computation of leading anomalous dimensions 
and leading three-point couplings in supergravity is now an eigenvalue problem \cite{Aprile:2018efk},
\begin{align}
\ \mathbf{c}_{\vec{\tau}}^{(0)}  \mathbf{c}_{\vec{\tau}} ^{(0)T} = \mathbf{I}_{\vec{\tau}}\qquad;\qquad
\mathbf{c}_{\vec{\tau}}^{(0)}  {\pmb \eta_{\vec{\tau}}}^{(0)}  \mathbf{c}_{\vec{\tau}}^{(0)T} = \mathbf{N}^{(0)}_{\vec{\tau}}
\end{align}
where we have defined
\begin{align}
\mathbf{c}_{\vec{\tau}}^{(0)} =\mathbf{L}^{-\frac{1}{2}}_{\vec{\tau}} \mathbf{C}^{(0)}_{\vec{\tau}} \qquad;\qquad
 \mathbf{N}_{\vec{\tau}}^{(0)}= \mathbf{L}^{-\frac{1}{2}}_{\vec{\tau}}\mathbf{M}_{\vec{\tau}}^{(0)}\mathbf{L}^{-\frac{1}{2}}_{\vec{\tau}}
\end{align}
Notice that $\mathbf{L}_{\vec{\tau}}$ 
is diagonal.\footnote{See explicit formula in \eqref{disco_FT_formula}.}
The tree-level anomalous dimensions $\eta^{(0)}$ are the eigenvalues of 
$\mathbf{N}_{\vec{\tau}}^{(0)}$ and were computed in \cite{Aprile:2018efk}, 
\begin{gather}
\tfrac{1}{2} \eta^{(0)}_{pq} = -\frac{ \delta^{(8)}_{\tau,l,[aba]}}{ (l-a+2p-2- \frac{1+(-1)^{a+l}}{2} )_6} \notag \\[.2cm]
\delta^{(8)}_{\tau,l,[aba]}=(\tfrac{\tau}{2}\pm\tfrac{1\pm1+2+2a+b}{2})(\tfrac{\tau}{2}\pm\tfrac{1\pm1+b}{2})(\tfrac{\tau}{2}+l+1\pm\tfrac{1\pm1+2+2a+b}{2})(\tfrac{\tau}{2}+l+1\pm\tfrac{1\pm1+b}{2})
\label{delta8_factor}
\end{gather}

Remarkably, the anomalous dimensions are all rationals, and only depend on $p$, 
rather than the pair $(pq)$, so all operators in $R_{\vec{\tau}}$ with the same value of 
$p$ but varying $q$ have the same anomalous dimension. 
This brings to the conclusion that 
\begin{mdframed}
\begin{center}
the resolution of the operator mixing in tree-level supergravity is only partial!
\end{center}
\end{mdframed}
To help visualising the partial degeneracy we introduce the level-splitting label $m$ 
of an operator $\mathcal{K}_{pq}$ in $R_{\vec{\tau}}$, which measures the distance on 
the $p$ axis from the value $p_A=a+2$, 
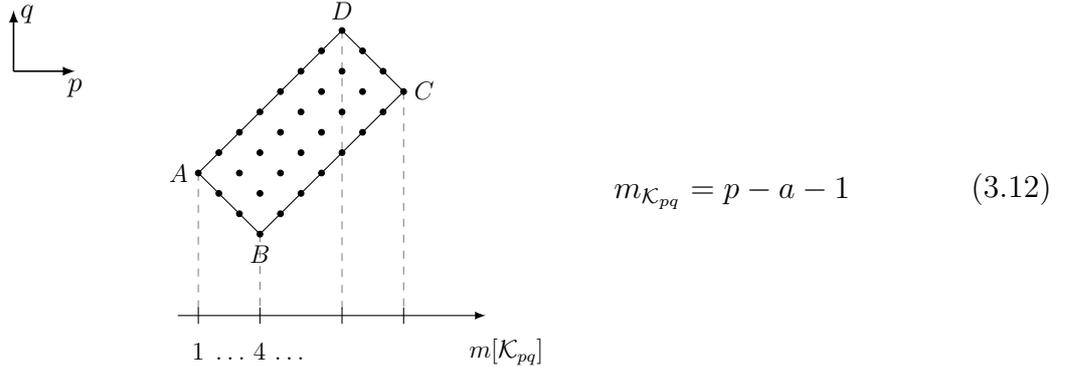
\be
\begin{array}{c}
\begin{tikzpicture}[scale=.54]
%
%\draw[step=2cm,gray,very thin] (-4,2) grid (8,8);
%
\def\prop{.5}
\def\shifthor{\prop*2}
\def\ptuno{(\prop*2-\shifthor,\prop*8)}
\def\ptdue{(\prop*5-\shifthor,\prop*5)}
\def\pttree{(\prop*9-\shifthor,\prop*15)}
\def\ptquattro{(\prop*12-\shifthor,\prop*12)}

%%%%%%%%%%%%%%%%%%

\draw[gray,dashed] (\prop*2-\shifthor,\prop*8) --(\prop*2-\shifthor,\prop*1);	
\draw[] (\prop*2-\shifthor,\prop*0.6) -- (\prop*2-\shifthor,\prop*1.4);	
%su A
\node[scale=.8] (m1) at (\prop*2-\shifthor,\prop*2.2) {};
\node[scale=.8] [below of=m1] {$1$}; 
% ancora su A
\draw[gray,dashed] (\prop*5-\shifthor,\prop*5) --(\prop*5-\shifthor,\prop*1);	
\draw[] (\prop*5-\shifthor,\prop*0.6) -- (\prop*5-\shifthor,\prop*1.4);	
%su B
\node[scale=.8] (m2) at (\prop*5-\shifthor,\prop*2.2) {};
\node[scale=.8] [below of=m2] {$\ldots\, 4\, \ldots$}; 
% ancora su B
\draw[gray,dashed] (\prop*9-\shifthor,\prop*15) --(\prop*9-\shifthor,\prop*1);
\draw[] (\prop*9-\shifthor,\prop*0.6) -- (\prop*9-\shifthor,\prop*1.4);		
%su C
\draw[gray,dashed] (\prop*12-\shifthor,\prop*12) --(\prop*12-\shifthor,\prop*1);	
\draw[] (\prop*12-\shifthor,\prop*0.6) -- (\prop*12-\shifthor,\prop*1.4);	
%su D
\draw[-latex] (\prop*1-\shifthor,\prop*1) -- (\prop*12+\shifthor,\prop*1);
\node[scale=.8] (etaa) at (\prop*12+1.5*\shifthor,\prop*2.2) {};
\node[scale=.8] [below of=etaa] {$m[\mathcal{K}_{pq}]$}; 
									
%%%%%%%%%%%%%%%%%%

%
%axis horizontal
\draw[-latex, line width=.6pt]		(\prop*1   -\shifthor-4,         \prop*14          -0.5*\shifthor)    --  (\prop*1  -\shifthor-2.5  ,   \prop*14-      0.5*\shifthor) ;
\node[scale=.8] (oxxy) at 			(\prop*1   -\shifthor-2.5,  \prop*16.5     -0.5*\shifthor)  {};
\node[scale=.9] [below of=oxxy] {$p$};
%
%axis vertical
\draw[-latex, line width=.6pt] 		(\prop*1   -\shifthor-4,     \prop*14       -0.5*\shifthor)     --  (\prop*1   -\shifthor-4,        \prop*17-      0.5*\shifthor);
\node[scale=.8] (oxyy) at 			(\prop*1   -\shifthor-2,   \prop*16.8   -0.5*\shifthor) {};
\node[scale=.9] [left of= oxyy] {$q$};
%
%rectangle
\draw[] 								\ptuno -- \ptdue;
\draw[black]							\ptuno --\pttree;
\draw[black]							\ptdue --\ptquattro;
\draw[]								\pttree--\ptquattro;
%
%		
%dots
%
\foreach \indeyc in {0,1,2,3}
\foreach \indexc  in {2,...,9}
\filldraw   					 (\prop*\indexc+\prop*\indeyc-\shifthor, \prop*6+\prop*\indexc-\prop*\indeyc)   	circle (.07);
%
%letters
%
\node[scale=.8] (puntouno) at (\prop*4-\shifthor,\prop*8) {};
\node[scale=.8]  [left of=puntouno] {$A$};   
\node[scale=.8] (puntodue) at (\prop*5-\shifthor,\prop*6+.5) {};
\node[scale=.8] [below of=puntodue]  {$B$}; 
\node[scale=.8] (puntoquattro) at (\prop*13-\shifthor,\prop*15) {};
\node[scale=.8] [below of=puntoquattro] {$C$};
\node[scale=.8] (puntotre) at (\prop*9-\shifthor,\prop*13) {};
\node[scale=.8] [above of=puntotre] {$D$}; 
%
%
%legend
%\node[scale=.84] (legend) at (15,5) {$\begin{array}{l}  
%													\displaystyle A=(a+2,a+b+2); \\[.1cm]
%													\displaystyle B=(a+1+\mu,a+b+3-\mu); \\[.1cm]
%													\displaystyle C=(a+\mu+t-1,a+b+1+t-\mu); \\[.1cm]
%													\displaystyle D=(a+t,a+b+t); \\[.1 cm] \end{array}$  };
%													

\node at  (13,\prop*7) {$m_{\mathcal{K}_{pq}}=p-a-1$};
													
\end{tikzpicture}
\end{array}
%\rule{1cm}{0pt}
%\notag
\ee
For each anomalous dimension labelled by $m$, the partial degeneracy is counted by 
the number of points on the $q$ axis. The left most corner of the rectangle $A=(p_A,q_A)$ 
corresponds to the most negative anomalous dimension. The partial degeneracy is 
bounded by the parameter $\mu$ introduced in \eqref{multiplicity}, but
notice that the level-splitting label $m$ and the parameter $\mu$ are not the same.

Because of the residual degeneracy, the eigenvalue problem on 
$R_{\vec{\tau}}\otimes R_{\vec{\tau}}$ is well-posed, but the leading order three-point 
functions are not uniquely fixed, since these are determined by the columns of $\mathbf{c}^{(0)}_{\vec{\tau}}$  
which by definition  are the eigenvectors of $\mathbf{N}^{(0)}_{\vec{\tau}}$. 
If the anomalous dimension is degenerate, only a certain hyperplane is singled out, 
whose dimension is given by the partial degeneracy of $\eta(m)$. If the anomalous dimension 
is non degenerate, this dimension is unity and a unique vector is singled out. In any case we can fix a 
basis of eigenvectors and provide an orthogonal decomposition of $\mathbb{R}^{d}$,
\begin{align}\label{deco_map}
\mathbb{V}^{}_{\vec{\tau},1 }\oplus \mathbb{V}^{}_{\vec{\tau},\,2}\oplus  \ldots   \simeq \mathbb{R}^{d}
\end{align}
where $\mathbb{V}^{}_{\vec{\tau},m}$ span the hyperplane labelled by $\eta^{(0)}(m_{\cal K})$.
Obviously $d=\mu(t-1)$ counts the total number of operators, as explained around \eqref{multiplicity}.

A fundamental remark about the tree level anomalous dimensions in supergravity
came from \cite{Caron-Huot:2018kta}. In that paper it was   
recognised that the denominator can be written as $(l_{10}+1)_6$, 
i.e.~as a pochhammer of an effective ten-dimensional spin, defined by
\be
\label{10dspin}
l_{10} = l + a +2\,m_{\mathcal{K}_{(pq)}}
- \tfrac{1+(-1)^{a+l}}{2}-1\,.
\ee 
Then, it was pointed out that an accidental ten-dimensional conformal symmetry 
governs the physics at tree level in supergravity, and it explains the pattern of residual degeneracy.

\subsubsection{Consequences of the tree-level hidden conformal symmetry}

Some additional facts about the way the hidden conformal symmetry at tree level works, and about the 
meaning of $l_{10}$, will be important in our discussion. For convenience of the reader we 
spell them out in this section. 

As shown in \cite{Caron-Huot:2018kta}, all $AdS_5\times S^5$ tree level correlators 
$\mathcal{A}_{\vec{p}}(U,V,\tilde U,\tilde V)$ in position space,\footnote{Our conventions are given
in appendix \ref{CONVENTIONS}.} can be obtained by Taylor expanding  a generating function $\mathcal{G}$, 
which corresponds to a 10d version of the $2222$ correlator, namely $\mathcal{G}(U_{10},V_{10})=U_{10}^4 \mathcal{A}_{2222}(U_{10},V_{10})$, 
where $U_{10}$ and $V_{10}$ are 10d cross ratios,  rather than $AdS_5\times S^5$ cross ratios. 
A nice way to represent this expansion is to use operators
$\widehat{\mathcal{D}}_{\vec{p}}$ such that, directly on $AdS_5\times S^5$, we have
\be\label{iniSimon}
\mathcal{A}_{\vec{p}}(U,V,\tilde U,\tilde V)= \widehat{\mathcal{D}}_{\vec{p}}\,\Big[ U^4 \overline{D}_{2422}(U,V)\Big]\qquad;\qquad \mathcal{A}_{2222}(U,V)= \overline{D}_{2422}(U,V).
\ee
These operators were found explicitly in \cite{Aprile:2020luw}, 
\be\label{operatorDhat}
\widehat{\cal D}^{}_{\vec{p}}=\frac{ 1}{(U \tilde U)^{2}} \sum_{\tilde s,\tilde t} 
\left( \frac{\tilde U}{U}\right)^{\!\!{\tilde s}+2} \left(\frac{\tilde V}{V}\right)^{\!\!{\tilde t}} \widehat{\cal D}^{(0,0,0)}_{\vec{p},(\tilde s,\tilde t)} \widehat{\cal D}^{(c_s,c_t,c_u)}_{\vec{p},(\tilde s,\tilde t)}
\ee
where
%As in \cite{Aprile:2020luw} we can find its explicit expression, 
\be\label{operatorD}
\!\!\!\!\begin{array}{rl}
\widehat{\cal D}^{(a,b,c)}_{\vec{p},(\tilde s,\tilde t)} =&  {\displaystyle \frac{ (U\partial_U +1-\theta - \tilde s -a)_{\tilde s+a}}{(-)^a (\tilde s+a)!}} %\times  \\[.3cm]
								%\ \ \ \ \ \ \ &\times   
								 {\displaystyle \frac{ (V\partial_V +1- \tilde t -b)_{\tilde t+b}}{(-)^b (\tilde t+b)!}}
								  %\prod_{c=\{0, c_u\}}    
								  {\displaystyle \frac{ (U\partial_U+V\partial_V)_{\tilde u+c}}{ (\tilde u+c)!}}
								 
\end{array}
\ee
The result of \eqref{operatorDhat} on the Mellin amplitude is the covariantisation $\mathcal{M}_{\vec{p}}=\mathcal{M}_{2222}({\bf s,t,u})$, 
which at tree level is indeed exact. As we discussed in the previous section, this convariantisation 
is not exact when we consider $\alpha'$ stringy corrections to the tree level amplitude, since it only gives the leading flat space amplitude in the large $p$ limit. 

The generating function $\mathcal{G}$ can be interpreted, within 10d effective field theory, 
as the correlator of four fields $\Phi$ of dimensions $\Delta_{10}=4$, and in particular is a 10d conformal integral. This hidden conformal symmetry 
then implies that $\mathcal{G}_{}$ should have an OPE expansion  in which the exchanged primary fields 
are schematically of the form $\Phi\partial^{l_{10}}\Phi$, and  therefore should have a natural expansion in 10d blocks 
at the unitarity bound, i.e.~$\Delta_{10}-l_{10}=8$. By construction, this expansion descends to 
$\mathcal{A}_{2222}(U,V)$. In particular we find
\be\label{Simon1111}
U^4 \mathcal{A}_{2222}(U,V)\Big|_{\log U}\!\!=\!
\sum_{l_{10}=0,2,4,\ldots }\frac{ 4\,\Gamma[l_{10}+4]^2 }{ \Gamma[2l_{10}+7]} ~_2F_1[ 4+l_{10},4+l_{10};8+2l_{10}; P]
\ee
where $_2F_1[ \ldots ; P]$ is a single 10d block at the unitarity bound in our conventions, 
\be
\,_2F_1\Big[\,^{4+l_{}\,,\,4+l_{}}_{\ \ 8+2l_{}}; P\Big]= \sum_{n\ge 0} \frac{ (4+l_{})_n (4+l_{})_n }{ n! (8+2l_{})_n  } 
\sum_{k=0}^{n+l_{} } \frac{ (1+k)_{n+l_{}-k}  (4)_{n+l_{}-k}  }{  (4+k)_{n+l_{}-k} (1)_{n+l_{}-k}  }\,x_1^{4+k}x_2^{4+l_{}+n-k}
\ee
with $x_1 x_2=U_{}$ and $(1-x_1)(1-x_2)=V_{}$. 

The expansion in \eqref{Simon1111} will then propagate to 
$\mathcal{A}_{\vec{p}}=\sum_{l_{10}} {\tt Gammas}\ \widehat{\mathcal{D}}_{\vec{p}}\,\Big[  \,_2F_1[ \ldots ; P] \Big]$, thanks to the relation in \eqref{iniSimon}.

The simplest correlator where we can appreciate the consequences of  \eqref{Simon1111} is the $2222$ correlator. It is indeed well known that 
$\mathcal{A}_{2222}(U,V)$ has a 4d conformal block expansion \cite{Dolan:2004iy}, in the ${\tt Blc}_{\tau,l}$ defined in \eqref{Blc_def}, 
which is a double expansion in twist $\tau\ge 4$ and even spin $l$.  The crucial observation is that a given 10d block 
$_2F_1[ \ldots ; P]$ labelled by $l_{10}$, contains `multiplets' of ${\tt Blc}_{\tau,l}$ with a range of  $\frac{\tau-4}{2} \leq l\leq l_{10}$. 
More precisely,  
\be\label{usefulexpa}
\,_2F_1\Big[\,^{4+l_{10}\,,\,4+l_{10}}_{\ \ \ 8+2l_{10}}; P\Big]=\ U^2\!\!\sum_{\tau=4,6,\ldots }\,\sum_{\iota=0}^{\frac{\tau-4}{2} }  A_{\tau,l_{10},\iota} \,{\tt Blc}_{\tau,l_{10}-2\iota}
\ee
where
\begin{align}
A_{\tau,l,\iota}= &\ \frac{ l!(7+2l)!  (\frac{\tau}{2})!^2}{(3+l)!^2(6+l)! \tau!}\, \frac{ (4+2\iota)! }{ i! (2+\iota)!} \times \frac{(1-2\iota+l)(2-2\iota+l+\tau)}{12}\times \\
&
\ \frac{(6-2\iota+2l)!   (1-2\iota+l+\frac{\tau}{2})!^2   (\tau-2\iota)! (2-\iota+l+\frac{\tau}{2})! (3-\iota+l+\frac{\tau}{2})!}{ 
(1-\iota+l)!(3-\iota+l)!(\frac{\tau}{2}-\iota-2)!(\frac{\tau}{2}-\iota)!(2-4\iota+2l+\tau)!(4-2\iota+2l+\tau)!} \notag
\end{align}
Note $A_{4,l,0}=1$. 

For the singlet channel $[000]$, where \eqref{usefulexpa} applies with no modifications, since $\widehat{\cal D}_{2222}$ is trivial, 
the index $\iota$ in \eqref{usefulexpa} is indeed running over the rectangle \eqref{ir} from right to left.
In fact, in this case the rectangle collapses to a line as the figure below shows
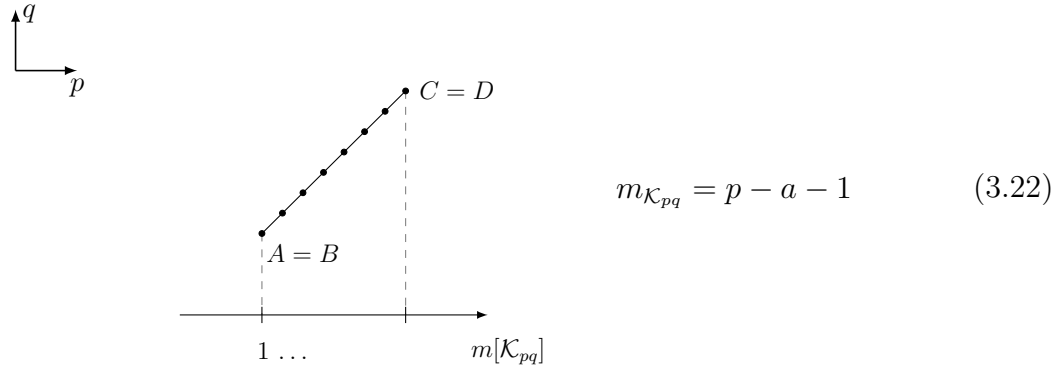
\be
\begin{array}{c}
\begin{tikzpicture}[scale=.54]
%
%\draw[step=2cm,gray,very thin] (-4,2) grid (8,8);
%
\def\prop{.5}
\def\shifthor{\prop*2}
\def\ptuno{(\prop*2-\shifthor,\prop*8)}
\def\ptdue{(\prop*5-\shifthor,\prop*5)}
\def\pttree{(\prop*9-\shifthor,\prop*15)}
\def\ptquattro{(\prop*12-\shifthor,\prop*12)}

%%%%%%%%%%%%%%%%%%

%\draw[gray,dashed] (\prop*2-\shifthor,\prop*8) --(\prop*2-\shifthor,\prop*1);	
%\draw[] (\prop*2-\shifthor,\prop*0.6) -- (\prop*2-\shifthor,\prop*1.4);	
%su A
%\node[scale=.8] (m1) at (\prop*2-\shifthor,\prop*2.2) {};
%\node[scale=.8] [below of=m1] {$1$}; 
% ancora su A
\draw[gray,dashed] (\prop*5-\shifthor,\prop*5) --(\prop*5-\shifthor,\prop*1);	
\draw[] (\prop*5-\shifthor,\prop*0.6) -- (\prop*5-\shifthor,\prop*1.4);	
%su B
\node[scale=.8] (m2) at (\prop*6-\shifthor,\prop*2.2) {};
\node[scale=.8] [below of=m2] {$1\, \ldots$}; 
% ancora su B
%\draw[gray,dashed] (\prop*9-\shifthor,\prop*15) --(\prop*9-\shifthor,\prop*1);
%\draw[] (\prop*9-\shifthor,\prop*0.6) -- (\prop*9-\shifthor,\prop*1.4);		
%su C
\draw[gray,dashed] (\prop*12-\shifthor,\prop*12) --(\prop*12-\shifthor,\prop*1);	
\draw[] (\prop*12-\shifthor,\prop*0.6) -- (\prop*12-\shifthor,\prop*1.4);	
%su D
\draw[-latex] (\prop*1-\shifthor,\prop*1) -- (\prop*12+\shifthor,\prop*1);
\node[scale=.8] (etaa) at (\prop*12+1.5*\shifthor,\prop*2.2) {};
\node[scale=.8] [below of=etaa] {$m[\mathcal{K}_{pq}]$}; 
									
%%%%%%%%%%%%%%%%%%

%
%axis horizontal
\draw[-latex, line width=.6pt]		(\prop*1   -\shifthor-4,         \prop*14          -0.5*\shifthor)    --  (\prop*1  -\shifthor-2.5  ,   \prop*14-      0.5*\shifthor) ;
\node[scale=.8] (oxxy) at 			(\prop*1   -\shifthor-2.5,  \prop*16.5     -0.5*\shifthor)  {};
\node[scale=.9] [below of=oxxy] {$p$};
%
%axis vertical
\draw[-latex, line width=.6pt] 		(\prop*1   -\shifthor-4,     \prop*14       -0.5*\shifthor)     --  (\prop*1   -\shifthor-4,        \prop*17-      0.5*\shifthor);
\node[scale=.8] (oxyy) at 			(\prop*1   -\shifthor-2,   \prop*16.8   -0.5*\shifthor) {};
\node[scale=.9] [left of= oxyy] {$q$};
%
%rectangle
%\draw[] 								\ptuno -- \ptdue;
%\draw[black]							\ptuno --\pttree;
%
\draw[black]							\ptdue --\ptquattro;
%\draw[]								\pttree--\ptquattro;
%
%		
%dots
%
\foreach \indeyc in {3}
\foreach \indexc  in {2,...,9}
\filldraw   					 (\prop*\indexc+\prop*\indeyc-\shifthor, \prop*6+\prop*\indexc-\prop*\indeyc)   	circle (.07);
%
%letters
%
%\node[scale=.8] (puntouno) at (\prop*4-\shifthor,\prop*8) {};
%\node[scale=.8]  [left of=puntouno] {$A$};   
%
\node[scale=.8] (puntodue) at (\prop*7-\shifthor,\prop*6+.5) {};
\node[scale=.8] [below of=puntodue]  {$A=B$}; 
\node[scale=.8] (puntoquattro) at (\prop*14.5-\shifthor,\prop*15) {};
\node[scale=.8] [below of=puntoquattro] {$C=D$};
%							
%\node[scale=.8] (puntotre) at (\prop*9-\shifthor,\prop*13) {};
%\node[scale=.8] [above of=puntotre] {$D$}; 
%
%
%legend
%\node[scale=.84] (legend) at (15,5) {$\begin{array}{l}  
%													\displaystyle A=(a+2,a+b+2); \\[.1cm]
%													\displaystyle B=(a+1+\mu,a+b+3-\mu); \\[.1cm]
%													\displaystyle C=(a+\mu+t-1,a+b+1+t-\mu); \\[.1cm]
%													\displaystyle D=(a+t,a+b+t); \\[.1 cm] \end{array}$  };
%													

\node at  (13,\prop*7) {$m_{\mathcal{K}_{pq}}=p-a-1$};
													
\end{tikzpicture}
\end{array}
%\rule{1cm}{0pt}
%\notag
\ee
and $A_{\tau,l,\iota}$ is related to the three-point couplings $C_{22 \mathcal{K}_{(pq),\vec{\tau} } }$, which are just 
a part of the matrix  $\mathbf{C}^{(0)}$ on $R_{\tau,l,[000]}$, defined in \eqref{def_three_pt}.

In the more general case, the action of $\widehat{\cal D}_{\vec{p}}$\, on 10d blocks, re-expanded in ${\tt Blc}_{\tau,l}$, 
yields the projectors relative to the orthogonal decomposition of the space of three-point couplings, 
$\mathbb{V}^{}_{\vec{\tau},1 }\oplus \mathbb{V}^{}_{\vec{\tau},\,2}\oplus  \ldots   \simeq \mathbb{R}^{d}$, 
as  %of the two-particle operators, 
explained in \eqref{deco_map}.
Reconstructing the degeneracy, we fix conventions w.r.t.~the level splitting label so that 
\be
%\label{10dspin}
l_{10} = l + a +2\,m_{\mathcal{K}_{(pq)}}
- \tfrac{1+(-1)^{a+l}}{2}-1\,.
\ee 
which, as showed in \eqref{delta8_factor}, is the only quantity distinguishing among the anomalous dimensions $\eta^{(0)}_{pq}$.

%====================================================================================
%=========================================

\subsection{STRINGY eigenvalue problem}\label{STRINGY_eige_sec}

%===========================================
%====================================================================================

The general claim is that genus zero STRINGY corrections to the supergravity amplitude will lift the residual degeneracy. 
The way this happens at $(\alpha')^5$ was discussed in \cite{Drummond:2020dwr}. 
In this section we generalise that discussion to all orders in $\alpha'$, covering the whole 
spectrum of two-particle operators at genus zero.

The core of our reasoning is to notice that the VS amplitude in the $\alpha'$ expansion is
 polynomial in the Mellin variables, and therefore at fixed order $(\alpha')^{n+3}$
contributes to finitely many spins in the block decomposition \cite{Heemskerk:2009pn}.
The \emph{maximal} spin $l$ we can measure is controlled by the greatest exponent of the variable $t$, 
while we keep $s$ fixed and we substituted the constraint for $u$.
This is found in $\mathcal{M}^{flat}_{n,n}$, thus can be covariatised, 
and we can work directly with an inequality for $l_{10}$,
\begin{align}
{\bf s}^n+ {\bf t}^n + {\bf u}^n \qquad\quad \longleftrightarrow \qquad\quad  l_{10}\leq n\quad,\quad n\in\mathbb{N}\ {\rm even}
\end{align}
The fact that $n$ is only even has to do with the constraint ${\bf u}=-{\bf s}-{\bf t}-4$.
For the same reason $l_{10}$ only takes even values. A more precise argument is given in appendix \ref{more_10dspin}.

For the superblock decomposition of a correlator $p_1p_2p_3p_4$ the inequality above means that we will get 
average CFT data stratified as
\be
a+l=n,n-1,n-2,\ldots 0
\ee
In terms of physical operators there is still a sum to take into account. Consider  
schematically
\be\label{example_stringy_probl}
\mathcal{A}_{p_1p_2p_3p_4}\Big|_{[aba],\tau,l} \ \ \simeq \ \ \sum_{(pq)\in R_{\vec{\tau}} } C_{p_1p_2 \mathcal{K}_{\tau,l,[aba],(pq) }  }  C_{\mathcal{K}_{\tau,l,[aba],(pq) } p_3p_4}
\ee
%This sum is tied with the matrix representation we explained in the previous section.
At tree level in supergravity the two-particle operators ${\cal K}_{\vec{\tau},(pq)}$ are only distinguished by 
$l_{10}$ given in \eqref{10dspin}, or equivalently by the level splitting label $m$, if $a$ and $l$ have been fixed.
Therefore it is preferable to write \eqref{example_stringy_probl} as
\be\label{ulteriore_expla}
\mathcal{A}_{p_1p_2p_3p_4}\Big|_{[aba],\tau,l} \ \ \simeq \ \ \sum_{m}\left[ \sum_{q} C_{p_1p_2 \mathcal{K}_{\tau,l,[aba],(pq) }  }  C_{\mathcal{K}_{\tau,l,[aba],(pq) } p_3p_4}\right]
\ee
In the flat space limit,  $l_{10}\leq n$, and for each value of $a+l$ we read off the bound $m\leq m^*$ where
\be
\begin{array}{r | c | c | c | c | c}
a+l=& n& n-1 & n-2& n-3 & \ldots\\[.2cm]
\hline
\rule{0pt}{.5cm}
m^*=&1 & 1 & 2 &  2 & \ldots 
\end{array}
\ee
Operators labelled by $m\leq m^*$ are the ones should survive if we now take the limit of large twist 
in their CFT data, which is controlled by the large $s$ limit in the Mellin amplitude.
In this paper we will assume that the inequality $l_{10}\leq n$ holds in $AdS_5\times S^5$, thus not only in a limit, 
but for all $\tau$ and $b$. 
As a result, we obtain the following bound on the $AdS_5\times S^5$ spectrum visible by ${\cal V}_n(\alpha')^{n+3}$
\begin{align}\label{level_splitting_formula}
\ \ \ m\leq m^*\qquad;\qquad m^*=\frac{n-(a+l)-\frac{1-(-1)^{a+l}}{2} }{2}+1\qquad;\qquad n\in\mathbb{N}\ {\rm even}
\end{align}
For the case $m^*=1$ and $a=0$ we will be able to prove
that our assumption is indeed correct  in section \ref{sec_rank_equal1_anadim}. 
Results for this case then naturally generalise to $a+l=n,n-1$ with $a\neq 0$ and will go under 
the name of $rank=1$ solution. For $m^*=2$ our working assumption is supported by a previous computation in \cite{Drummond:2020dwr}. 
Let us anticipate that the simple fact that we will be able to solve infinitely many rank constraints in $\tau$ and $b$ 
with finitely many parameters in our ansatz for the VS amplitude will be quite reassuring.

We understood that  $\mathcal{V}_{n}(\alpha')^{n+3}$ contributes to the CFT data 
of  all two-particle operators with  $m\leq m^*$.  But very importantly, the operators at the 
edge $m=m^*$ are the ones that get a contributions \emph{for the first time}. 

We will now translates the discussion so far into formulae by considering the OPE.
The OPE will be referred to an orthonormal basis in supergravity, as in \eqref{deco_map}, for which
we have the relation
\be
span\big(\,{\tt columns\ of\ }\mathbf{c}^{(0)}_{\vec{\tau}}\,\big)\simeq \Big[ \mathbb{V}^{}_{\vec{\tau},1 }, \mathbb{V}^{}_{\vec{\tau},\,2},  \ldots  \Big] 
\ee
The most general constraints from the OPE are discussed in appendix \ref{more_OPE}. The ones we 
need in this section can be written as the following matrix equation, 
\begin{mdframed}
\begin{gather}
 \mathbb{V}^T_{\vec{\tau},m} 
 \Big(\mathbf{c}^{(0)}{\pmb \eta}^{(n+3)}\mathbf{c}^{(0)T}+ \mathbf{D}^{(n+3)}\mathbf{N}^{(0)}\ +\ \mathbf{N}^{(0)} \mathbf{D}^{(n+3)T}\Big)  \mathbb{V}_{\vec{\tau},m'}
 = \mathbb{V}^T_{\vec{\tau},m} \mathbf{N}^{(n+3)}  \mathbb{V}_{\vec{\tau},m'} \notag\\
\ \forall m\ge 1 \qquad;\qquad   \forall m' \ge m^* 
\label{master_f_OPE}
\end{gather}
where the matrix of three-point couplings has been rotated to
\begin{align}
\mathbf{D}^{(k)}_{}= 
\mathbf{L}^{-\frac{1}{2}}\, \mathbf{C}^{(k)}\, \mathbf{c}^{(0)T}
\end{align}
\end{mdframed}
The matrix of rotated three-point couplings ${\bf D}^{(n+3)}$ has a block structure depicted below,
\be
\begin{tikzpicture}[scale=.9]
\def\step{.9};

\filldraw[red!20] (0,0) rectangle (2*\step,-2*\step);
\filldraw[green!20] (2*\step,0) rectangle (3*\step,-2*\step);
\filldraw[green!20] (0,-2*\step) rectangle (2*\step,-3*\step);

\draw[step=\step,gray] (0,0) grid (3,-3);
\draw[gray] (0,-3*\step) -- (0,-6*\step);
\draw[gray] (3*\step,0) -- (6*\step,0);
\draw[gray] (3*\step,-3*\step) -- (6*\step,-3*\step);
\draw[gray] (3*\step,-3*\step) -- (3*\step,-6*\step);

%\draw (.5*\step,-.5*\step) node[scale=1.5] {$0$};
\draw (1.5*\step,-1.5*\step) node[rotate=-45,scale=1.8] {$\ldots$};
\draw (2.5*\step,-2.5*\step) node[scale=1.5] {$0$};

\draw (1.5*\step,-4.5*\step) node[scale=4] {$0$};
\draw (4.5*\step,-4.5*\step) node[scale=4] {$0$};
\draw (4.5*\step,-1.5*\step) node[scale=4] {$0$};

\draw  (.5*\step,+.5*\step) node[scale=1] {$\mathbb{V}_{\vec{\tau},1}$};
\draw  (1.5*\step,+.5*\step) node[scale=1] {$\ldots$};
\draw  (2.5*\step,+.5*\step) node[scale=1] {$\mathbb{V}_{\vec{\tau},m^*}$};
\draw  (4.3*\step,+.5*\step) node[scale=1] {$\mathbb{V}_{\vec{\tau},m^*+1}\ \ldots$};
%\draw  (4.5*\step,+.5*\step) node[scale=1] {$\ldots$};

\draw  (+6.*\step,-.5*\step) node[scale=1] {$\mathbb{V}_{\vec{\tau},1}$};
\draw  (+6.*\step,-1.5*\step) node[scale=1] {$\vdots$};
\draw  (+6.2*\step,-2.5*\step) node[scale=1] {$\mathbb{V}_{\vec{\tau},m^*}$};
\draw  (+6.4*\step,-3.5*\step) node[scale=1] {$\mathbb{V}_{\vec{\tau},m^*+1}$};
\draw  (+6.*\step,-4.5*\step) node[scale=1] {$\vdots$};

\draw  (-2*\step,-3*\step) node[scale=1] {$\left(\mathbf{D}^{(n+3)}_{\vec{\tau}}\right)_{mm'}=$};

%space
\draw  (+8.2*\step,-3.5*\step) node[scale=1] {$\rule{.8cm}{0pt}$};

\end{tikzpicture}
\ee
The block structure of ${\bf D}$ goes together with the obvious diagonal structure 
of the matrix of anomalous dimensions ${\pmb \eta}$. Moreover, 
\begin{center}
\begin{minipage}{15cm}
1)~In the subspaces $\mathbb{V}_{\vec{\tau},m\ge 1}\otimes \mathbb{V}_{\vec{\tau},m> m^*}$ 
both ${\bf D}^{(n+3)}$ and ${\pmb \eta}^{(n+3)}$  vanish
under the assumption that the operators $\mathcal{K}_{(pq),\vec{\tau}}$ with $m> m^*$ are decoupled at that order. \\[.3cm]
2)~In the subspace $\mathbb{V}_{\vec{\tau}, m^*}\otimes \mathbb{V}_{\vec{\tau},m\leq m^*}$ (green part) of ${\bf D}^{(n+3)}$ is anti-symmetric. 
\end{minipage}
\end{center}
The content of the red part will be addressed in appendix \ref{more_OPE}. 

Combining the information from the OPE on the l.h.s.~of \eqref{master_f_OPE} with 
the r.h.s.~determined from the superblock decomposition of $\mathcal{V}_n$, 
we find, in correspondence of the previous items, 
\begin{itemize}
\item[1)] 
The rank constraints,
\be\label{rank_constraints_formula}
\mathbb{V}^T_{\vec{\tau},m}\, \mathbf{N}_{\vec{\tau}}^{(n+3)}\,  \mathbb{V}_{\vec{\tau},m'}=0 \qquad;\qquad \forall m'>m^* \quad;\quad \forall m\ge 1
\ee
\item[2)]
The the level splitting problem, written as the eigenvalue problem of 
\be\label{lev_spl_mat_1}
\mathbf{E}^{(n+3)}_{\vec{\tau},m^*}=  \left(  \mathbf{v}^T_{\vec{\tau},I}\, \mathbf{N}_{\vec{\tau}}^{(n+3)}\,\mathbf{v}_{\vec{\tau},J} \right)_{I,J} \qquad;\qquad  \mathbf{v}_{\vec{\tau},I}\in \mathbb{V}_{\vec{\tau},m^*} 
\ee
\end{itemize}
The matrix $\mathbf{E}^{}_{\vec{\tau},m^*}$ is a square matrix of dimension $dim(\,\mathbb{V}_{\vec{\tau},m^*})$.\footnote{Given 
$m^*$ and $a+l$ we will avoid the label $(n+3)$ from now on, since $n$ follows from \eqref{level_splitting_formula}.}
Its eigenvalues are the new corrections to the tree level anomalous dimensions of the operators $\mathcal{K}_{(pq),\vec{\tau}}$, with level splitting label $m^*$, 
\begin{align}\label{STRINGY_eigenvalues_problem}
{\pmb \eta}^{}_{\vec{\tau}}\Bigg|_{\mathbb{V}_{\vec{\tau},m^*}}={\tt eigenvalues}[\, \mathbf{E}^{}_{m^*,\vec{\tau}}\,]
\end{align}
and will provide the lift of the partial degeneracy of tree level supergravity. This is always a zeta-odd 
valued function, i.e.~$\sim\zeta_{n+3}(\alpha')^{n+3}$. The eigenvectors of $\mathbf{E}^{}$ single out 
particular directions on the hyperplane $\mathbb{V}_{\vec{\tau},m^*}$, 
and the full three point functions are given by
\begin{align}\label{STRINGY_eigenvector_problem}
{\bf c}^{(0)}_{\vec{\tau}}\Bigg|_{\mathbb{V}_{\vec{\tau},m^*}}=\ \mathbb{V}_{\vec{\tau},m^*}\cdot\, {\tt eigenvectors}[\, \mathbf{E}^{}_{m^*,\vec{\tau}}\ ]\
\end{align}
where the ${\tt eigenvectors}$ are taken to be orthonormal. In this way the computation of the matrix ${\bf c}^{(0)}$ is complete, 
and the spectrum of operators at genus zero is fully unmixed.

Finally, let us point out a consequence of the relation $a+l\sim n-m^*$. 
The value of $n$ here sets the order of the $\alpha'$  expansion, therefore, an operator with fixed level-splitting 
label $m^*$, in a given $su(4)$ channel $[aba]$,  but varying spin $l$, receive for the first time a correction  
to its supergravity anomalous dimension at order $\sim (\alpha')^{m^*+a+l}$. To study operators with 
large spin $l$ in the same $R_{\vec{\tau}}$ we then have to look at high orders in perturbation theory!! 
\begin{mdframed}
\begin{center}
The level splitting problem is \emph{not} a problem of fixed order in the $\alpha'$ expansion. 
\end{center}
\end{mdframed}

Before entering the details of how we impose the constraints \eqref{rank_constraints_formula}, 
let us discuss some simple cases to let the reader familiarise with 
our various statements.

%=================================================================================
%==============================================================

\subsubsection{$rank$ formula}\label{rank_section}

%==============================================================
%=================================================================================

It is nice to understand the rank constraints in \eqref{rank_constraints_formula} as a sort of exclusion plot, i.e.~we 
know how many eigenvectors of tree level supergravity are in the kernel of $\mathbf{N}^{(n+3)}$, at given order in $\alpha'$,
therefore we know that the ones not in the kernel give us the rank.

To fix ideas consider a \emph{generic} rectangle $R_{\vec{\tau}}$. 
The table below shows for the first few orders  in the  $(\alpha')^{}$ expansion, and varying values of $a+l$, 
the expected rank of $\mathbf{N}^{(n+3)}$ and the position of the edge $m^*$,
\be\label{tabella_ranghi}
\begin{array}{c|l|l|c}
(\alpha')^3  	   & a+l=0 	  	& { rank}={ 1} 			& 	m^*=1\\[.15cm]
\hline
%\\
\rule{0pt}{.7cm}
(\alpha')^5     & a+l=2,1	 & { rank}={ 1} 			& 	m^*=1\\
		  & a+l=0 		 & { rank}=3=1+{ 2} 		& 	m^*=2\\[.1cm]
\hline		  
%\\	
\rule{0pt}{.7cm}	
(\alpha')^7     & a+l=4,3 	& { rank}={ 1}         		 & 	m^*=1\\
		  & a+l=2,1 	& { rank}=3=1+{ 2} 		 & 	m^*=2\\  
		  & a+l=0 		& { rank}=6=1+2+{ 3} 		& 	m^*=3\\   
\end{array}
\ee

The $rank=1$ problem is kind of independent and corresponds to $a+l=n,n-1$. It is actually 
not unmixing any residual degeneracy, rather it gives a starting point. There is only one operator 
for any $R_{\vec{\tau}}$ corresponding to the operator labelled by the left most corner, 
which we defined by $A$, i.e.~the one highlighted in red in the figure below 
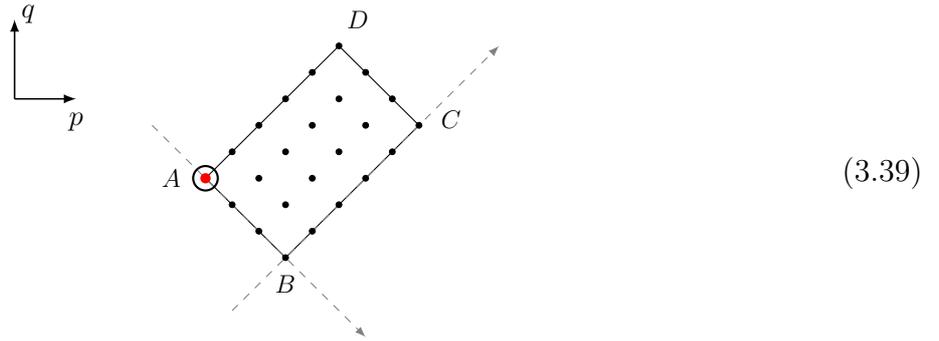
\be
\begin{array}{c}
\begin{tikzpicture}[scale=.54]
%
%\draw[step=2cm,gray,very thin] (-4,2) grid (8,8);
%
\def\prop{.65}
\def\shifthor{\prop*2}
\def\ptuno{(\prop*2-\shifthor,\prop*8)}
\def\ptdue{(\prop*5-\shifthor,\prop*5)}
\def\pttree{(\prop*7-\shifthor,\prop*13)}
\def\ptquattro{(\prop*10-\shifthor,\prop*10)}
%
%axis horizontal
\draw[-latex, line width=.6pt]		(\prop*1   -\shifthor-4,         \prop*12          -0.5*\shifthor)    --  (\prop*1  -\shifthor-2.5  ,   \prop*12-      0.5*\shifthor) ;
\node[scale=.8] (oxxy) at 			(\prop*1   -\shifthor-2.5,  \prop*13.7     -0.5*\shifthor)  {};
\node[scale=.9] [below of=oxxy] {$p$};
%
%axis vertical
\draw[-latex, line width=.6pt] 		(\prop*1   -\shifthor-4,     \prop*12       -0.5*\shifthor)     --  (\prop*1   -\shifthor-4,        \prop*15-      0.5*\shifthor);
\node[scale=.8] (oxyy) at 			(\prop*1   -\shifthor-2,   \prop*15.2   -0.5*\shifthor) {};
\node[scale=.9] [left of= oxyy] {$q$};
%
%rectangle
\draw[] 								\ptuno -- \ptdue;
\draw[black]							\ptuno --\pttree;
\draw[black]							\ptdue --\ptquattro;
\draw[]								\pttree--\ptquattro;
\draw[-latex,gray, dashed]					(\prop*0-\shifthor,\prop*10) --(\prop*8-\shifthor,\prop*2);
\draw[-latex,gray, dashed]					(\prop*3-\shifthor,\prop*3) --(\prop*13-\shifthor,\prop*13);
%		
%dots
%
\foreach \indeyc in {0,1,2,3}
\foreach \indexc  in {2,...,7}
\filldraw   					 (\prop*\indexc+\prop*\indeyc-\shifthor, \prop*6+\prop*\indexc-\prop*\indeyc)   	circle (.07);
\filldraw[red,thick]  (\prop*2+\prop*0-\shifthor, \prop*6+\prop*2-\prop*0) circle (.1);
\draw[black,thick]  (\prop*2+\prop*0-\shifthor, \prop*6+\prop*2-\prop*0) circle (.3);
%
%letters
%
\node[scale=.8] (puntouno) at (\prop*4-1.5*\shifthor,\prop*8) {};
\node[scale=.8]  [left of=puntouno] {$A$};   
\node[scale=.8] (puntodue) at (\prop*5-\shifthor,\prop*5.5+.5) {};
\node[scale=.8] [below of=puntodue]  {$B$}; 
\node[scale=.8] (puntoquattro) at (\prop*11.2-\shifthor,\prop*12.5) {};
\node[scale=.8] [below of=puntoquattro] {$C$};
\node[scale=.8] (puntotre) at (\prop*7.7-\shifthor,\prop*11.7) {};
\node[scale=.8] [above of=puntotre] {$D$}; 
\node at  (8.5,\prop*13) {\phantom{space}};
%legend
%\node[scale=.84] (legend) at (11,5) {$\begin{array}{l}  
%													\displaystyle A=(2,8); \\[.1cm]
%													\displaystyle B=(5,5); \\[.1cm]
%													\displaystyle C=(12,12); \\[.1cm]
%													\displaystyle D=(9,15); \\[.1 cm] \end{array}$  };
\end{tikzpicture}
\end{array}
\ee

When $a+l=n-2,n-3$ operators with level splitting label $m=2$ are visible, 
so as in the table let's say that  $\mathbf{N}_{\vec{\tau}}$ has $rank=3$.
The three operators in question are simply the ones labelled by $A$ 
and the pair $A+(1,1)$ and $A+(1,-1)$.
In the figure below, the pair is encircled in blue,
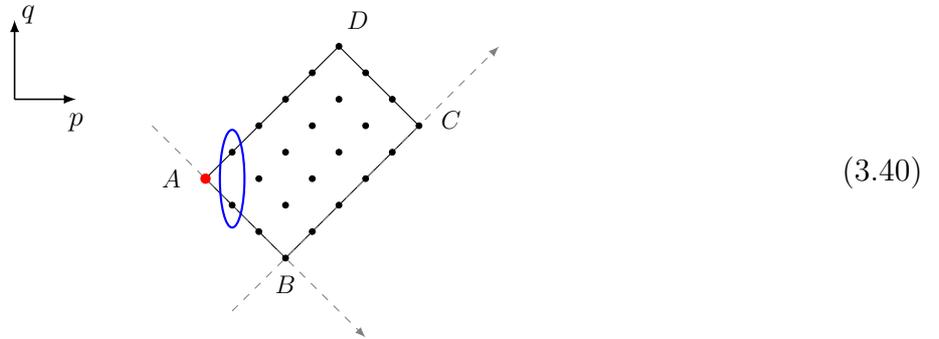
\be
\begin{array}{c}
\begin{tikzpicture}[scale=.54]
%
%\draw[step=2cm,gray,very thin] (-4,2) grid (8,8);
%
\def\prop{.65}
\def\shifthor{\prop*2}
\def\ptuno{(\prop*2-\shifthor,\prop*8)}
\def\ptdue{(\prop*5-\shifthor,\prop*5)}
\def\pttree{(\prop*7-\shifthor,\prop*13)}
\def\ptquattro{(\prop*10-\shifthor,\prop*10)}
%
%axis horizontal
\draw[-latex, line width=.6pt]		(\prop*1   -\shifthor-4,         \prop*12          -0.5*\shifthor)    --  (\prop*1  -\shifthor-2.5  ,   \prop*12-      0.5*\shifthor) ;
\node[scale=.8] (oxxy) at 			(\prop*1   -\shifthor-2.5,  \prop*13.7     -0.5*\shifthor)  {};
\node[scale=.9] [below of=oxxy] {$p$};
%
%axis vertical
\draw[-latex, line width=.6pt] 		(\prop*1   -\shifthor-4,     \prop*12       -0.5*\shifthor)     --  (\prop*1   -\shifthor-4,        \prop*15-      0.5*\shifthor);
\node[scale=.8] (oxyy) at 			(\prop*1   -\shifthor-2,   \prop*15.2   -0.5*\shifthor) {};
\node[scale=.9] [left of= oxyy] {$q$};
%
%rectangle
\draw[] 								\ptuno -- \ptdue;
\draw[black]							\ptuno --\pttree;
\draw[black]							\ptdue --\ptquattro;
\draw[]								\pttree--\ptquattro;
\draw[-latex,gray, dashed]					(\prop*0-\shifthor,\prop*10) --(\prop*8-\shifthor,\prop*2);
\draw[-latex,gray, dashed]					(\prop*3-\shifthor,\prop*3) --(\prop*13-\shifthor,\prop*13);
%		
%dots
%
\foreach \indeyc in {0,1,2,3}
\foreach \indexc  in {2,...,7}
\filldraw   					 (\prop*\indexc+\prop*\indeyc-\shifthor, \prop*6+\prop*\indexc-\prop*\indeyc)   	circle (.07);
\filldraw[red,thick]  (\prop*2+\prop*0-\shifthor, \prop*6+\prop*2-\prop*0) circle (.1);
\draw[blue,thick]  (\prop*2+\prop*1-\shifthor, \prop*6+\prop*2-\prop*0) ellipse (.3 and 1.2);
%
%letters
%
\node[scale=.8] (puntouno) at (\prop*4-1.5*\shifthor,\prop*8) {};
\node[scale=.8]  [left of=puntouno] {$A$};   
\node[scale=.8] (puntodue) at (\prop*5-\shifthor,\prop*5.5+.5) {};
\node[scale=.8] [below of=puntodue]  {$B$}; 
\node[scale=.8] (puntoquattro) at (\prop*11.2-\shifthor,\prop*12.5) {};
\node[scale=.8] [below of=puntoquattro] {$C$};
\node[scale=.8] (puntotre) at (\prop*7.7-\shifthor,\prop*11.7) {};
\node[scale=.8] [above of=puntotre] {$D$}; 
\node at  (8.5,\prop*13) {\phantom{space}};
%legend
%\node[scale=.84] (legend) at (11,5) {$\begin{array}{l}  
%													\displaystyle A=(2,8); \\[.1cm]
%													\displaystyle B=(5,5); \\[.1cm]
%													\displaystyle C=(12,12); \\[.1cm]
%													\displaystyle D=(9,15); \\[.1 cm] \end{array}$  };
\end{tikzpicture}
\end{array}
\ee
Now consider that the operator labelled by $A$ 
receives the first correction to its anomalous dimension from the $rank=1$ problem, therefore, 
the $rank=3$ problem will add a second correction to $A$, which we will not study.\footnote{ 
This correction depends on ambiguities which are not fixed within the bootstrap.}
Instead, operators labelled by $A+(1,1)$ and $A+(1,-1)$, are the ones to look at
since these are degenerate in SUGRA, and it is the first time they receive a correction. 

The reasoning for the $rank=6$ example at~$a+l=n-4,n-5$ is very similar, and we would conclude that this 
is responsible for the lifting the SUGRA degeneracy among $A+(2,2)$, $A+(2,0)$ and $A+(2,-2)$, i.e.~the ones with level splitting label $m=3$. \footnote{The $rank=6$ problem 
will add another correction to $A$, $A+(1,1)$ and $A+(1,-1)$ which again gets a contribution ambiguous within the bootstrap.}

In our table above we assumed a large rectangle $R_{\vec{\tau}}$ to start with, 
therefore for small values of $m$ we only explored operators labelled by points in 
between $A$ and $B$. The level splitting for operators lying on the right of $B$
takes place for values of $n$ and $a+l$ as in \eqref{level_splitting_formula},
but one has to pay attention to the actual numerical value of  the rank. %${rank}\ \mathbf{N}_{\vec{\tau}}$.
Graphically there are two situations,
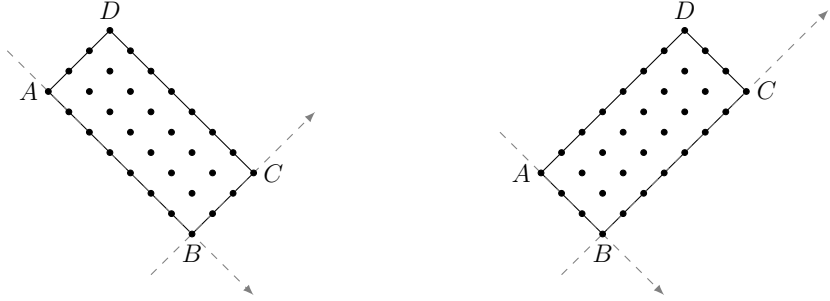
\be
\begin{tikzpicture}[scale=.54]
\def\prop{.5}
\def\shifthor{\prop*20}
\def\ptuno{(-\prop*2-\shifthor,\prop*12)}
\def\ptdue{(\prop*5-\shifthor,\prop*5)}
\def\pttree{(\prop*1-\shifthor,\prop*15)}
\def\ptquattro{(\prop*8-\shifthor,\prop*8)}
%
%
%rectangle
\draw[] 								\ptuno -- \ptdue;
\draw[black]							\ptuno --\pttree;
\draw[black]							\ptdue --\ptquattro;
\draw[]								\pttree--\ptquattro;
\draw[-latex,gray, dashed]					(-\prop*4-\shifthor,\prop*14) --(\prop*8-\shifthor,\prop*2);
\draw[-latex,gray, dashed]					(\prop*3-\shifthor,\prop*3) --(\prop*11-\shifthor,\prop*11);
%		
%dots
%
\foreach \indeyc in {0,1,2,3}
\foreach \indexc  in {0,...,7}
\filldraw   					 (-\prop*2+\prop*\indexc+\prop*\indeyc-\shifthor,\prop*12-\prop*\indexc+\prop*\indeyc)  	circle (.07);
%
%letters
%
\node[scale=.8] (puntouno) at (-\prop*0-\shifthor,\prop*12) {};
\node[scale=.8]  [left of=puntouno] {$A$};   
\node[scale=.8] (puntodue) at (\prop*5-\shifthor,\prop*6+.5) {};
\node[scale=.8] [below of=puntodue]  {$B$}; 
\node[scale=.8] (puntoquattro) at (\prop*6-\shifthor,\prop*8) {};
\node[scale=.8] [right of=puntoquattro] {$C$};
\node[scale=.8] (puntotre) at (\prop*1-\shifthor,\prop*13) {};
\node[scale=.8] [above of=puntotre] {$D$}; 
%
%
%=================================================================
%
%
\def\shifthor{\prop*0}
\def\ptuno{(\prop*2-\shifthor,\prop*8)}
\def\ptdue{(\prop*5-\shifthor,\prop*5)}
\def\pttree{(\prop*9-\shifthor,\prop*15)}
\def\ptquattro{(\prop*12-\shifthor,\prop*12)}
%
%
%rectangle
\draw[] 								\ptuno -- \ptdue;
\draw[black]							\ptuno --\pttree;
\draw[black]							\ptdue --\ptquattro;
\draw[]								\pttree--\ptquattro;
\draw[-latex,gray, dashed]					(\prop*0-\shifthor,\prop*10) --(\prop*8-\shifthor,\prop*2);
\draw[-latex,gray, dashed]					(\prop*3-\shifthor,\prop*3) --(\prop*16-\shifthor,\prop*16);
%		
%dots
%
\foreach \indeyc in {0,1,2,3}
\foreach \indexc  in {2,...,9}
\filldraw   					 (\prop*\indexc+\prop*\indeyc-\shifthor, \prop*6+\prop*\indexc-\prop*\indeyc)   	circle (.07);
%
%letters
%
\node[scale=.8] (puntouno) at (\prop*4-\shifthor,\prop*8) {};
\node[scale=.8]  [left of=puntouno] {$A$};   
\node[scale=.8] (puntodue) at (\prop*5-\shifthor,\prop*6+.5) {};
\node[scale=.8] [below of=puntodue]  {$B$}; 
\node[scale=.8] (puntoquattro) at (\prop*13-\shifthor,\prop*15) {};
\node[scale=.8] [below of=puntoquattro] {$C$};
\node[scale=.8] (puntotre) at (\prop*9-\shifthor,\prop*13) {};
\node[scale=.8] [above of=puntotre] {$D$};

\notag
\end{tikzpicture}
\ee
and the general formula is
\begin{align}
{rank}\ \mathbf{N}^{}_{\vec{\tau}}\Big|_{m^* }  
=\#\{  (p,q){\rm\ with\ }\ 2+a\leq p\leq a+m^*+1\} 
\end{align}
where the r.h.s.~is simply counting the points in $R_{\vec{\tau}}$ of the form $(p,q)$ with $p\leq a+m^*+1$. 
%

%=======================================================================================
\subsubsection{Tailoring the bootstrap program}\label{tailoring_sec}
%=======================================================================================

Our bootstrap algorithm, begins by taking a crossing invariant ansatz for  $\mathcal{V}_n$, 
i.e.~the one we built in section \ref{VS_section}, and computing the matrices $\mathbf{N}_{\vec{\tau}}^{(n+3)}$, 
as function of the parameters in the ansatz. Then, we impose the rank constraints 

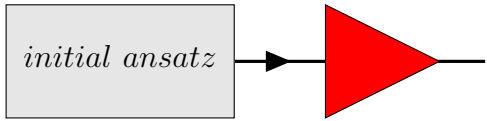
\be\label{rank_constr_tailored}
\begin{tikzpicture}
\def\inix{0};
\def\iniy{0};
\def\stepx{3};
\def\stepy{1.5};

% ARROWS
\draw[-,very thick] (\inix+1.2*\stepx,\iniy+0.5*\stepy)  --  (\inix+2.1*\stepx,\iniy+0.5*\stepy) ;
\draw[-triangle 45,very thick] (\inix+0.5*\stepx,\iniy+0.5*\stepy) --  (\inix+1.25*\stepx,\iniy+0.5*\stepy) ;

\filldraw[gray!20,draw=black] (\inix,\iniy) rectangle (\inix+\stepx,\iniy+\stepy); 
\draw[very thick]  (\inix+0.5*\stepx,\iniy+0.55*\stepy) node[scale=1] {$initial\ ansatz$};

\filldraw[red,draw=black]  (\inix+1.4*\stepx,\iniy) --  (\inix+1.9*\stepx,\iniy+0.5*\stepy) --  (\inix+1.4*\stepx,\iniy+\stepy) -- cycle;
\draw[very thick]  (\inix+1.7*\stepx,\iniy-0.4*\stepy) node[scale=1] {$\mathbb{V}^T_{\vec{\tau},m}\, \mathbf{N}_{\vec{\tau}}^{(n+3)}\,  \mathbb{V}_{\vec{\tau},m'}=0$};
\draw[very thick]  (\inix+3.3*\stepx,\iniy-0.4*\stepy) node[scale=1] {$\forall m'>m^* \quad;\quad \forall m\ge 1.$};

\end{tikzpicture}
%\label{boot_scheme}
\ee
Notice that a necessary intermediate step here is to compute a basis of orthonormal eigenvectors
of the supergravity matrix $\mathbf{N}^{(0)}_{\vec{\tau}}$, which we borrow from  \cite{Aprile:2018efk}. 
Many details about how to perform the superblock decomposition are given in appendix \ref{more_OPE}.

Equations \eqref{rank_constr_tailored} are linear in the parameters $\vec{\ltr} $ of the ansatz and can be rearranged as a linear system of the form $\mathscr{L}\cdot\,\vec{\ltr} =\vec{f}$,
where on the r.h.s.~we have put the covariantised flat space contribution, which is known. 
We repeat our procedure for many rectangles $R_{\vec{\tau}}$ until the solution of the linear system saturates. A convenient way to do so
is to consider first a selection of quantum numbers $\tau$ and $[aba]$ with $a+l=n$, then add the results from 
another selection of quantum numbers $\tau$ and $[aba]$ with $a+l=n-1$, and keep going until $a+l=0$. 
In principe we can take infinite values of $\tau$ and $b$. In practise we have taken finitely many for each $[aba]$, 
and we have seen the system saturating already at half way.

In the table below we summarise how many independent conditions are imposed from the rank constraints,
\be\label{table_coeff_out}
\rule{.7cm}{0pt}
\begin{array}{c|l|l}
 			&initial\ ansatz & rank\ constraints  \\
			\hline
			\rule{0pt}{.8cm}
{\cal V}_2 		& \rule{1cm}{0pt}	6	       &   	\rule{1cm}{0pt}		4\\[.1cm]
{\cal V}_3 		& \rule{1cm}{0pt}	18	       &   	\rule{1cm}{0pt}		14 \\[.1cm]
{\cal V}_4 		& \rule{1cm}{0pt}	44	       &   	\rule{1cm}{0pt}		34 \\[.1cm]
{\cal V}_5 		& \rule{1cm}{0pt}	98	       &  	\rule{1cm}{0pt}		82 \\[.1cm]
%{\cal V}_{6,6} 		& \rule{1cm}{0pt} 208 (-17)  \    		& 	 \rule{1cm}{0pt}		176\\
%{\cal V}_{6,4} 		& \rule{1cm}{0pt} 208 (-17)  \    		& 	 \rule{1cm}{0pt}		176\\

\end{array}
\ee
The number of initial parameters is the one counted by using the table in \eqref{tabella_H}, where $\mathcal{M}_{n,n}^{flat}$ is assumed.
As $n$ increases the number of new crossing invariants $\mathcal{H}_{n,(\ell,n-\ell)}$ grows as well, and moreover,
more spin structures are turned on, as it is the case  in flat space (see discussion in appendix \ref{more_10dspin}).
The first case in which more than one spin structure is turned on in flat space is at $(\alpha')^9$, 
i.e.~the spin six contribution $({\bf s}^6+{\bf t}^6+{\bf u}^6)$ and the spin four $({\bf s}^3+{\bf t}^3+{\bf u}^3)^2$ contribution.
In this case there are two different problems, 
\be\label{table_coeff_out}
\begin{array}{c|l|l}
 			&initial\ ansatz & rank\ constraints  \\
			\hline
			\rule{0pt}{.8cm}
%{\cal V}_2 		& \rule{1cm}{0pt}	6	       &   	\rule{1cm}{0pt}		4\\[.1cm]
%{\cal V}_3 		& \rule{1cm}{0pt}	18	       &   	\rule{1cm}{0pt}		14 \\[.1cm]
%{\cal V}_4 		& \rule{1cm}{0pt}	44	       &   	\rule{1cm}{0pt}		34 \\[.1cm]
%{\cal V}_5 		& \rule{1cm}{0pt}	98	       &  	\rule{1cm}{0pt}		82 \\[.1cm]
{\cal V}_{6,spin=6} 		& \rule{1cm}{0pt} 208 (-17)  \    		& 	 \rule{1cm}{0pt}		176\\
{\cal V}_{6,spin=4} 		& \rule{1cm}{0pt} 208 (-17)  \    		& 	 \rule{1cm}{0pt}		176\\

\end{array}
\ee
%This explains intuitively why the number of initial parameters in ${\cal V}_6$ essentially double compared to ${\cal V}_{5}$.
%Moreover, each spin comes with its own set of rank constraints. 
%We will come back on this point shortly on the use of the parameter $q$.
%
%
%We used $\mathcal{M}_{6,6}^{flat}=({\bf s}^6+{\bf t}^6+{\bf u}^6) + {q}\, ({\bf s}{\bf t} {\bf u})^2$. 
where for computational simplicity we also fixed a particular gauge.\footnote{We
set $span( \mathcal{H}_{2,1,0} )$ to zero and we only kept the terms with no $\Sigma$ in $span(\mathcal{H}_{3})$ 
and set to zero the others. In total we used a gauge with 17 parameters set to zero, as given in the ancillary file.} 
In both cases we have found the same number of constraints, as shown in the ancillary file.
Notice that since ${\cal V}_{6}$ is the completion of two spin structures, rather than one, 
the number of new crossing invariants in ${\cal V}_6$ essentially doubles compared to ${\cal V}_{5}$, see the counting in table \eqref{tabella_H}.
%The parameter $q$ is introduced for later convenience.

In some cases we can exploit the OPE even further, especially if we can set something to zero. 
For example, at $(\alpha')^{6,7}$, we can look at the subspace of operators with $a+l=0$ and $m=2$, 
consisting of two degenerate operators at tree level in SUGRA. These operators were at the edge of the 
$(\alpha')^5$ contribution, thus the amplitude ${\cal V}_2(\alpha')^5$ unmixes them and returns 
well defined three-point coupling in the ${\bf c}^{(0)}_{\vec{\tau}}$ matrix, 
let's say $[{\bf c}^{(0)}_{i=1,2}]_{\tau,0,[0b0],m=2}$ orthonormal.\footnote{Explicitly written in appendix \ref{singlet_eigenvectors}.} From the OPE follows that  
\ba\label{extra_condition}
\rule{.5cm}{0pt}
0= \Bigg[ [{\bf c}^{(0)}_{1}]^T
 \Big(\mathbf{c}^{(0)}{\pmb \eta}^{(k)}\mathbf{c}^{(0)T}+ \mathbf{D}^{(k)}\mathbf{N}^{(0)}\ +\ \mathbf{N}^{(0)} \mathbf{D}^{(k)T}\Big) [{\bf c}^{(0)}_{2}]\Bigg]_{\tau,0,[0b0],m=2}\!\!\!\!\quad;\qquad k=6,7\notag\\
 \ea
because both ${\bf D}^{(6)}$ and ${\bf D}^{(7)}$ are just anti-symmetric at this order. Therefore,
\be
0=\Bigg[[{\bf c}^{(0)}_{1}]^T\, \mathbf{N}^{(k)}\,  [{\bf c}^{(0)}_{2}]\Bigg]_{\tau,0,[0b0],m=2} \qquad;\qquad \forall \tau,b,\quad;\quad k=6,7
\ee
This is a new condition in addition to the rank constraints, which we expect to saturate for all values of $b$ and $\tau$.  
For $(\alpha')^6$  we find that no independent constraint is added, while at $(\alpha')^7$ we find a new relation among free parameters.  
This is reasonable because the operators we are using here are strictly below the edge of $(\alpha')^7$, but not for 
$(\alpha')^6$. At order $(\alpha')^8$ we find two more constraints, and we checked instead that some of the new data 
from unmixing operators at the edge of $(\alpha')^7$ is automatically implemented, similarly to the behaviour 
between $(\alpha')^6$ and $(\alpha')^5$. We have attached the results in the ancillary file.

%========================================================================================
\paragraph{Uniqueness of the CFT data at the edge.}~\\[-.2cm]
%========================================================================================

The reformulation of the rank constraints as a linear system $\mathscr{L}\cdot\,\vec{\ltr} =\vec{f}$, 
where $f$ comes solely from the flat space contribution, explains how $\mathcal{M}^{flat}_{n,n}$ propagates 
into $\mathcal{V}_n$. If $f$ is determined uniquely, the solution consists of a particular one, supplemented by ${\tt ker}\mathscr{L}$. 
The particular solution, which depends both on $\mathscr{L}$ and $f$, is the most interesting part for the CFT data.

For $n$ even, we look at the one and only one term with greatest 10d spin in the covariantised 
flat space amplitude, ${\bf s}^n+{\bf t}^n+{\bf u}^n$. Since ${\bf u}=-{\bf s}-{\bf t}-4$ with ${\bf s}=s+\tilde s$ 
and ${\bf t}=t+\tilde t$, this contribution is not homogeneous in $s,\tilde s,t,\tilde t$ and  
it contributes to the CFT data at the edge at $m=m^*$ for all values of $a+l\leq n$.
Therefore $f$ is unambiguous, and we expect that after solving the linear system the CFT data at the edge at
 $m=m^*$ is fixed uniquely. This will indeed be the case.

For $n$ odd or any contribution in the flat space amplitude given by products of amplitudes at previous orders,
the reasoning above does not go through since the ansatz will contain at least two types of terms 
contributing to the same 10d spin, meaning that $f$ is ambiguous. A nice example is $(\alpha')^9$ 
which contains both ${\bf s}^6+{\bf t}^6+{\bf u}^6$ and $({\bf s}{\bf t}{\bf u})^2$. But the first one has $l_{10}=6$, 
and it is the first time that such a value of the 10d spin appears in the $\alpha'$ expansion, 
while the second one has $l_{10}=4$, so it will mix with ${\bf s}^4+{\bf t}^4+{\bf u}^4$ present in the ansatz.
The CFT data at the edge of $(\alpha')^9$ is the one corresponding to $l_{10}=6$, 
for which the term $f$ is uniquely determined by ${\bf s}^6+{\bf t}^6+{\bf u}^6$.
Experimentally, we checked that  if we introduce a parameter $q$ to deform the spin six problem
as ${\bf s}^6+{\bf t}^6+{\bf u}^6+q({\bf s t u})^2$,
the CFT data at the edge is independent of $q$, as it should.

In fact, in no way the free parameters lefts in the ansatz after imposing the rank 
constraints affect the computation of the level splitting matrix 
\be\label{lev_spl_mat}
\mathbf{E}^{(n+3)}_{\vec{\tau},m^*}=  \left(  \mathbf{v}^T_{\vec{\tau},I}\, \mathbf{N}_{\vec{\tau}}^{(n+3)}\,\mathbf{v}_{\vec{\tau},J} \right)_{I,J} \qquad ;\qquad \mathbf{v}_{\vec{\tau},I}\in \mathbb{V}_{\vec{\tau},m^*} 
\ee
In practise, even though $\mathbf{N}_{\vec{\tau}}^{}$ still depends on free parameters, when we project on $\mathbb{V}_{\vec{\tau},m^*}$ 
they cancel out, as they should.
For concreteness, consider again the amplitude at $(\alpha')^5$, 
\begin{align}
\mathcal{V}_2&= \ (\Sigma-1)_3 \mathcal{M}_{2,0}+(\Sigma-1)_{4}\ltr_{4,1}^{} \left( {\bf s}\tilde{\bf s} + {\bf t}\tilde{\bf t}+ {\bf u}\tilde{\bf u} \right)+(\Sigma-1)_{5}( {\bf s}^2 + {\bf t}^2+{\bf u}^2 )\ ;\\[.3cm]
&\mathcal{M}_{2,0}=\ \ltr_{3,1}^{} \Sigma^2+ \ltr_{3,2}^{}(c_s^2+c_t^2+c_u^2) + \ltr_{3,3}^{}\left( \tilde{\bf s}^2+\tilde{\bf t}^2+{\bf \tilde{u}}^2 \right)+ \ltr_{3,4}^{} \Sigma  + \ltr_{3,5}^{}\ \notag
\end{align}
The rank constraints give four relations for six coefficients, namely
\be%\label{rank_constraints_V2}
\ltr_{4,1}^{}=-5\qquad;\qquad \ltr_{3,3}^{}=5\qquad;\qquad \ltr_{3,2}^{}-\ltr_{3,1}^{}=11\qquad;\qquad \ltr_{3,4}=0
\ee
It is simple to confirm with computer algebra that both $\mathbf{E}^{(5)}_{\vec{\tau},2}$ at $a+l=0$, and the CFT data at $m^*=1$ with $a+l=2,1$, 
do not depend on the two remaining free parameters, despite the fact that the amplitude at this point still does. 
Because of this property,  an idea would be that the rank constraints are solved by an ansatz which 
is as close as possible to a homogeneous polynomial, 
which is further decomposed in powers of $\Sigma$. 
This of course is not the true amplitude but 
it would be enough to compute the level splitting matrices \eqref{lev_spl_mat}. 
We tried this experiment and at $(\alpha')^{7,9}$ we found that 
only the terms with $\checkmark$ are needed to saturate the rank constraints.
\be\label{tabella_minimal_amplitude}
\begin{array}{c|c|c|c}
 (\alpha')^7  			&1  		& \Sigma^1 	& \Sigma^2	\\
 \hline 
 \rule{0pt}{.8cm}
{\rm degree\ 4\  polynomial}	 &      \checkmark  & \checkmark 	&\checkmark         \\[.1cm]
{\rm degree\ 3\  polynomial}		 &      \checkmark  &  	\checkmark		&         			
\end{array}
\qquad;\qquad
\begin{array}{c|c|c|c|c}
 (\alpha')^9  			&1  			    & \Sigma^1 	       & \Sigma^2	    &  \Sigma^3   \\
 \hline 
 \rule{0pt}{.8cm}
{\rm degree\ {6}\  polynomial}		 &      \checkmark  & \checkmark 	&\checkmark          &   \checkmark       \\[.1cm]
{\rm degree\ {5}\  polynomial}		 &      \checkmark  &  \checkmark	& \checkmark	     &        \\[.1cm]
{\rm degree\ {4}\  polynomial}		 &      \checkmark  &  \checkmark	& 	     &       
\end{array}\quad
\ee
We then computed the level splitting matrices with random values of the other parameters, multiple times, and checked extensively
that none of them was affecting the final result for the level splitting problem.

In this regard we believe that the uniqueness of the CFT data at the edge at $m^*$  
strongly suggests that a preferred sub-amplitude exists, and we just have to 
look at the greater picture \cite{NEWPAUL}.

%====================================================================================
\subsection{All anomalous dimensions at $rank=1$ and $a+l=n,n-1$}\label{sec_rank_equal1_anadim}
%====================================================================================

%Before exploring the level splitting problem in the next section, 
In this section
we study in isolation the case of $rank=1$ at $a+l=n,n-1$,
because it can be solved \emph{independently} at all orders in $\alpha'$.
We will then be able to demonstrate  various properties of the $m^*=1$ 
anomalous dimensions w.r.t.~the quantum numbers $\vec{\tau}$ analytically.
These properties will provide the starting point for the analysis of the 
more general characteristic polynomial when $m^*>1$ in the next section.

When the matrix $\mathbf{N}^{(n+3)}_{\vec{\tau}}$ has $rank=1$ 
only the operator on the left most corner of $R_{\vec{\tau}}$ is getting a correction to its CFT data. 
\be
\begin{array}{c}
\begin{tikzpicture}[scale=.5]
%
%\draw[step=2cm,gray,very thin] (-4,2) grid (8,8);
%
\def\prop{.65}
\def\shifthor{\prop*2}
\def\ptuno{(\prop*2-\shifthor,\prop*8)}
\def\ptdue{(\prop*5-\shifthor,\prop*5)}
\def\pttree{(\prop*7-\shifthor,\prop*13)}
\def\ptquattro{(\prop*10-\shifthor,\prop*10)}
%
%axis horizontal
\draw[-latex, line width=.6pt]		(\prop*1   -\shifthor-4,         \prop*12          -0.5*\shifthor)    --  (\prop*1  -\shifthor-2.5  ,   \prop*12-      0.5*\shifthor) ;
\node[scale=.8] (oxxy) at 			(\prop*1   -\shifthor-2.5,  \prop*13.7     -0.5*\shifthor)  {};
\node[scale=.9] [below of=oxxy] {$p$};
%
%axis vertical
\draw[-latex, line width=.6pt] 		(\prop*1   -\shifthor-4,     \prop*12       -0.5*\shifthor)     --  (\prop*1   -\shifthor-4,        \prop*15-      0.5*\shifthor);
\node[scale=.8] (oxyy) at 			(\prop*1   -\shifthor-2,   \prop*15.2   -0.5*\shifthor) {};
\node[scale=.9] [left of= oxyy] {$q$};
%
%rectangle
\draw[] 								\ptuno -- \ptdue;
\draw[black]							\ptuno --\pttree;
\draw[black]							\ptdue --\ptquattro;
\draw[]								\pttree--\ptquattro;
\draw[-latex,gray, dashed]					(\prop*0-\shifthor,\prop*10) --(\prop*8-\shifthor,\prop*2);
\draw[-latex,gray, dashed]					(\prop*3-\shifthor,\prop*3) --(\prop*13-\shifthor,\prop*13);
%		
%dots
%
\foreach \indeyc in {0,1,2,3}
\foreach \indexc  in {2,...,7}
\filldraw   					 (\prop*\indexc+\prop*\indeyc-\shifthor, \prop*6+\prop*\indexc-\prop*\indeyc)   	circle (.07);
\filldraw[red,thick]  (\prop*2+\prop*0-\shifthor, \prop*6+\prop*2-\prop*0) circle (.1);
\draw[black,thick]  (\prop*2+\prop*0-\shifthor, \prop*6+\prop*2-\prop*0) circle (.3);
%
%letters
%
\node[scale=.8] (puntouno) at (\prop*4-1.5*\shifthor,\prop*8) {};
\node[scale=.8]  [left of=puntouno] {$A$};   
\node[scale=.8] (puntodue) at (\prop*5-\shifthor,\prop*5.5+.5) {};
\node[scale=.8] [below of=puntodue]  {$B$}; 
\node[scale=.8] (puntoquattro) at (\prop*11.2-\shifthor,\prop*12.5) {};
\node[scale=.8] [below of=puntoquattro] {$C$};
\node[scale=.8] (puntotre) at (\prop*7.7-\shifthor,\prop*11.7) {};
\node[scale=.8] [above of=puntotre] {$D$}; 
\node at  (8.5,\prop*13) {\phantom{space}};
\end{tikzpicture}
\end{array}
\ee
The only relevant quantity to compute is the anomalous dimension, since 
the correction to the three-point function is a vanishing one-by-one anti-symmetric matrix.  
The first set of $m^*=1$ anomalous dimensions are found for $a+l=n$ even, in the following channels
\begin{align}
(\alpha')^{n+3}\quad;\quad n=0,2,4\ldots\qquad;\qquad [aba]\quad;\quad \begin{array}{llll} a=& n,& n-1, &\ldots   \\ l= &0,&1,& \ldots \end{array} 
\end{align}
The second set of $m^*=1$ anomalous 
dimensions are found for $a+l=n-1$ odd, in the following channels, 
\begin{align}
\rule{.3cm}{0pt}(\alpha')^{n+3}\quad;\quad\  n=2,4\ldots\qquad;\qquad \quad [aba]\ ;\quad \begin{array}{llll} a=& n-1,& n-2, &\ldots   \\ l= &0,&1,& \ldots \end{array} 
\end{align}

To extract the anomalous dimension 
consider the matrix $\mathbf{A}^{(n+3)}_{\vec{\tau} }=
\mathbf{M}^{(n+3)}_{\vec{\tau} }\cdot\,\,\mathbf{L}_{\vec{\tau} }^{-1}$,  with the same conventions as in section \ref{rew_sec_SUGRA}, and
notice that being $rank=1$ its minimal polynomial is $\mathbf{A}^{(n+3)}_{\vec{\tau}}\cdot
\mathbf{A}^{(n+3)}_{\vec{\tau}}=\eta^*_{\vec{\tau} }\,\mathbf{A}^{(n+3)}_{\vec{\tau}}$. 
Therefore,
\begin{align}
\mathbf{M}^{(n+3)}_{\vec{\tau}}\cdot \mathbf{L}_{\vec{\tau}}^{-1}\cdot \mathbf{M}^{(n+3)}_{\vec{\tau}} = \eta^*_{\vec{\tau}}\ \mathbf{M}^{(n+3)}_{\vec{\tau}}
\end{align}
is a true equation for each component. 
Since  $\mathbf{L}_{\vec{\tau}}$ is diagonal, we arrive at are 
\begin{align}\label{component_rank1}
\eta^*_{\vec{\tau}}=
\frac{1}{\left(\mathbf{M}^{(n+3)}_{\vec{\tau}}\right)}_{p_1p_2,p_3p_4  }\!\!\!\!\times \ \ \sum_{r,s}\  
\frac{ \left(\mathbf{M}^{(n+3)}_{\vec{\tau}}\right)_{p_1p_2,rs} \left( \mathbf{M}^{(n+3)}_{\vec{\tau}} \right)_{rs,p_3p_4}  }{ \left(\mathbf{L}_{\vec{\tau}}\right)_{rs,rs} } 
\end{align} 
Instead of computing $\mathbf{M}^{(n+3)}$ explicitly from the VS amplitude, going back to 
the discussion in section \ref{rew_sec_SUGRA}, the idea here is to obtain $\eta^*_{\vec{\tau}}$  
by solving a problem of $rank=1$ matrices built out of $\overline{D}$ functions 
of the correct form,  which has the case $(\alpha')^3$ and the $[0b0]$ channels as starting point. 
The crucial point is the fact that since $\eta^*_{\vec{\tau}}$ cannot depend on $p_1p_2p_3p_4$, 
whatever algebra takes place on the r.h.s.~of \eqref{component_rank1} it must be such that the 
dependence on the external charges cancels out. 

We will find that there is a natural (probably unique) and very non trivial way of solving \eqref{component_rank1}. 
The details are explained in great details in appendix \ref{rank1_app}. A posteriori, we can then show that in order for 
$\eta^*_{\vec{\tau}}$ to be independent from $p_1p_2p_3p_4$ we need the VS amplitude to 
be such that \footnote{For the definition of the factor ${\tt Y}_{[aba]}$ see \eqref{formulaBMichele}. 
%
%\be
%\!\!
%{\tt Y}_{[aba],\vec{p}}=\frac{(\Sigma-2)! b!(b+1)!(2+a+b)  }{  \Gamma[ \pm \frac{p_1-p_2}{2} + \frac{b+2}{2}]  
%\Gamma[\frac{p_1+p_2}{2} + \frac{b+2}{2}]   \Gamma[\frac{p_1+p_2}{2} - \frac{b+2a+2}{2}]  
%\Gamma[ \pm \frac{p_3-p_4}{2} + \frac{b+2}{2}] \Gamma[\frac{p_3+p_4}{2} + \frac{b+2}{2} ] \Gamma[\frac{p_3+p_4}{2} - \frac{b+2a+2}{2} ] }\notag
%\ee
} 
\be\label{proof_aln}
\mathcal{V}_{n}\Big|_{[aba],l=n-a}={2}{\texttt{Bin}[\substack{n\\ l}]}\zeta_{n+3}\, (\Sigma-1)_{3+l}\, {\tt Y}_{[aba]}\times \overline{D}_{p_1+2+l,p_2+2,p_3+2,p_4+2+l} 
\ee
This is a valuable result because allows us to deduce information about 
the the homogeneous top terms $\tilde{u}^{d_1} \tilde{t}^{d_2} t^l$ with $d_1+d_2=a$ and $a+l=n$ in $\mathcal{V}_{n}$. 
The case $a=0$ only comes from $\mathcal{M}_{n,n}^{flat}$, but the decomposition 
of the amplitude onto the $su(4)$ channel  $[aba]$ for $a\neq 0$ and spin $l=n-a$ 
mixes up the various strata of ${\cal V}_n$, and these have to recombine (smartly) a 
pochhammer $(\Sigma - 1 - a)_{a + l + 3}$ in order to reproduce the r.h.s.~of \eqref{proof_aln}.

The solution of the $rank=1$ problem gives us the following representation of the anomalous dimensions
\begin{align}
\label{intro_thesums}
\!\!\!\!\!\tfrac{1}{2}\eta^*_{\vec{\tau}}= &\,\zeta_{n+3} \frac{n!}{(a+1)!(l+1)!}\, \frac{ (-)^l \,\delta^{(8)}_{\tau,l,[aba]}}{ (b+1)_{2a+3} (\tau+1)_{2l+3} }\times\\
&
\ \sum_{(rs)}\, \bigg[\ \frac{ rs}{ (1+\delta_{rs}) }\prod_{m=1}^{1+l} \left( (\tfrac{\tau}{2}+m)^2 -(\tfrac{r\pm s}{2} )^2 \right) 
\prod_{m=1}^{1+a} \left( (\tfrac{b}{2}+m)^2 -(\tfrac{r\pm s}{2} )^2 \right)\bigg] \times \texttt{factor}_{a+l}(r,s)
\notag
\end{align}
where
\begin{align}
\texttt{factor}_{n}=1\qquad;\qquad\texttt{factor}_{n-1}=\frac{(r^2- s^2)^2}{8} \left[ \frac{1}{\tau(\tau+2l+4)}-\frac{1}{b(b+2a+2)}\right]
\end{align}
where the sum $\sum_{r,s}$ goes over $R_{\vec{\tau}}$. 
A posteriori, we checked that \eqref{intro_thesums} agrees with the direct computation
\be
\eta^*_{\vec{\tau}}=  \mathbb{V}^T_{\vec{\tau},m^*=1}\, {\bf N}^{(n+3)}_{\vec{\tau}}\, \mathbb{V}_{\vec{\tau},m^*=1}
\ee
where $\mathbb{V}_{\vec{\tau},1}$ consists of the most negative singlet eigenvector 
of ${\bf N}^{(0)}_{\vec{\tau}}$, which we wrote (almost for all cases) in appendix \ref{singlet_eigenvectors}, 
and ${\bf N}^{(n+3)}$ is computed by using the amplitudes presented in section \ref{sec_solutions}. 

Our computation 
thus shows that there is a one-to-one correspondence between the projection of 
the amplitude in \eqref{proof_aln} and $\mathbb{V}_{\vec{\tau},1}$. Similarly for $a+l=n-1$.
Moreover, formula $\eqref{intro_thesums}$ suggests very strongly the presence of the symmetry 
\begin{align}\label{my_symm}
\tau\leftrightarrow b\qquad;\qquad a\leftrightarrow l
\end{align} 
In order to infer how the anomalous dimension behave under \eqref{my_symm} 
we have to resum, since the sum over $(rs)$ still depends (implicitly) on $b$.
Very nicely, the result is the following

\paragraph{$a+l=n$ even.}
\be
\label{anomdimnnmin1}
\eta^*_{\vec{\tau}}=
-2\times \zeta_{n+3}\frac{n!(n+4)!}{(2n+8)!}\  \delta^{(8)}_{[aba],\tau,l}\ 
\left(\frac{\tau}{2}-\frac{b+2a+2}{2}\right)_{\!\!n+3} \left(\frac{\tau}{2}+\frac{b+2}{2}\right)_{\!\!n+3}
\ee

The anomalous dimension is \emph{odd} under the symmetry \eqref{my_symm}, i.e. the first pochhammer 
flips depending on the value of $a+l+3$, therefore if $a+l$ is even there are an odd numbers of terms, 
thus we pick a negative sign. The second pochhammer is invariant.   

\paragraph{$a+l=n-1$ odd.}
\be
\label{anomdimnnmin1poly}
\eta^*_{\vec{\tau}}= -\mathcal{F}_{\vec{\tau},n}\times \frac{\tau(\tau+2l+4)-b(b+2a+4)}{4}
\ee
where we defined 
\be\label{def1_F}
{\cal F}_{\vec{\tau},n}\equiv+{2}\times\zeta_{n+3}\frac{n!(n+4)!}{(2n+8)!}\  \delta^{(8)}_{[aba],\tau,l}\ 
\left(\frac{\tau}{2}-\frac{b+2a+2}{2}\right)_{\!\!a+l+3} \left(\frac{\tau}{2}+\frac{b+2}{2}\right)_{\!\!a+l+3} 
\ee

\paragraph{Observations.}%~\\
\begin{center}
\begin{minipage}{15cm}
%\begin{itemize}
%\item[1.]
1)~Notice that for both $a+l=n$ and $a+l=n-1$, the total degree in twist of $\eta^*$ is unchanged. 
The $-1$ in the odd case is regained by the $\frac{\tau(\tau+2l+4)-b(b+2a+4)}{4}$ contribution. 
This polynomial is more compactly $T-B$, where
\be\label{def1_TB}
T\equiv\tfrac{1}{4}\tau(\tau+2l+4)\qquad;\qquad B\equiv \tfrac{1}{4}b(b+2a+4)
\ee

%\item[2.]
2)~The anomalous dimension are \emph{odd} under the symmetry $T\leftrightarrow B$ and $l\leftrightarrow a$.  \\

%\item[2.]
3)~Both anomalous dimensions in \eqref{anomdimnnmin1} and \eqref{anomdimnnmin1poly} are negative definite 
for physical values of $\vec{\tau}$, and for given value of $n$ can be written solely in terms of $T$ and $B$, $a$ and $l$.  \\

%\item[3.]
4)~Upon factoring out $\mathcal{F}$, we find unity when $a+l=n$ even, and $T-B$ when $a+l=n-1$ odd. 
Notice that if we assume $T$ is present, then we know that $B$ is also present because the flat space 
amplitude cannot distinguish $\tau$ from $b$, and $a$ from $l$, thus they have to appear on equal footing 
at leading order. This is equivalent to saying that the flat space Mellin amplitude only depends on ${\bf s}=
s+\tilde s$ and ${\bf t}=t+\tilde t$. We infer in this way that the limit from $AdS_5\times S^5$ to flat space limit 
is more properly the limit in which $T$ and $B$ scale in the same way and are large.  
%\end{itemize}
\end{minipage}
\end{center}
~
With these infomations at our disposal, we are now ready to study the splitting  of degenerate (long) 
two-particle operators at tree level in supergravity.

%=======================================================================================
%=========================================

\subsection{Level splitting}

%=========================================
%=======================================================================================

The dual of the level splitting problem is a quantum mechanical problem about the bulk 
S-matrix of the four point scattering process. The accidental 10d conformal symmetry of the supergravity contribution allows 
us to think of this scattering process as if it was  taking place at the bulk saddle point given by 
the large $p$ expansion of the correlator \cite{Aprile:2020luw}, where the only geometry that matters is flat space. 
% WE SHOWED THAT THE MELLIN AMPLITUDE IS GIVEN BY AN EVALUATION FORMULA
Adding $\alpha'$ corrections to this picture adds curvature effects of the actual $AdS_5\times S^5$ 
background, lifts the symmetry, and delocalises the bulk point. Since the curvature is sourced 
by the Ramond flux, we can think of the breaking of the accidental symmetry as an analog of the Stark/Zeeman effect. 

It will be convenient to define a rescaled anomalous dimension, following the discussion for 
the case $m^*=1$ of the previous section, 
\ba
\label{def2_F}
\eta^*_{\vec{\tau},m}&=&{\cal F}_{\tau,l,[aba],n}\ \tilde{\eta}_{\vec{\tau},m}
\ea
The factor ${\cal F}$ is precisely the one in \eqref{def1_F},
\be%\label{def1_F}
{\cal F}_{\tau,l,[aba],n}\equiv+{2}\times\zeta_{n+3}\frac{n!(n+4)!}{(2n+8)!}\  \delta^{(8)}_{[aba],\tau,l}\ \left(\frac{\tau}{2}-\frac{b+2a+2}{2}\right)_{\!\!a+l+3} \left(\frac{\tau}{2}+\frac{b+2}{2}\right)_{\!\!a+l+3} 
\ee
and the notation $\eta^*$ will always refer to the edge $m=m^*$ of a rectangle $R_{\vec{\tau}}$.
But we shall keep the label $m$ unspecified,
because when we focus on a given amplitude ${\cal V}_n$ 
various values of $m^*$ are accessible, depending on $a+l=n,n-1,\ldots,0$, as we 
discussed in section \ref{rank_section}.

%\begin{mdframed}
The information about the new anomalous dimensions is carried by the characteristic polynomial 
\be
\mathcal{P}^*_{\vec{\tau},m}\ =\ \frac{(-)^{m}}{({\cal F}_{\vec{\tau},n})^{m}} \det\left[\, \mathbf{E}^{}_{\vec{\tau},m} -\eta^*_{\vec{\tau},m}\,  \mathbf{1}\, \right]
\ee
where the matrix ${\bf E}$ is the level splitting matrix \eqref{lev_spl_mat}.
%\end{mdframed}

The simplest observation we can make about $\eta^{*}$ has to do with 
the flat space contribution in the bold font variables ${\bf s,t,u}$, which is blind to the level splitting, since 
the hidden symmetry is restored.  We can access this limit by taking the twist $\tau$ to be large in 
the anomalous dimensions, then we expect the polynomial to covariantise, 
and collapse in such a way that all roots are equal, 
\be\label{flat_space_chp}
\mathcal{P}_{\vec{\tau},m}\quad \substack{ \phantom{spa} \\ \displaystyle \longrightarrow \\ {\tau \gg 1}}\quad  \left(\,\tilde{\eta}+(T-B)^{n-a-l}\right)^{m} \ +\ \ldots
\ee
with the variables $T$ and $B$ as in \eqref{def1_TB}. 

We know the
exponent of the term $(T-B)$ after comparing $\tilde\eta$ with the anomalous dimension 
for the $rank=1$ problem. This is simply $\mathcal{F}_{[aba],\tau,l,n}$ with $a+l=n$, 
i.e.~the formula we gave in \eqref{anomdimnnmin1}. When we increase the values of $m^*>1$, 
equivalently we decrease the value of $a+l$ w.r.t.~$n$,  
the mismatch in powers of $T$ is precisely $n-a-l$.

The flat space limit \eqref{flat_space_chp} tells us what is 
the maximum degree in $T$ and $B$ of the coefficients  in $\tilde{\eta}$ of the characteristic polynomial
\ba
\mathcal{P}^*_{\vec{\tau},m}&=&\tilde{\eta}^{\,m}  +K_{m,1}(T,B,a,l)\,\tilde{\eta}^{\,m-1} +\ldots +K_{m,m}(T,B,a,l)  \notag\\[.2cm] 
deg[K_{m,j}]&\leq& j\times \left(2 m-2+\tfrac{1}{2}(1-(-1)^{a+l})\right) 
\label{deg_of_K_intro}
\ea

For $m^*=2$ there are only $K_{j=1,2}$, and we know how to compute the roots
of a degree two polynomial. For $m^*\ge 3$ we can only deal  with the %characteristic polynomial and the analytic 
the properties of the coefficients $K_j(T,B,a,l)$ w.r.t.~the quantum numbers, and we  will do so in the next section \ref{sec_mstar3}.

%==============================================================================
%===============================================

\subsubsection{$m^*=2$ operators at all orders in $\alpha'$}\label{sec_mstar2}

%===============================================
%==============================================================================

In this section we study the level splitting of $m^*=2$ operators
with $a+l=n-2$ even first, and then $a+l=n-3$ odd. 
Given the simplicity of the degree two characteristic polynomial in these cases, 
we will be able to include explicitly all orders in $\alpha'$.

%======================================================================================
\paragraph{$a+l=n-2$ even.}~\\[-.2cm]
%======================================================================================

Long story short: We used our results at $(\alpha')^{5,7,9}$ to gather data for $a+l=0,2,4$, respectively.
Let us quote an example for concreteness,
\be
{\bf E}_{\tau=12,l=0,[040]}=
\left(\begin{array}{cc} 
-\frac{8070480000}{7} 			&		\frac{118800000\sqrt{187}}{7} \\[.2cm]
\frac{118800000\sqrt{187}}{7} 		&		-\frac{8624880000}{7}
\end{array}\right)\qquad
\ee
The quantum numbers $b$ and $\tau$ are arbitrary in principle, subject only to the bound $\tau\ge b+2a+4$, 
thus we fitted first the characteristic polynomial as functions of $T$ and $B$, keeping $a+l$ fixed. 
Collecting all pairs $(a,l)$ we then looked at the dependence on $a$ and $l$.
For $m^*=2$ we had the bonus of looking directly to the roots, rather than the individual coefficients 
of the characteristic polynomial. This was fruitful because suggested the following representation of the 
characteristic polynomial, 
\be\label{db_dege_ch_poly}
\mathcal{P}^*_{\vec{\tau},2}=(\tilde{\eta}+r )^2+ (\tilde{\eta}+r )\gamma_{2,1} + \gamma_{2,0} 
\ee
where
\begin{align}
\gamma_{2,1}&=-\frac{(n+2)(n+3)}{2n+5}\Big(B(2l+5) + (2a+5)T-  (a+2)(l+2) \Big)\\[.2cm]
\gamma_{2,0}&= +\frac{(n+2)^2(n+3)^2}{2n+5}  BT  
%r&= (T-B)^2+  B(2+l) +(2+a)T 
\end{align}
and the shift is
\begin{align}
r&= (T-B)^2+  B(2+l) +(2+a)T 
\end{align}
The square root responsible for splitting the anomalous dimensions 
does not depend on $r$, and is quite simple $\pm(\gamma_{2,1}^2-4\gamma_{2,0})^{\frac{1}{2}}$.

We will now switch to the explicit form, 
\be\label{stand_poly_char_mstar2}
\mathcal{P}^*_{\vec{\tau},2}=\tilde{\eta}^2 + K_{2,1}(T,B,a,l) \tilde{\eta}+ K_{2,2}(T,B,a,l)
\ee
and look for additional properties. 
The first observation is that 
\be
K_{2,j}(T,B,a,l)=K_{2,j}(B,T,l,a)\qquad ;\qquad j=1,2
\ee
and in fact the rescaled anomalous dimension $\tilde{\eta}$ is even under the symmetry. 
%and it is particularly nice. 

The second observation is about the covariantised flat space limit in $T$ and $B$. This is manifest in 
the parameterisation \eqref{db_dege_ch_poly},  and to see it scale $\eta\rightarrow \epsilon^2 \eta$ 
and $(B,T)\rightarrow \epsilon(B,T)$,  and take the limit $\epsilon$ large. At leading order, 
\be
\mathcal{P}^*_{\vec{\tau},2}(\epsilon^2 \tilde{\eta},\epsilon B,\epsilon T)\Big|_{\epsilon^4}=(\tilde{\eta}+(T-B)^2 )^2
\ee
where the term $(T-B)$ comes just from the shift by $r$.
This collapsed polynomial has indeed two equal roots. 
A nice experiment is to go beyond the leading term, and see how the anomalous dimensions split, 
since we know they will split. The $\epsilon$ expansion reads,
\begin{align}
\mathcal{P}^*_{\vec{\tau},2}(\epsilon^2 \tilde{\eta},\epsilon B,\epsilon T)&=\epsilon^4\Bigg[ \tilde\eta^2_{flat} - 
\frac{1}{\epsilon} \tilde\eta_{flat} \left[ T+B +\frac{n^2+3n+1}{2n+5}( T(2a+5)+(2l+5)B) \right] +\ldots \Bigg] \notag\\
\tilde\eta_{flat}&=\tilde{\eta}+(T-B)^2
\end{align}
Remarkably, keeping the first correction we find the solutions
\be
O(\epsilon^{-1})\qquad;\qquad \tilde\eta_{flat}=0\qquad;\qquad \tilde\eta_{flat}=\frac{1}{\epsilon}\Big[ T+B +\ldots \Big]
\ee
We learn from this formula that as we move away from flat space, the degeneracy 
is lifted sequentially, one of the two roots is still at the flat space locus, while the other is shifted.

Our next observation has to do with a factorisation in the coefficient $K_{2,2}$, which is not manifest 
in \eqref{db_dege_ch_poly}, but it becomes apparent in \eqref{stand_poly_char_mstar2} upon replacing $n=a+l+2$. 
Very nicely we find
\begin{align}
&
\!\!\!K_{2,2}(B,T,a,l)=\\
&
\rule{.8cm}{0pt}(\tfrac{\tau+b}{2})_{} (\tfrac{\tau+b}{2}+a+l+4)_{} (\tfrac{\tau-b}{2}-a-2)_{}(\tfrac{\tau-b}{2}+l+2)_{} \times \tilde{K}_{2,2}(T,B,a,l)\quad \notag
\end{align}
for a non factorisable $\tilde{K}_{2,2}$ such that $deg[ \tilde{K}_{2,2}]\leq 2$, as expected from \eqref{deg_of_K_intro}.

Notice that $K_{2,2}$ above vanishes precisely at $\tau=b+2a+4$, which the minimum value of $\tau$ for a two-particle operator.
Recall now that at the minimum twist, the rectangle collapses to a single line with $-45^\circ$ orientation
and there is no residual degeneracy in this case (see section \ref{rew_sec_SUGRA}). 
The partial degeneracy for $m^*=2$ will start showing up at $\tau=b+2a+6$. 
For example,
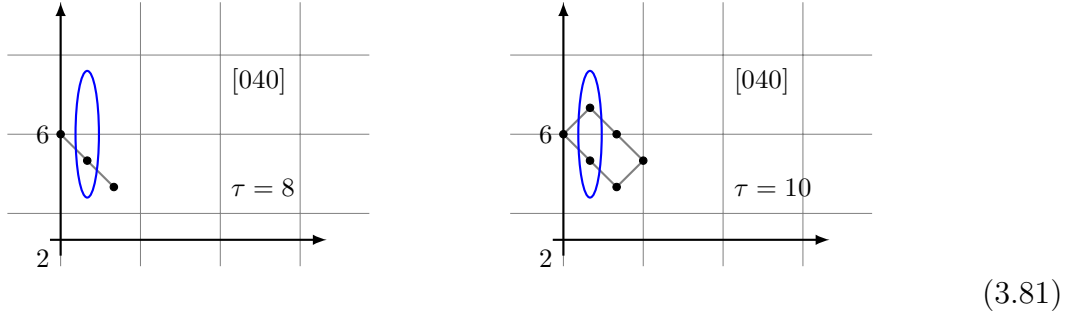
\be
\begin{array}{c}
\label{example_rank_reduction}
\begin{minipage}{\textwidth}
    	%\centering
	%
	%
	%
	% FIG UNO
	%
	%
	%
\rule{2.cm}{0pt}	
  	\begin{minipage}{.4\textwidth}
			\begin{tikzpicture}[scale=.7]

			\draw[step=1.5cm,gray,very thin] (-1,.5) grid (5.8,5.5);
			\def\prop{.5}
			\def\shifthor{\prop*2}
%

    				%rectangle [040]
			 \draw[black!50,thick] (\prop*2-\shifthor,\prop*6) --  (\prop*4-\shifthor,\prop*4);

			\foreach \indeyc in {0,1,2}
			\foreach \indexc  in {2}
			\filldraw   					 (\prop*\indexc+\prop*\indeyc-\shifthor, \prop*4+\prop*\indexc-\prop*\indeyc)   	circle (.07);

			\draw[thick, -latex] (\prop*2-1.2*\shifthor, \prop*2) --  (\prop*12-\shifthor, \prop*2);
			\draw[thick, -latex] (\prop*2-\shifthor, \prop*2-.3) --  (\prop*2-\shifthor, \prop*11);
			\draw (\prop*2-\shifthor, \prop*2-.35) node[left] {\footnotesize 2};
			\draw (\prop*2-\shifthor, \prop*6) node[left] {\footnotesize 6};
			
			\draw (\prop*8-\shifthor, \prop*4) node[right] {\footnotesize $\tau=8$};
			\draw (\prop*8-\shifthor, \prop*8) node[right] {\footnotesize [040]};
			
			\draw[blue,thick]  (\prop*3+\prop*0-\shifthor, \prop*6+\prop*1-\prop*1) ellipse (.22 and 1.2);

\end{tikzpicture}
\end{minipage}
%\ \ 
%
%
  	\begin{minipage}{.4\textwidth}
			\begin{tikzpicture}[scale=.7]

			\draw[step=1.5cm,gray,very thin] (-1,.5) grid (5.8,5.5);
			\def\prop{.5}
			\def\shifthor{\prop*2}
%

    				%rectangle [040]
			 \draw[black!50,thick] (\prop*2-\shifthor,\prop*6) --  (\prop*4-\shifthor,\prop*4);
 			 \draw[black!50,thick] (\prop*2-\shifthor,\prop*6) -- (\prop*3-\shifthor,\prop*7);
  			 \draw[black!50,thick] (\prop*4-\shifthor,\prop*4) -- (\prop*5-\shifthor,\prop*5);
 			 \draw[black!50,thick]  (\prop*5-\shifthor,\prop*5) -- (\prop*3-\shifthor,\prop*7);

			\foreach \indeyc in {0,1,2}
			\foreach \indexc  in {2,3}
			\filldraw   					 (\prop*\indexc+\prop*\indeyc-\shifthor, \prop*4+\prop*\indexc-\prop*\indeyc)   	circle (.07);

			\draw[thick, -latex] (\prop*2-1.2*\shifthor, \prop*2) --  (\prop*12-\shifthor, \prop*2);
			\draw[thick, -latex] (\prop*2-\shifthor, \prop*2-.3) --  (\prop*2-\shifthor, \prop*11);
			\draw (\prop*2-\shifthor, \prop*2-.35) node[left] {\footnotesize 2};
			\draw (\prop*2-\shifthor, \prop*6) node[left] {\footnotesize 6};
			
			\draw (\prop*8-\shifthor, \prop*4) node[right] {\footnotesize $\tau=10$};
			\draw (\prop*8-\shifthor, \prop*8) node[right] {\footnotesize [040]};
			
			\draw[blue,thick]  (\prop*3+\prop*0-\shifthor, \prop*6+\prop*1-\prop*1) ellipse (.22 and 1.2);

\end{tikzpicture}
\end{minipage}
\end{minipage}
\end{array}
\ee%nd{center}
When there is no degeneracy, the two particle operator with label $m^*=2$ is already identified by 
the SUGRA eigenvalue problem, therefore the $\alpha'$ correction is linear and is obtained  by the 
following direct computation,
\be\label{reduction_mstar2}
\eta^*_{\vec{\tau}}\Big|_{\tau=b+2a+4}=\  \mathbb{V}^T_{\vec{\tau},2}\, \cdot {\bf N}^{(n+3)}_{\vec{\tau}}\, \cdot\mathbb{V}_{\vec{\tau},2}\Big|_{\tau=b+2a+4}
\ee
where $\mathbb{V}_{b+2a+4,2}$ consists of a single eigenvector.  Let us emphasise that there is 
\emph{no} $2\times 2$ level splitting matrix corresponding to this case. Remarkably what we find by looking 
at the characteristic polynomial, and forcing $\tau=b+2a+4$ is
\be\label{rank_red_mstar2}
\mathcal{P}^*_{\vec{\tau},2}\Big|_{\tau=b+2a+4}= \tilde\eta \left( \tilde\eta+ \gamma_{2,1}+2r \right)\Big|_{\tau=b+2a+4}
\ee
Thus, one root of the polynomial goes to zero, and upon inspection the other root precisely coincides 
with the rescaled anomalous dimension from \eqref{reduction_mstar2}!

We interpret  the above phenomenon as follows. Because the characteristic polynomial is analytic 
in the quantum numbers, we can think of its roots  as the anomalous dimensions of 
two analytically continued operators. The reduction in \eqref{rank_red_mstar2} 
shows the decoupling of one of the two operators, when physically only one operator exists 
in the theory.  A priori there would be no reason to expect the non zero root to correctly reproduce
the rescaled anomalous dimension of the physical operator, since there is really no $2\times 2$ level 
splitting matrix at $\tau=b+2a+4$. Quite surprisingly we find that it does it, here and in all other examples that we will check. \\[1.2cm]

%======================================================================================
\paragraph{Unmixed three point couplings.}~\\[-.2cm]
%======================================================================================

The three point couplings of the newly identified two-particle operators are given by the columns of the ${\bf c}^{0}_{\vec{\tau}}$ matrix, 
as explained around \eqref{STRINGY_eigenvector_problem}, namely
\begin{align}\label{mstar2_3pt}
{\bf c}^{(0)}_{\vec{\tau}}\Bigg|_{\mathbb{V}_{\vec{\tau},2}}=\ \mathbb{V}_{\vec{\tau},2}\cdot\, {\tt eigenvectors}[\, \mathbf{E}^{}_{2,\vec{\tau}}\ ]\
\end{align}
where the ${\tt eigenvectors}$ are taken to be orthonormal. This formula simply means that 
the three point couplings are given by taking an orthonormal basis for $\mathbb{V}_{\vec{\tau},2}$, 
from the SUGRA eigenvalue problem,  then solve the STRINGY eigenvalue problem in that basis, 
and use the STRINGY eigenvectors to fix the residual freedom on $\mathbb{V}_{\vec{\tau},2}$.\footnote{Differently 
from the characteristic polynomial, which is computable for any $m^*$, 
the computation of \eqref{mstar2_3pt} requires knowledge of the roots. Thus,
the three-point couplings will remain somewhat implicit/numerical in the general case $m^*\ge 3$. }

The general form of the three-point couplings is
\be
\begin{tikzpicture}
\def\step{.6}

\draw[thick,gray] (.25*\step,.3*\step)-- (.25*\step-.25,.3*\step)  -- (.25*\step-.25,-6.6*\step)  --  (.25*\step,-6.6*\step);
\draw[thick,gray] (1.75*\step,.3*\step)-- (1.75*\step+.25,.3*\step)  -- (1.75*\step+.25,-6.6*\step)  --  (1.75*\step,-6.6*\step);

\filldraw[green!20,draw=black] (.4*\step,0) rectangle (1.6*\step,-1.5*\step);
\filldraw[green!20,draw=black] (.4*\step,-1.6*\step) rectangle (1.6*\step,-1.5*\step-1.6*\step);

\draw[] (1*\step,-3.7*\step) node {$\vdots$};

\filldraw[green!20,draw=black] (.4*\step,-3*1.6*\step) rectangle (1.6*\step,-1.5*\step-3*1.6*\step);

\draw[] (1*\step,-.75*\step) node {${\cal T}_1$};
\draw[] (1*\step,-2.25*\step) node {${\cal T}_2$};
\draw[] (1*\step,-5.5*\step) node {${\cal T}_{\mu}$};

\draw[] (9.2*\step,-3.25*\step) node {${\cal T}_{\beta,\vec{\tau}}={\tt Table}\Big[\,\ldots\, ,\{i,1,t-1\}\Big]$};

\draw[] (-2.8*\step,-3.25*\step) node {${\tt cln}\Big({\bf c}^{(0)}_{\vec{\tau}}\Big)=$};

%\draw (0,-2*\step) node[scale=5] {(};
\end{tikzpicture}
\ee
We will now label the new three-point couplings at $m^*=2$ with $\pm$ signs, 
\begin{align}\label{formulazza_3pt}
&
\rule{2cm}{0pt}{\cal T}^{\pm}_{\beta,\vec{\tau}}=\sqrt{ {\cal N}_{\vec{\tau},\beta}\times
\frac{ (\frac{\tau+b}{2}-\beta+2)_{\beta-2}  (\frac{\tau-b}{2}+l+3)_{\beta-2} }{  (\frac{\tau-b}{2}+\mu-2)_{\beta+2-\mu}  (\frac{\tau+b}{2}+l+3-\beta)^{}_{\beta+2-\mu}} 
 }\times \widetilde{\cal T}_{\beta,\vec{\tau}}^{\pm} 
\\[.3cm]
& 
\!\!{\cal N}_{\vec{\tau},\beta}=\frac{1}{(\tau-1)_{2l+7} } \frac{(\frac{\tau+b}{2}+a+l+5)_{-a-\mu} }{ (\frac{\tau-b}{2}-a-2)_{+a+\mu}}
\left\{ \begin{array}{lc} 
(\frac{\tau+b}{2}) (\frac{\tau-b}{2}+l+2) ; 			&				 \beta=1\\[.2cm]
1 										& 				  2\leq \beta\leq \mu-1 \\[.2cm]
(\frac{\tau-b}{2}+\mu-1)(\frac{\tau+b}{2}-\mu+l+3) ;	 &				 \beta=\mu
  \end{array}\right. \notag 
\end{align}
where
\be\label{formula_calT}
\!\!\!\!\!\!\!\!\!\!\!\widetilde{\cal T}_{\beta,\vec{\tau}}^{\pm}= {\tt Table}\Big[ \sigma_{\beta,i} \left[(\tfrac{\tau+b}{2}+a+i+2)_{l+1}(\tfrac{\tau-b}{2}-i-a)_{l+1}\!
\Bigg(\widetilde{\cal P}_{\beta,1}(T,I)\pm  \frac{ \widetilde{\cal P}_{\beta,2}(T,I) }{ \sqrt{ \gamma_{2,1}^2-4 \gamma_{2,0}} }\Bigg)^{}\right]^{\!\!\frac{1}{2}}\!\!\!,\,\{i,1,t-1\} \Big]
\ee
with $\sigma^2=1$ and ${\cal P}_{\beta}$ polynomials in the variables $T$ and $I\equiv i(i+b+2a+2)$, 
containing non factorisable pieces in most of the cases.

The general form of the polynomials $\widetilde{\cal P}_{\beta,1}$ and $\widetilde{\cal P}_{\beta,2}$ 
is of course complicated. Ultimately, they come from combining two eigenvalue problems, because 
of the very definition of ${\bf c}^{(0)}$ in \eqref{mstar2_3pt}.
For example, in the $su(4)$ channel $[040]$ and $l=0$ we find
\begin{align}
&\rule{.8cm}{0pt}
\widetilde{\cal P}_{\beta,1}\!=\!\!\left[ \begin{array}{c}
\scalebox{.8}{
$440 (-270 (9 + I)^2 + 3 (9 + I) (103 + 7 I) T - 
   2 (89 + 9 I) T^2 + 5 T^3)$ }\\
\scalebox{.8}{
  $88 T (-27 (505 + I (130 + 9 I)) + 
   12 (609 + I (158 + 11 I)) T - 16 (62 + 9 I) T^2 + 
   40 T^3)$} \\
\scalebox{.8}{
   $132 (-72 (5 + I)^2 + (5 + I) (323 + 55 I) T - 
   6 (53 + 9 I) T^2 + 15 T^3)$}
   \end{array}\right]\notag\\
\label{formulas_040_3pt} \\
 &
\!\!\!\!\!\!\!\!\!\!\!\!\!\!\!\!\!\!\!\!\!\!\widetilde{\cal P}_{\beta,2}\!=\!\!\left[\!\!\!\!\begin{array}{c}
\scalebox{.8}{
$1760 (-9720 (9 + I)^2 + 54 (9 + I) (161 + 9 I) T - 
   3 (3539 + I (550 + 19 I)) T^2 + 2 (427 + 45 I) T^3 - 
   25 T^4)$ }\\
\scalebox{.8}{
  $ -352 T (972 (305 + I (50 + I)) + 
   27 (-3379 + I (-246 + 29 I)) T + 
   12 (2755 + I (674 + 53 I)) T^2 - 16 (292 + 45 I) T^3 + 
   200 T^4)$} \\
\scalebox{.8}{
   $528 (2592 (5 + I)^2 - 
   36 (5 + I) (313 + 53 I) T - (4147 + 
      I (1654 + 211 I)) T^2 + 6 (247 + 45 I) T^3 - 75 T^4)$}
   \end{array}\!\!\!\right]\notag
\end{align}

Rather than looking for a general formula, in the following it will be more illuminating to
discuss features of the three-point couplings related to the flat space limit and the rank reduction, by 
making a parallel with the discussion about the characteristic polynomial. 

In \eqref{formulas_040_3pt}, the degree in $T$ of $\widetilde{\cal P}_{\beta,2}$ is one power higher than $\widetilde{\cal P}_{\beta,1}$, but
what enters the three-point couplings is the combination $\widetilde{\cal P}_{\beta,2}\times(\gamma_{2,1}^2-4 \gamma_{2,0})^{\frac{1}{2}}$. 
The square root precisely lowers the degree by one in the regime of large $T$. In fact,
%Therefore, let us compute the flat space 
%limit and see what happens,
\footnote{Notice that our normalisation $\mathcal{N}$ 
extracts a factor that we understood to be present always, for the first and the last block, 
i.e.~$\beta=1,\mu$, otherwise all components of $\widetilde{\mathcal{P}}$ will scale the same. } 
\begin{align}
%\!\!\!\!\!
\lim_{T\gg 1}\ \widetilde{\cal P}_{\beta,1}\rightarrow&\left[\!\begin{array}{c} 
\scalebox{.8}{
$+2200 T^3 -(78320 + 7920 I )T^2+O(T)$}\\
\scalebox{.8}{
$+3520 T^4 -(87296 +12672 I)T^3+O(T^2)$}\\
\scalebox{.8}{
$+1980 T^3 - (41976 + 7128 I)T^2 + O(T)$}
\end{array}\!\right]\notag \qquad\\
\label{example_3ptcoup}
\\
\lim_{T\gg 1}\ \frac{\widetilde{\cal P}_{\beta,2}}{ \scalebox{.8}{$4\sqrt{ (36 + 5 T)^2-288 T}$} } \rightarrow&\left[\!\begin{array}{c} 
\scalebox{.8}{
$-2200 T^3 +(78320 + 7920 I )T^2+O(T)$}\\
\scalebox{.8}{
$-3520 T^4 +(87296 +12672 I)T^3+O(T^2)$}\\
\scalebox{.8}{
$-1980 T^3 + (41976 + 7128 I)T^2 + O(T)$}
\end{array}\!\right] \qquad \notag
\end{align}
When we add/subtract \eqref{example_3ptcoup} to  build ${\cal T}^{\pm}$ we find that in the flat space limit ${\cal T}^{+}_{\beta,\vec{\tau}}$ 
vanishes at leading and subleading order (the next one is non trivial), while ${\cal T}^{-}_{\beta,\vec{\tau}}$  survives.  

Next we would like to see what happens 
 when we go to the minimum twist.\footnote{We thank Pedro Vieira for motivating this investigation.} 
Reconsider our previous picture, which was suited for the example we are illustrating here,  
%$[040]$ and $l=0$,
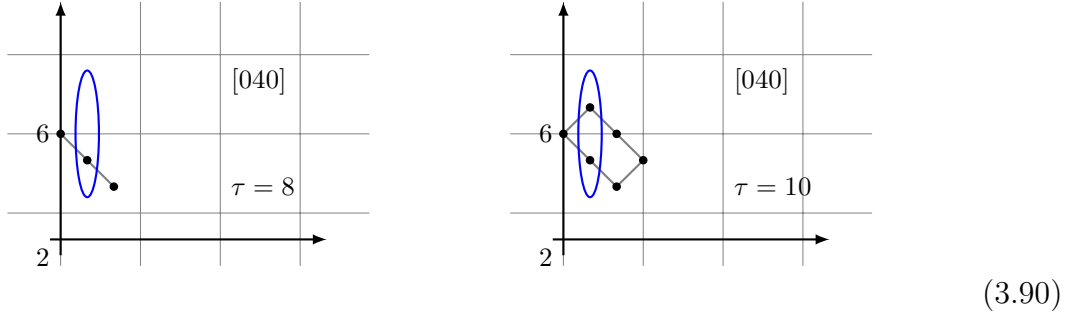
\be%gin{center}
\begin{array}{c}
\label{example_rank_reduction_complicated}
\begin{minipage}{\textwidth}
    	%\centering
	%
	%
	%
	% FIG UNO
	%
	%
	%
\rule{2.cm}{0pt}	
  	\begin{minipage}{.4\textwidth}
			\begin{tikzpicture}[scale=.7]

			\draw[step=1.5cm,gray,very thin] (-1,.5) grid (5.8,5.5);
			\def\prop{.5}
			\def\shifthor{\prop*2}
%

    				%rectangle [040]
			 \draw[black!50,thick] (\prop*2-\shifthor,\prop*6) --  (\prop*4-\shifthor,\prop*4);
% 			 \draw[blue!50,thick] (\prop*2-\shifthor,\prop*6) -- (\prop*9-\shifthor,\prop*13);
%  			 \draw[blue!50,thick] (\prop*4-\shifthor,\prop*4) -- (\prop*11-\shifthor,\prop*11);
% 			 \draw[blue!50,thick]  (\prop*9-\shifthor,\prop*13) -- (\prop*11-\shifthor,\prop*11);

			%
			%
			\foreach \indeyc in {0,1,2}
			\foreach \indexc  in {2}
			\filldraw   					 (\prop*\indexc+\prop*\indeyc-\shifthor, \prop*4+\prop*\indexc-\prop*\indeyc)   	circle (.07);

			%\foreach \indeyc in {0}
			%\foreach \indexc  in {2,...,9}
			%\filldraw   					 (\prop*\indexc+\prop*\indeyc-\shifthor, \prop*6+\prop*\indexc-\prop*\indeyc)   	circle (.07);

			\draw[thick, -latex] (\prop*2-1.2*\shifthor, \prop*2) --  (\prop*12-\shifthor, \prop*2);
			\draw[thick, -latex] (\prop*2-\shifthor, \prop*2-.3) --  (\prop*2-\shifthor, \prop*11);
			\draw (\prop*2-\shifthor, \prop*2-.35) node[left] {\footnotesize 2};
			\draw (\prop*2-\shifthor, \prop*6) node[left] {\footnotesize 6};
			
			\draw (\prop*8-\shifthor, \prop*4) node[right] {\footnotesize $\tau=8$};
			\draw (\prop*8-\shifthor, \prop*8) node[right] {\footnotesize [040]};
			
			\draw[blue,thick]  (\prop*3+\prop*0-\shifthor, \prop*6+\prop*1-\prop*1) ellipse (.22 and 1.2);

\end{tikzpicture}
\end{minipage}
%\ \ 
%
%
  	\begin{minipage}{.4\textwidth}
			\begin{tikzpicture}[scale=.7]

			\draw[step=1.5cm,gray,very thin] (-1,.5) grid (5.8,5.5);
			\def\prop{.5}
			\def\shifthor{\prop*2}
%

    				%rectangle [040]
			 \draw[black!50,thick] (\prop*2-\shifthor,\prop*6) --  (\prop*4-\shifthor,\prop*4);
 			 \draw[black!50,thick] (\prop*2-\shifthor,\prop*6) -- (\prop*3-\shifthor,\prop*7);
  			 \draw[black!50,thick] (\prop*4-\shifthor,\prop*4) -- (\prop*5-\shifthor,\prop*5);
 			 \draw[black!50,thick]  (\prop*5-\shifthor,\prop*5) -- (\prop*3-\shifthor,\prop*7);

			\foreach \indeyc in {0,1,2}
			\foreach \indexc  in {2,3}
			\filldraw   					 (\prop*\indexc+\prop*\indeyc-\shifthor, \prop*4+\prop*\indexc-\prop*\indeyc)   	circle (.07);

			%\foreach \indeyc in {0}
			%\foreach \indexc  in {2,...,9}
			%\filldraw   					 (\prop*\indexc+\prop*\indeyc-\shifthor, \prop*6+\prop*\indexc-\prop*\indeyc)   	circle (.07);

			\draw[thick, -latex] (\prop*2-1.2*\shifthor, \prop*2) --  (\prop*12-\shifthor, \prop*2);
			\draw[thick, -latex] (\prop*2-\shifthor, \prop*2-.3) --  (\prop*2-\shifthor, \prop*11);
			\draw (\prop*2-\shifthor, \prop*2-.35) node[left] {\footnotesize 2};
			\draw (\prop*2-\shifthor, \prop*6) node[left] {\footnotesize 6};
			
			\draw (\prop*8-\shifthor, \prop*4) node[right] {\footnotesize $\tau=10$};
			\draw (\prop*8-\shifthor, \prop*8) node[right] {\footnotesize [040]};
			
			\draw[blue,thick]  (\prop*3+\prop*0-\shifthor, \prop*6+\prop*1-\prop*1) ellipse (.22 and 1.2);

\end{tikzpicture}
\end{minipage}
\end{minipage}
\end{array}
\ee%nd{center}
For the characteristic polynomial at the minimum twist %accordingly with the number of operators.
we understood the appearance of a vanishing root (out of two)
as a form of decoupling of one of the two analytically continued operators.
For the three-point couplings we expect something different to happen.
Continuing with our example \eqref{formulas_040_3pt}, we find
\be
\!\!\!\!\!
\widetilde{\cal P}_{\beta,1}\Bigg|_{\tau=8}=\left[\!\begin{array}{c} 
\scalebox{.8}{
$102960 (-1 + i)^2 (7 + i)^2 $}\\
\scalebox{.8}{
$6177600 (-1 + i)^2 (7 + i)^2$}\\
\scalebox{.8}{
$164736 (-1 + i)^2 (7 + i)^2$}
\end{array}\!\right] %\notag\\
\label{example_rank_red_3pt}
%\\
\quad;\quad
\frac{ \widetilde{\cal P}_{\beta,2}}{  \scalebox{.8}{$4\sqrt{ (36 + 5 T)^2-288 T}$} }\Bigg|_{\tau=8}=\left[\!\!\begin{array}{c} 
\scalebox{.8}{
$-102960  (-1 + i)^2 (7 + i)^2 $}\\ %54362880
\scalebox{.8}{
$-6177600  (-1 + i)^2 (7 + i)^2$}\\ %3261772800
\scalebox{.8}{
$-164736  (-1 + i)^2 (7 + i)^2$} %86980608
\end{array}\!\right] 
\ee
Only the $i=1$ component exists at the minimum twist, thus 
both polynomials vanish independently and we conclude that the two analytically 
continued operators decouple.

To understand the physics of the three-point decoupling, let us start again, this time from 
a simpler case, i.e.~$[020]$ even spin, $\mu=2$. Varying the twist we would find the following picture
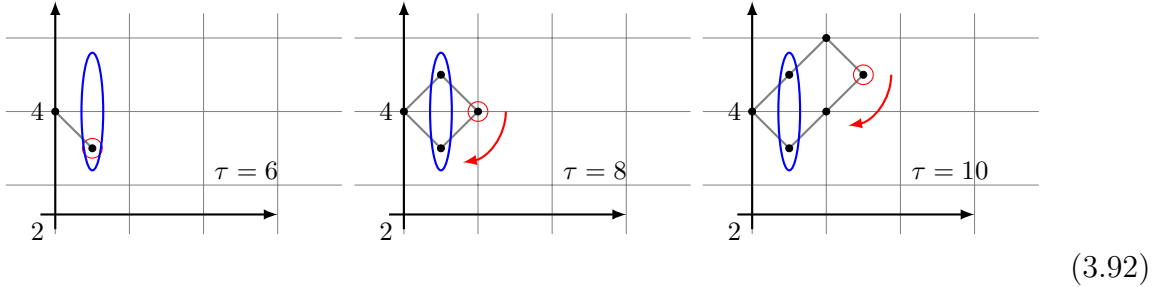
\be%gin{center}
\label{example_rank_reduction}
\begin{minipage}{\textwidth}
    	%\centering
	%
	%
	%
	% FIG UNO
	%
	%
	%
\rule{1.cm}{0pt}	
  	\begin{minipage}{.3\textwidth}
			\begin{tikzpicture}[scale=.65]

			\draw[step=1.5cm,gray,very thin] (-1,2) grid (5.8,6.5);
			\def\prop{.75}
			\def\shifthor{\prop*2}
%

 				%rectangle [040]
			 \draw[black!50,thick] (\prop*2-\shifthor,\prop*6) --  (\prop*3-\shifthor,\prop*5);
 %			 \draw[black!50,thick] (\prop*2-\shifthor,\prop*6) -- (\prop*3-\shifthor,\prop*7);
 % 			 \draw[black!50,thick] (\prop*4-\shifthor,\prop*6) -- (\prop*3-\shifthor,\prop*7);
 %			 \draw[black!50,thick]  (\prop*4-\shifthor,\prop*6) -- (\prop*3-\shifthor,\prop*5);

			\draw[red] (\prop*3-\shifthor,\prop*5) circle (.2);

			\foreach \indeyc in {0,1}
			\foreach \indexc  in {2}
			\filldraw   					 (\prop*\indexc+\prop*\indeyc-\shifthor, \prop*4+\prop*\indexc-\prop*\indeyc)   	circle (.07);

			%\foreach \indeyc in {0}
			%\foreach \indexc  in {2,...,9}
			%\filldraw   					 (\prop*\indexc+\prop*\indeyc-\shifthor, \prop*6+\prop*\indexc-\prop*\indeyc)   	circle (.07);

			\draw[thick, -latex] (\prop*2-1.2*\shifthor, \prop*3.2) --  (\prop*8-\shifthor, \prop*3.2);
			\draw[thick, -latex] (\prop*2-\shifthor, \prop*3.2-.3) --  (\prop*2-\shifthor, \prop*9);
			\draw (\prop*2-\shifthor, \prop*3.2-.35) node[left] {\footnotesize 2};
			\draw (\prop*2-\shifthor, \prop*6) node[left] {\footnotesize 4};
			
			\draw (\prop*8-2*\shifthor, \prop*4.4) node[right] {\footnotesize $\tau=6$};
			%\draw (\prop*8-\shifthor, \prop*3) node[right] {\footnotesize [040]};
			
			\draw[blue,thick]  (\prop*3+\prop*0-\shifthor, \prop*6+\prop*1-\prop*1) ellipse (.22 and 1.2);

\end{tikzpicture}
\end{minipage}
%\ \ 
%
\!\!\!\!\!\!\!\!
  	\begin{minipage}{.3\textwidth}
			\begin{tikzpicture}[scale=.65]

			\draw[step=1.5cm,gray,very thin] (-1,2) grid (5.8,6.5);
			\def\prop{.75}
			\def\shifthor{\prop*2}
%

    				%rectangle [040]
			 \draw[black!50,thick] (\prop*2-\shifthor,\prop*6) --  (\prop*3-\shifthor,\prop*5);
 			 \draw[black!50,thick] (\prop*2-\shifthor,\prop*6) -- (\prop*3-\shifthor,\prop*7);
  			 \draw[black!50,thick] (\prop*4-\shifthor,\prop*6) -- (\prop*3-\shifthor,\prop*7);
 			 \draw[black!50,thick]  (\prop*4-\shifthor,\prop*6) -- (\prop*3-\shifthor,\prop*5);

			\draw[red] (\prop*4-\shifthor,\prop*6) circle (.2);
			
			%\draw[red] (\prop*4-\shifthor,\prop*6) circle (.2);
			\draw [thick,-latex,red] (\prop*4.75-\shifthor,\prop*6) arc (0:-80:30pt);

			\foreach \indeyc in {0,1}
			\foreach \indexc  in {2,3}
			\filldraw   					 (\prop*\indexc+\prop*\indeyc-\shifthor, \prop*4+\prop*\indexc-\prop*\indeyc)   	circle (.07);

			%\foreach \indeyc in {0}
			%\foreach \indexc  in {2,...,9}
			%\filldraw   					 (\prop*\indexc+\prop*\indeyc-\shifthor, \prop*6+\prop*\indexc-\prop*\indeyc)   	circle (.07);

			\draw[thick, -latex] (\prop*2-1.2*\shifthor, \prop*3.2) --  (\prop*8-\shifthor, \prop*3.2);
			\draw[thick, -latex] (\prop*2-\shifthor, \prop*3.2-.3) --  (\prop*2-\shifthor, \prop*9);
			\draw (\prop*2-\shifthor, \prop*3.2-.35) node[left] {\footnotesize 2};
			\draw (\prop*2-\shifthor, \prop*6) node[left] {\footnotesize 4};
			
			\draw (\prop*8-2*\shifthor, \prop*4.4) node[right] {\footnotesize $\tau=8$};
			%\draw (\prop*8-\shifthor, \prop*3) node[right] {\footnotesize [040]};
			
			\draw[blue,thick]  (\prop*3+\prop*0-\shifthor, \prop*6+\prop*1-\prop*1) ellipse (.22 and 1.2);

\end{tikzpicture}
\end{minipage}
%\ \ 
%
\!\!\!\!\!\!\!\!
  	\begin{minipage}{.3\textwidth}
			\begin{tikzpicture}[scale=.65]

			\draw[step=1.5cm,gray,very thin] (-1,2) grid (5.8,6.5);
			\def\prop{.75}
			\def\shifthor{\prop*2}
%

    				%rectangle [040]
			 \draw[black!50,thick] (\prop*2-\shifthor,\prop*6) --  (\prop*3-\shifthor,\prop*5);
 			 \draw[black!50,thick] (\prop*2-\shifthor,\prop*6) -- (\prop*4-\shifthor,\prop*8);
  			 \draw[black!50,thick] (\prop*4-\shifthor,\prop*8) -- (\prop*5-\shifthor,\prop*7);
 			 \draw[black!50,thick]  (\prop*3-\shifthor,\prop*5)-- (\prop*5-\shifthor,\prop*7);

			\draw[red] (\prop*5-\shifthor,\prop*7) circle (.2);
			\draw [thick,-latex,red] (\prop*5.75-\shifthor,\prop*7) arc (0:-80:30pt);

			\foreach \indeyc in {0,1}
			\foreach \indexc  in {2,3,4}
			\filldraw   					 (\prop*\indexc+\prop*\indeyc-\shifthor, \prop*4+\prop*\indexc-\prop*\indeyc)   	circle (.07);

			%\foreach \indeyc in {0}
			%\foreach \indexc  in {2,...,9}
			%\filldraw   					 (\prop*\indexc+\prop*\indeyc-\shifthor, \prop*6+\prop*\indexc-\prop*\indeyc)   	circle (.07);

			\draw[thick, -latex] (\prop*2-1.2*\shifthor, \prop*3.2) --  (\prop*8-\shifthor, \prop*3.2);
			\draw[thick, -latex] (\prop*2-\shifthor, \prop*3.2-.3) --  (\prop*2-\shifthor, \prop*9);
			\draw (\prop*2-\shifthor, \prop*3.2-.35) node[left] {\footnotesize 2};
			\draw (\prop*2-\shifthor, \prop*6) node[left] {\footnotesize 4};
			
			\draw (\prop*8-2*\shifthor, \prop*4.4) node[right] {\footnotesize $\tau=10$};
			%\draw (\prop*8-\shifthor, \prop*3) node[right] {\footnotesize [040]};
			
			\draw[blue,thick]  (\prop*3+\prop*0-\shifthor, \prop*6+\prop*1-\prop*1) ellipse (.22 and 1.2);

\end{tikzpicture}
\end{minipage}
\end{minipage}
\ee%nd{center}
The red circle is pinning the operator which
together with ${\cal K}_{m^*=1}$ is a singlet eigenvector of the SUGRA eigenvalue problem. 
This operator has its own analytic trajectory  and the arrow indicates that the three-point coupling 
goes analytically in twist, from right to left. Its three-point couplings are studied 
in appendix \ref{singlet_eigenvectors}.  When we move from $\tau=8$ to $\tau=6$ the red colored operator 
takes the place of an $m^*=2$ operator, but already in SUGRA it is not the analytic continuation of the pair 
of degenerate operators, which therefore has to decouple. %This is what happens in formulae  and as it is the case in  \eqref{example_rank_reduction_complicated}, 
Again, when $i=1$ the three-point couplings vanish at the minimum twist.\footnote{Explicit 
formulae for this case are given in appendix \ref{singlet_eigenvectors}}, The example in \eqref{example_rank_reduction_complicated}
is more complicated, since it comes with $\mu=3$ to begin with, but the fate of the pair at $m^*=2$ pair 
at the minimum twist is the same.\footnote{In \eqref{example_rank_reduction_complicated} %the singlets 
%two-particle operator have a clear pattern, but 
the operator in the middle at $\tau=8$ will come from the reduction of the $m^*=3$ operators.}

The three-point decoupling is thus stronger compared to the decoupling in the characteristic polynomial, which 
in this sense is quite smart because it retains information 
about all physical operators. 
%operators will remain in the physical spectrum, and in fact returns its anomalous dimension. 

%======================================================================================
\paragraph{$a+l=n-3$ odd.}~\\[-.2cm]
%======================================================================================

The case $a+l=n-3,$ generalises in a simple way the previous case. 
We will focus mainly on the characteristic polynomial, which we can write as 
\be\label{db_dege_ch_poly_odd}
\mathcal{P}^*_{\vec{\tau},2}=(\tilde{\eta}+r^{} )^2+ (\tilde{\eta}+r )\gamma^{}_{2,1} + \gamma^{}_{2,0} 
\ee
in terms of a new shift 
\begin{align}
r&= (T-B)^3+(T-B)(B(3l+7)+(3a+7)T) +(a-l)( B(l+2)+(a+2)T)
\end{align}
and new coefficients
\begin{align}
\gamma_{2,1}&=-\frac{(n+2)(n+3)}{2n+5}(T-B)\Big(B(2l+7) + (2a+7)T-  3 al-7(a+l)-16 \Big)\\[.2cm]
\gamma_{2,0}&=+\frac{(n+2)^2(n+3)^2}{2n+5} (T-B)^2\Big( 3 BT-B(l+2)-(a+2)T\Big)
\end{align}

The shift by $r$ makes manifest the flat space limit, which this times goes with 
\be
\mathcal{P}^*(\epsilon^3 \tilde{\eta},\epsilon T,\epsilon B)\Big|_{\epsilon^6}= \big(\tilde{\eta}_{flat}\big)^2 \qquad;\qquad \tilde{\eta}_{flat}=\tilde{\eta}+(T-B)^3
\ee
The power of $(T-B)$ is one more compared to $a+l=n-2$ even. In general $T>B$ 
therefore there is no ambiguity with odd powers. This odd power remind us that in this case
the rescaled anomalous dimension $\tilde{\eta}$ is odd under symmetry $T\leftrightarrow B$ 
and $a\leftrightarrow l$, which implies on the polynomial
\be
K_{2,j}(T,B,a,l)=(-)^j K_{2,j}(B,T,l,a)\qquad;\qquad j=1,2
\ee

As in the previous case, the splitting of the anomalous dimensions away 
from $\tilde{\eta}_{flat}=0$ is sequential, and the rank reduction at the minimum twist 
decouples one of the two analytically continued operators, and reproduces the anomalous dimension of the physical operator.

%=======================================================================================
\section{General properties of the characteristic polynomial}\label{sec_mstar3}
%=======================================================================================

The characteristic polynomial $\mathcal{P}^*_{\vec{\tau},m}$ associated to the level splitting problem 
is a novel and very intriguing object.  With the level splitting matrix defined  in \eqref{lev_spl_mat_1}, and 
the normalisation ${\cal F}$ in introduced in \eqref{def1_F}, 
\be
\mathcal{P}^*_{\vec{\tau},m}\ =\ \frac{(-)^{m}}{({\cal F}_{\vec{\tau},n})^{m}} \det\left[\, \mathbf{E}^{}_{\vec{\tau},m} -\eta^*_{\vec{\tau},m}\,  \mathbf{1}\, \right]
\qquad;\qquad \eta^*_{\vec{\tau},m}={\cal F}_{\vec{\tau},n}\ \tilde{\eta}_{\vec{\tau},m}
\ee
This object nicely packages the CFT data from the $AdS_5\times S^5$ VS amplitude
which lifts the partial degeneracy of the SUGRA anomalous dimensions. But not only that.

Analyticity of $\mathcal{P}^*_{\vec{\tau},m}$ w.r.t.~$T,B,a,l$ might be obvious for $m^*=2$ , since the level splitting matrix is just $2\times 2$.
However, this is not so intuitive in more complicated cases as
\be
\frac{ {\bf E}_{\vec{\tau}} }{ {\cal F}_{\vec{\tau},6} }\Bigg|_{ \substack{ \tau=14\\ l=2, [040] }  } =
\left(\begin{array}{ccc}
\scalebox{.9}{$-\frac{6945359904}{499}$} 						&\scalebox{.9}{$\frac{48965850432}{499} \sqrt{\frac{69}{41735}}$} 		&  \scalebox{.9}{$-1524096\sqrt{ \frac{ 690690}{4165153}}$}\\[.2cm]
\scalebox{.9}{$\frac{48965850432}{499}\sqrt{ \frac{69}{41735} } $}	& \scalebox{.9}{$-\frac{337620067080624}{20825765}$} 				& \scalebox{.9}{$\frac{ 33255693072}{8347}\sqrt{ \frac{2002}{499} }$} \\[.2cm]
\scalebox{.9}{$-1524096\sqrt{\frac{ 690690}{4165153}}$} 			&\scalebox{.9}{$\frac{ 33255693072}{8347}\sqrt{ \frac{2002}{499} }$}	 &	\scalebox{.9}{$ -\frac{183139846560}{8347}$} 
\end{array}\right)
\ee
and the characteristic polynomial will certainly not be analytic if the square roots remain in the final result. 
For general $m^*$ there is a short computation 
we can do to actually see what determines the analytic properties of $\mathcal{P}^*_{}$ and it uses the 
known formula,\footnote{For 
example, see \url{https://en.wikipedia.org/wiki/Characteristic_polynomial}.}
\be
K_{j}=\frac{(-)^j}{j!}\det \left[\begin{array}{lcccc}  {\tt tr}{\bf E} & j-1 & 0 & \ldots  & \ldots \\[.2cm]
													{\tt tr}{\bf E}^2 & {\tt tr}{\bf E} & j-2 & 0 & \ldots   \\ 
														\ \ \ \vdots  \\  
														\ \ \ \vdots  & & & & \vdots \\
														 		&	 & 		& 		&    1\\ 
													 {\tt tr}{\bf E}^j & {\tt tr}{\bf E}^{j-1} & \ldots &  \ldots & {\tt tr}{\bf E} \end{array}\right]\quad\ ;\quad\ 
													 \mathcal{P}^*_{\vec{\tau},m}=\tilde{\eta}^{\,m}  +\sum_{j=1}^{m} K_{m,j}\, \tilde{\eta}^{m-j}
\ee
The analytic properties of $K_{m,j}(T,B,a,l)$ then follow from those of ${\tt tr}\,{\bf E}^k$. 
Let us consider $k=1$, since the general case will be analogous.  
From the definition of the level splitting matrix in \eqref{lev_spl_mat_1}, we find 
\be
\qquad
{\tt tr}\,{\bf E}_{\vec{\tau},m}= {\tt tr}\left[   {\bf M}_{\vec{\tau}} \,
				{\bf L}_{\vec{\tau}}^{-\frac{1}{2} } {\bf P}_{\vec{\tau},m} {\bf L}_{\vec{\tau}}^{-\frac{1}{2} }\right] \qquad;\qquad {\bf P}_{\vec{\tau},m}= \left( \sum_{I=1}^{m} {\bf v}_{I}{\bf v}_{I}^T\right)_{1\leq i,j\leq \mu(t-1)}
\ee
where ${\bf P}_{m}$ is the projector onto the hyperplane $\mathbb{V}_{\vec{\tau},m}$ 
spanned by the vectors ${\bf v}_I$. 
Analyticity of ${\tt tr}\,{\bf E}_{\vec{\tau},m}$ will hold if both ${\bf M}_{\vec{\tau}}$ 
and the combination involving ${\bf P}_{\vec{\tau},m}$ are analytic.\footnote{The difference w.r.t.~the SUGRA eigenvalue problem is the 
projector ${\bf P}_{\vec{\tau},m}$, i.e.~in SUGRA we would find the resolution of the identity, 
rather than the sum from $1$ to $m$. In fact, the SUGRA anomalous dimensions at tree level 
are also the eigenvalues of ${\bf M}{\bf L}^{-1}$, as shown in \cite{Aprile:2019rep}.}
By definition ${\bf M}_{\vec{\tau}}$ collects the superblock 
decomposition of the VS amplitude on $R_{\vec{\tau}}\otimes R_{\vec{\tau}}$, thus is analytic in $\vec{\tau}$ 
when the superblock decomposition is analytic.
Now, notice that the combination %${\bf L}_{\vec{\tau}}^{-\frac{1}{2} } {\bf P}_{\vec{\tau},m} {\bf L}_{\vec{\tau}}^{-\frac{1}{2} }$ 
involving ${\bf P}_{\vec{\tau},m}$ is also analytic if the `square' of three-point couplings is. Indeed we can rewrite it as
\be
\Big({\bf L}_{\vec{\tau}}^{-\frac{1}{2} } {\bf P}_{\vec{\tau},m} {\bf L}_{\vec{\tau}}^{-\frac{1}{2} }\Big)_{ij}= 
\Big({\bf L}^{-1}_{\vec{\tau}}\Big)_{ii} \left[ \sum_{I=1}^{m} \Big( {\bf C}^{(0)}_{\vec{\tau}}\Big)_{iI} \Big({\bf C}^{(0)T}_{\vec{\tau}}\Big)_{Ij} \right]\Big({\bf L}^{-1}_{\vec{\tau}}\Big)_{jj} 
\ee
since  $span\big( {\bf v}_I\big)\simeq span\Big(\,\big({\bf L}_{\vec{\tau}}^{-\frac{1}{2} } {\bf C}^{(0)}_{\vec{\tau}}\big)_{iI}\Big)$, 
up to unitary transformations on the hyperplane. 

The domain of definition of $\mathcal{P}_m^*(T,B,a,l)$ is the physical domain 
of existence of the level splitting matrix ${\bf E}_{m,\vec{\tau}}$, 
\ba\label{physical_domain_E}
\ \ \ \tau\ge b+2a+4+2(m-1)\qquad;\qquad \left\{ \begin{array}{lcl} \ b\ge 2m-2\ &{\rm if\ }&\ a+l\ even\ \\[.2cm] \ b\ge 2m-1\ &{\rm if\ }&\ a+l\ odd \end{array}\right. 
\ea
In relation to this, analyticity in $\vec{\tau}$ is now quite important because allows us to think 
about the roots of the characteristic polynomial as the STRINGY anomalous dimensions of 
analytically continued two-particle operators, outside the physical domain of definition. 
In this sense our experiments on the $m^*=2$ problem had two amazing outcomes. 
Firstly, we learned that the new anomalous dimensions
start splitting sequentially as we move away from the flat space limit.
Secondly, we learned that as the space of physical operators reduces, 
the characteristic polynomial reduces as well, factorising a zero root each time.  
Already for $m^*=2$ we saw that the non vanishing root carries the correct information 
about the physical spectrum of operators, which is not an obvious feature. 
We will refer to this process as `rank reduction'. 
 This phenomenon is quite beautiful and yet to be fully understood.

We will now generalise both the sequential splitting and the rank reduction 
for arbitrary $m\ge 2$, and we will demonstrate 
that they hold for the case $m^*=3,4$, i.e.~the first cases for which the roots of ${\cal P}^*$ are not explicit.
It will be convenient to use the notation
\ba\label{repeat_deg_char_P}
\!\!K_{m,j}&=& \sum_{ 0\leq x,y \leq deg} T^x\, \Big({\bf K}_{m,j}(a,l)\Big)_{xy}\, B^y
\ea
introducing the matrix ${\bf K}_{m,j}$. 
Recall $deg[K_{m,j}]\leq j\times \left(2 m-2+\tfrac{1}{2}(1-(-1)^{a+l})\right)$.

We computed ${\cal P}^*_{m=3,4}$ as function of $T$ and $B$ without imposing
any constraint to start with. The results are attached in an ancillary file. Then, 
we checked that both the sequential splitting and the rank reduction hold. 
Thus, we repeated the computation the other way round. What happens on ${\bf K}_{m,j}$ is quite instructive, 
and is anticipated by the following picture,
\be
\!\!\!\begin{array}{c}
\begin{tikzpicture}[scale=.9]
\def\step{.7};

\draw[gray] (0,0) rectangle (9*\step,-9*\step);

%vertical lines
 \draw[gray] (0,0) -- (0,-9*\step) ;
\foreach \x in {1,...,9}
    \draw[gray] (\x*\step,0) -- (\x*\step,-9*\step+\x*\step-\step) ;

%horizontal lines
 \draw[gray] (0,0) -- (9*\step,0) ;
\foreach \y in {1,...,9}
    \draw[gray] (0,-1*\y*\step) -- (9*\step-\y*\step+\step,-1*\y*\step) ;

%rectangles red
\foreach \x in {0,...,8}
    \filldraw[red!70] (0+\x*\step+.05,-9*\step+\x*\step+.05) rectangle  (0+\x*\step+\step-.05,-9*\step+\x*\step+\step-.05);

\foreach \x in {0,...,7}
    \filldraw[red!40] (0+\x*\step+.05,-8*\step+\x*\step+.05) rectangle  (0+\x*\step+\step-.05,-8*\step+\x*\step+\step-.05);

\foreach \x in {0,...,6}
    \filldraw[red!20] (0+\x*\step+.05,-7*\step+\x*\step+.05) rectangle  (0+\x*\step+\step-.05,-7*\step+\x*\step+\step-.05);

%rectangles green

\filldraw[blue!70] (0+.05,-\step+.05) rectangle (\step-.05,0-.05);

\foreach \x in {0,1}
    \filldraw[blue!40] (\x*\step+.05,-2*\step+\x*\step+.05) rectangle (\x*\step+\step-.05,-2*\step+\x*\step+\step-.05);

\foreach \x in {0,1,2}
    \filldraw[blue!20] (\x*\step+.05,-3*\step+\x*\step+.05) rectangle (\x*\step+\step-.05,-3*\step+\x*\step+\step-.05);

%grading B
\draw (+0.5*\step,+0.6*\step) node[scale=1] {$B^0$};
\draw (+1.5*\step,+0.6*\step) node[scale=1] {$B^1$};
\draw (+2.5*\step,+0.6*\step) node[scale=1] {$\ldots$};
\draw (+6.3*\step,+0.6*\step) node[scale=1] {$\ldots$};
\draw (+7.8*\step,+0.6*\step) node[scale=1] {$B^{deg-1}$};
\draw (+9.3*\step,+0.6*\step) node[scale=1] {$B^{deg}$};

%grading T
\draw (-0.7*\step,-0.5*\step) node[scale=1] {$T^0$};
\draw (-0.7*\step,-1.5*\step) node[scale=1] {$T^1$};
\draw (-0.7*\step,-2.5*\step) node[scale=1] {$\vdots$};
\draw (-0.7*\step,-6.5*\step) node[scale=1] {$\vdots$};
\draw (-0.4*\step,-7.5*\step) node[scale=1] {$T^{deg-1}$};
\draw (-0.7*\step,-8.5*\step) node[scale=1] {$T^{deg}$};

%\draw (2.4*\step,-2.4*\step) node[scale=1.5,rotate=45] {unconstrained};

\draw (6.5*\step,-6.5*\step) node[scale=5] {$0$};
\draw  (-2.7*\step,-4.5*\step) node[scale=1] {$\Big(\mathbf{K}^{}_{m,j}\Big)_{xy}=$};

%space
\draw  (+11.1*\step,-3.5*\step) node[scale=1] {$\rule{.1cm}{0pt}$};

\end{tikzpicture}
\label{figure_constraints}
\end{array}
\ee\\
The flat space limit determines the entries on the diagonal. Then, constraints from the 
sequential splitting impose relations spreading on the other diagonals {\color{red} in red}. 
Constraints from rank reduction instead impose relations spreading from the left top corner {\color{blue} in blue}.
The rest of ${\bf K}_{m,j}$ can only be determined by looking at the characteristic polynomial 
from the actual computation of the VS amplitude. 

We will find that the symmetry $T\leftrightarrow B$ and $a\leftrightarrow l$ holds 
beyond the flat space approximation, and in fact, it relates ${\bf K}$ to its transposed with 
swapped parameters, 
\ba\label{duality_proposal}
{\bf K}_{m,j}(a,l)&=&\Big[{\bf K}_{m,j}(l,a)\Big]^{\rm T} \rule{1.6cm}{0pt} {a+l}\ {even} \\[.2cm]
{\bf K}_{m,j}(a,l)&=&\Big[(-)^{j} {\bf K}_{m,j}(l,a)\Big]^{\rm T} \qquad {a+l}\ {odd} 
\ea
Considering that the level splitting problem is uniquely determined within our bootstrap program, 
and assuming that we have been able to isolate a sub-amplitude inside the full VS amplitude in $AdS_5\times S^5$, 
responsible just for the level splitting problem, then we infer from \eqref{duality_proposal} that this subamplitude 
will have a non trivial duality in its Mellin representation reflecting the symmetry above. 
Moreover this duality will be non perturbative, since the level splitting problem is not a problem at fixed order in the $\alpha'$ expansion.

%Let us show now how the constraints work in formulae.

%======================================================================================
\paragraph{Sequential splitting away from flat space.}~\\[-.2cm]
%======================================================================================

The flat space limit formalises as follows 
\be
\mathcal{P}^*_{\vec{\tau},m}( \epsilon^{n-a-l} \tilde{\eta}, \epsilon T,\epsilon B )\Big|_{ \epsilon^{m(n-a-l)} }=  \left(\,\tilde{\eta}+(T-B)^{n-a-l}\right)^{m} 
\ee
Then the expansion away from the flat space limit is an expansion in the shifted anomalous dimension
\be
\eta_{flat}=\tilde{\eta}+(T-B)^{n-a-l}
\ee 
The sequential splitting is now the statement that the roots of the characteristic polynomial 
move away from the degenerate locus $\eta_{flat}=0$ one by one, sequentially for each 
extra term we keep in the $\epsilon$ expansion.  This means that $\mathcal{P}^*_{\vec{\tau},m}$ is such that
\begin{align}
&
\mathcal{P}^*_{\vec{\tau},m}( \epsilon^{n-a-l} \tilde{\eta}, \epsilon T,\epsilon B )=\epsilon^{m(n-a-l)}\Bigg[   \\
&
\rule{2cm}{0pt}
 \left(\eta_{flat}\right)^{m}  + \frac{1}{\epsilon} \left(\eta_{flat} \right)^{m-1} C_1(\tilde{\eta},B,T) +  \frac{1}{\epsilon^2} \left(\eta_{flat} \right)^{m-2} C_2(\tilde{\eta},B,T)  + \ldots  \Bigg] \notag
\end{align}
with generic $C_i$. The dependence on $a$ and $l$ is understood.

If we consider an ansatz for the coefficients $K_{m,j}$, we know the degree w.r.t.~to $B$ and $T$, given in \eqref{repeat_deg_char_P}, 
and then we know the diagonal entries of the  ${\bf K}_{m,j\ge 1}$, since these are determined by the flat space limit. 
The flat space limit is a universal constraint for any ${\bf K}_{m,j\ge 1}$, but is the only constraint for ${\bf K}_{m,1}$. 
The sequential splitting takes ${\bf K}_{m,1}$ as an input and moves forward. At the first step we find that the diagonal 
of ${\bf K}_{m,2}$, which is next-to-the-flat space limit, is determined by ${\bf K}_{m,1}$.  
Then, we find that the next-to- and next-to-next-to-the-flat space diagonals of ${\bf K}_{m,3}$ 
depend on ${\bf K}_{m,1}$, and ${\bf K}_{m,j=1,2}$, respectively, and so on so forth. 
The flow according to which the constraints move sequentially away from flat space, as we look 
to coefficients $K_{j>1}$, is represented by the shadowing in red in figure \eqref{figure_constraints}.

%======================================================================================
\paragraph{Rank reduction.}~\\[-.2cm]
%======================================================================================

%
The rank reduction is better phrased using  the concept of a `filtration'.\footnote{Filtration is the name for 
a sequence of sets $\{ S_i\}_{i\in\mathbb{N}}$ labelled by an integer, such that $S_{i}\subset S_{i+1}$.} 
Consider a rectangle $R_{\tau,l,[aba]}$ for fixed values of $l$ and $[aba]$ and varying twist.   
The minimum available twist is $\tau_{min}=b+2a+4$, but otherwise $\tau$ can grow unbounded. 
Then, the sequence
\begin{align}
R_{b+2a+4,l,[aba]}\,\subset\, R_{b+2a+6,l,[aba]}\,\subset\, R_{b+2a+8,l,[aba]}\, \subset\, \ldots
\end{align}
is a filtration. 
Graphically, %we would draw a figure like the one below, 

\begin{center}
\begin{minipage}{\textwidth}
   % 	\centering
	%
	%
	%
	% FIG UNO
	%
	%
	%
  	\begin{minipage}{.225\textwidth}
			\begin{tikzpicture}[scale=.54]

			\draw[step=1.5cm,gray,very thin] (-1,0) grid (6.3,7);
			\def\prop{.5}
			\def\shifthor{\prop*2}
%

    				%rectangle [040]
			 \draw[black!50,thick] (\prop*2-\shifthor,\prop*6) --  (\prop*4-\shifthor,\prop*4);
% 			 \draw[blue!50,thick] (\prop*2-\shifthor,\prop*6) -- (\prop*9-\shifthor,\prop*13);
%  			 \draw[blue!50,thick] (\prop*4-\shifthor,\prop*4) -- (\prop*11-\shifthor,\prop*11);
% 			 \draw[blue!50,thick]  (\prop*9-\shifthor,\prop*13) -- (\prop*11-\shifthor,\prop*11);

			%
			%
			\foreach \indeyc in {0,1,2}
			\foreach \indexc  in {2}
			\filldraw   					 (\prop*\indexc+\prop*\indeyc-\shifthor, \prop*4+\prop*\indexc-\prop*\indeyc)   	circle (.07);

			%\foreach \indeyc in {0}
			%\foreach \indexc  in {2,...,9}
			%\filldraw   					 (\prop*\indexc+\prop*\indeyc-\shifthor, \prop*6+\prop*\indexc-\prop*\indeyc)   	circle (.07);

			\draw[thick, -latex] (\prop*2-1.2*\shifthor, \prop*2) --  (\prop*12-\shifthor, \prop*2);
			\draw[thick, -latex] (\prop*2-\shifthor, \prop*2-.3) --  (\prop*2-\shifthor, \prop*14);
			\draw (\prop*2-\shifthor, \prop*2-.35) node[left] {\footnotesize 2};
			\draw (\prop*2-\shifthor, \prop*6) node[left] {\footnotesize 6};
			
			\draw (\prop*8-\shifthor, \prop*4) node[right] {\footnotesize $\tau=8$};
			%\draw (\prop*8-\shifthor, \prop*3) node[right] {\footnotesize [040]};
			
			\draw[blue,thick]  (\prop*4+\prop*0-\shifthor, \prop*6+\prop*1-\prop*1) ellipse (.22 and 1.5);

\end{tikzpicture}
\end{minipage}
\ \ 
  	\begin{minipage}{.225\textwidth}
			\begin{tikzpicture}[scale=.54]

			\draw[step=1.5cm,gray,very thin] (-1,0) grid (6.3,7);
			\def\prop{.5}
			\def\shifthor{\prop*2}
%

    				%rectangle [040]
			 \draw[black!50,thick] (\prop*2-\shifthor,\prop*6) --  (\prop*4-\shifthor,\prop*4);
 			 \draw[black!50,thick] (\prop*2-\shifthor,\prop*6) -- (\prop*3-\shifthor,\prop*7);
  			 \draw[black!50,thick] (\prop*4-\shifthor,\prop*4) -- (\prop*5-\shifthor,\prop*5);
 			 \draw[black!50,thick]  (\prop*5-\shifthor,\prop*5) -- (\prop*3-\shifthor,\prop*7);

			\foreach \indeyc in {0,1,2}
			\foreach \indexc  in {2,3}
			\filldraw   					 (\prop*\indexc+\prop*\indeyc-\shifthor, \prop*4+\prop*\indexc-\prop*\indeyc)   	circle (.07);

			%\foreach \indeyc in {0}
			%\foreach \indexc  in {2,...,9}
			%\filldraw   					 (\prop*\indexc+\prop*\indeyc-\shifthor, \prop*6+\prop*\indexc-\prop*\indeyc)   	circle (.07);

			\draw[thick, -latex] (\prop*2-1.2*\shifthor, \prop*2) --  (\prop*12-\shifthor, \prop*2);
			\draw[thick, -latex] (\prop*2-\shifthor, \prop*2-.3) --  (\prop*2-\shifthor, \prop*14);
			\draw (\prop*2-\shifthor, \prop*2-.35) node[left] {\footnotesize 2};
			\draw (\prop*2-\shifthor, \prop*6) node[left] {\footnotesize 6};
			
			\draw (\prop*8-\shifthor, \prop*4) node[right] {\footnotesize $\tau=10$};
			%\draw (\prop*8-\shifthor, \prop*3) node[right] {\footnotesize [040]};
			
			\draw[blue,thick]  (\prop*4+\prop*0-\shifthor, \prop*6+\prop*1-\prop*1) ellipse (.22 and 1.5);

\end{tikzpicture}
\end{minipage}
\ \ 
  	\begin{minipage}{.225\textwidth}
			\begin{tikzpicture}[scale=.54]

			\draw[step=1.5cm,gray,very thin] (-1,0) grid (6.3,7);
			\def\prop{.5}
			\def\shifthor{\prop*2}
%

    				%rectangle [040]
			 \draw[black!50,thick] (\prop*2-\shifthor,\prop*6) --  (\prop*4-\shifthor,\prop*4);
 			 \draw[black!50,thick] (\prop*2-\shifthor,\prop*6) -- (\prop*4-\shifthor,\prop*8);
  			 \draw[black!50,thick] (\prop*4-\shifthor,\prop*4) -- (\prop*6-\shifthor,\prop*6);
 			 \draw[black!50,thick]  (\prop*6-\shifthor,\prop*6) -- (\prop*4-\shifthor,\prop*8);

			\foreach \indeyc in {0,1,2}
			\foreach \indexc  in {2,3,4}
			\filldraw   					 (\prop*\indexc+\prop*\indeyc-\shifthor, \prop*4+\prop*\indexc-\prop*\indeyc)   	circle (.07);

			%\foreach \indeyc in {0}
			%\foreach \indexc  in {2,...,9}
			%\filldraw   					 (\prop*\indexc+\prop*\indeyc-\shifthor, \prop*6+\prop*\indexc-\prop*\indeyc)   	circle (.07);

			\draw[thick, -latex] (\prop*2-1.2*\shifthor, \prop*2) --  (\prop*12-\shifthor, \prop*2);
			\draw[thick, -latex] (\prop*2-\shifthor, \prop*2-.3) --  (\prop*2-\shifthor, \prop*14);
			\draw (\prop*2-\shifthor, \prop*2-.35) node[left] {\footnotesize 2};
			\draw (\prop*2-\shifthor, \prop*6) node[left] {\footnotesize 6};
			
			\draw (\prop*8-\shifthor, \prop*4) node[right] {\footnotesize $\tau=12$};
			%\draw (\prop*8-\shifthor, \prop*3) node[right] {\footnotesize [040]};
			
			\draw[blue,thick]  (\prop*4+\prop*0-\shifthor, \prop*6+\prop*1-\prop*1) ellipse (.22 and 1.5);

\end{tikzpicture}
\end{minipage}
\ \
  	\begin{minipage}{.225\textwidth}
			\begin{tikzpicture}[scale=.54]

			\draw[step=1.5cm,gray,very thin] (-1,0) grid (6.3,7);
			\def\prop{.5}
			\def\shifthor{\prop*2}
%

    				%rectangle [040]
			 \draw[black!50,thick] (\prop*2-\shifthor,\prop*6) --  (\prop*4-\shifthor,\prop*4);
 			 \draw[black!50,thick] (\prop*2-\shifthor,\prop*6) -- (\prop*5-\shifthor,\prop*9);
  			 \draw[black!50,thick] (\prop*4-\shifthor,\prop*4) -- (\prop*7-\shifthor,\prop*7);
 			 \draw[black!50,thick]  (\prop*7-\shifthor,\prop*7) -- (\prop*5-\shifthor,\prop*9);

			\foreach \indeyc in {0,1,2}
			\foreach \indexc  in {2,3,4,5}
			\filldraw   					 (\prop*\indexc+\prop*\indeyc-\shifthor, \prop*4+\prop*\indexc-\prop*\indeyc)   	circle (.07);

			%\foreach \indeyc in {0}
			%\foreach \indexc  in {2,...,9}
			%\filldraw   					 (\prop*\indexc+\prop*\indeyc-\shifthor, \prop*6+\prop*\indexc-\prop*\indeyc)   	circle (.07);

			\draw[thick, -latex] (\prop*2-1.2*\shifthor, \prop*2) --  (\prop*12-\shifthor, \prop*2);
			\draw[thick, -latex] (\prop*2-\shifthor, \prop*2-.3) --  (\prop*2-\shifthor, \prop*14);
			\draw (\prop*2-\shifthor, \prop*2-.35) node[left] {\footnotesize 2};
			\draw (\prop*2-\shifthor, \prop*6) node[left] {\footnotesize 6};
			
			\draw (\prop*8-\shifthor, \prop*4) node[right] {\footnotesize $\tau=14$};
			%\draw (\prop*8-\shifthor, \prop*3) node[right] {\footnotesize [040]};
			
			\draw[blue,thick]  (\prop*4+\prop*0-\shifthor, \prop*6+\prop*1-\prop*1) ellipse (.22 and 1.5);

\end{tikzpicture}
\end{minipage}
\end{minipage}
\end{center}

Consider now a value of the level splitting label in the filtration, say ${m}$. 
The figure above has $m=3$.  By varying the twist, the number of physical operators with that $m$ varies: 
It goes from $m$ operators in the domain  \eqref{physical_domain_E} for generic twist, 
down to a single physical operator at the minimum twist $\tau=b+2a+4$. 
In particular, we are always outside the domain of definition of the characteristic 
polynomial as long as $b+2a+4\leq\tau< b+2a+4+2(m-1)$.

Following on what happens at $m^*=2$, we should find that as decrease the twist below the domain of definition
the coefficients of the characteristic polynomial vanish in such a way to factor a zero root each time. The pattern is
\begin{align}\label{locus_rankreduction}
\begin{array}{ccl}
K_{m,m\phantom{-0}}=0 	& @ &  t=2,\,\ldots\,\ldots\, \ldots m\\
K_{m,m-1}=0 & @  & t=2,\,\ldots \ \ m-1\\
K_{m,m-2}=0 & @  & t=2,\,\ldots m-2\\
\vdots\ \ \ \ \ & & \ \ \ \vdots \\
K_{m,2}=0 & @ & t=2
\end{array}
\end{align}
which is solved by
\begin{align}\label{factori_rank_reductionV1}
&
\!\!\!\!K_{m,j}(T,B,a,l)= \\[.2cm]
&
(\tfrac{\tau+b}{2}-j+2)_{j-1} (\tfrac{\tau+b}{2}+a+l+4)_{j-1} (\tfrac{\tau-b}{2}-a-j)_{j-1}(\tfrac{\tau-b}{2}+l+2)_{j-1} \times \tilde{K}_{m,j}(T,B,a,l) \notag
\end{align}
where $\tilde{K}$ has reduced degree. 
The prefactor is the unique polynomial in $B$ and $T$ of minimal degree which 
vanishes at the locus \eqref{locus_rankreduction}.

In practise, the characteristic polynomial $\mathcal{P}^*_{\vec{\tau},m}$ always has $m$ roots, 
and since it is analytic we think of these roots as describing the anomalous dimensions of 
$m$ analytically continued operators. But as the rank of ${\bf N}_{\vec{\tau}}$ reduces, 
the number of physical operators changes. This is the picture given by the filtration. 
By experiment, analytically continued operators which do not correspond to physical 
operators localises on the vanishing roots.\footnote{This actually suggests that 
an alternative definition of this characteristic polynomial might exists such that it always lives on the
space of $m\times m$ matrices.} 
Considering that the bulk interpretation of the level splitting is the formation of energetically 
favourable bound states, exchanged in the S-matrix, we infer the bound $\tilde{\eta}\leq 0$. 
This would explain the degeneration of the roots onto $\eta=0$, that we see experimentally.

%========================================================================
\paragraph{A summary of experiments.}~\\[-.2cm]
%========================================================================

We have attached in an ancillary file the characteristic polynomials for 
$m^*=3$ and $a+l=0,1,2$, and the characteristic polynomial for $m^*=4$ and $a+l=0$. 

Apart from the properties mentioned in the previous section, we have found multiple 
factorisations, compared to the ones we could justify in \eqref{factori_rank_reductionV1} 
in relation to the rank reduction in twist. Consider the reduction factor, that we understood to 
be always present in order to reduced the coefficients $K_{m,j}$, namely
\be\label{Vanishing_facto}
\ \ \ {\tt R}_{j,\vec{\tau}}\equiv(\tfrac{\tau+b}{2}-j+2)_{j-1} (\tfrac{\tau+b}{2}+a+l+4)_{j-1} (\tfrac{\tau-b}{2}-a-j)_{j-1}(\tfrac{\tau-b}{2}+l+2)_{j-1}
\ee
in practise, we find the pattern
\be
\label{full_factorisation}
\!\!
\begin{array}{lll}
K_{m=2,2}={\tt R}_{2,\vec{\tau}}\times \tilde{K}_{2,2}\\[.2cm]
K_{m=3,2}={\tt R}_{2,\vec{\tau}}\times\tilde{K}_{3,2}\quad;\quad&
K_{m=3,3}={\tt R}_{3,\vec{\tau}}{\tt R}_{2,\vec{\tau}}\times\tilde{K}_{3,3}\\[.2cm]
K_{m=4,2}={\tt R}_{2,\vec{\tau}}\times\tilde{K}_{4,2}\quad;\quad&
K_{m=4,3}={\tt R}_{3,\vec{\tau}}{\tt R}_{2,\vec{\tau}}\times\tilde{K}_{4,3}\quad;\quad&
K_{m=4,4}={\tt R}_{4,\vec{\tau}}{\tt R}_{3,\vec{\tau}}{\tt R}_{2,\vec{\tau}}\times\tilde{K}_{4,4}\\[.2cm]
\end{array}
\ee
Therefore the various ${\tt R}_{m,j}$ appear multiple times, giving extra multiplicity to the 
individual factors in \eqref{Vanishing_facto}, and the $\tilde{K}_{m,j}(B,T,a,l)$ have reduced 
degree compared to $K_{m,j}$.

We shall now ask how the characteristic polynomial behaves when we vary the label $b$ of the $[aba]$ rep, 
keeping the level splitting label fixed. Consider for example the fate of the operators at 
$m^*=2$ across various $su(4)$ channels,  depicted in the figure below.
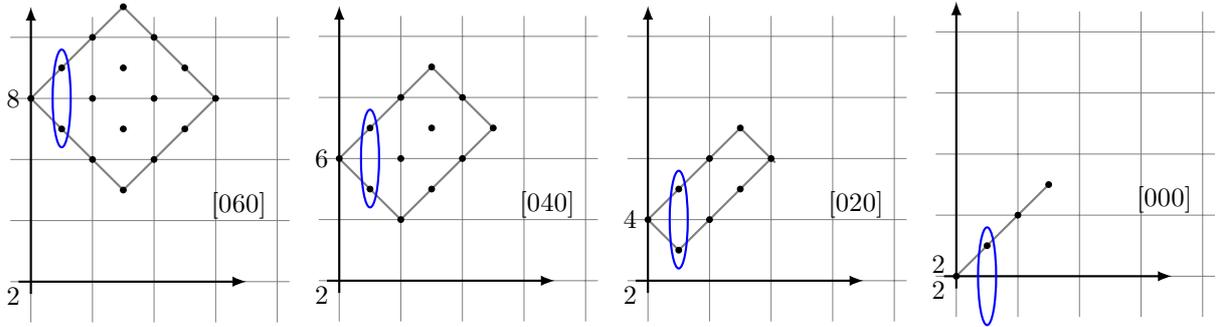
\begin{center}
\begin{minipage}{\textwidth}
   % 	\centering
	%
	%
	%
	% FIG UNO
	%
	%
	%
    	\begin{minipage}{.225\textwidth}
			\begin{tikzpicture}[scale=.54]

			\draw[step=1.5cm,gray,very thin] (-.5,.5) grid (6.3,8);
			\def\prop{.75}
			\def\shifthor{\prop*2}

   			 %rectangle [060]
 				 \draw[gray,thick] (\prop*2-\shifthor,\prop*8) --  (\prop*5-\shifthor,\prop*5);
 				 \draw[gray,thick] (\prop*2-\shifthor,\prop*8) -- (\prop*5-\shifthor,\prop*11);
				 \draw[gray,thick] (\prop*5-\shifthor,\prop*5) -- (\prop*8-\shifthor,\prop*8);
				 \draw[gray,thick]  (\prop*5-\shifthor,\prop*11) -- (\prop*8-\shifthor,\prop*8);

			\foreach \indeyc in {0,1,2,3}
			\foreach \indexc  in {2,...,5}
			\filldraw   					 (\prop*\indexc+\prop*\indeyc-\shifthor, \prop*6+\prop*\indexc-\prop*\indeyc)   	circle (.07);

			\draw[thick, -latex] (\prop*2-1.2*\shifthor, \prop*2) --  (\prop*9-\shifthor, \prop*2);
			\draw[thick, -latex] (\prop*2-\shifthor, \prop*2-.3) --  (\prop*2-\shifthor, \prop*11);
			\draw (\prop*2-\shifthor, \prop*2-.35) node[left] {\footnotesize 2};
			\draw (\prop*2-\shifthor, \prop*8) node[left] {\footnotesize 8};
			
			\draw (\prop*7.5-\shifthor, \prop*4.5) node[right] {\footnotesize [060]};

			\draw[blue,thick]  (\prop*3+\prop*0-\shifthor, \prop*6+\prop*3-\prop*1) ellipse (.22 and 1.2);

\end{tikzpicture}
\end{minipage}
\ \ 
  	\begin{minipage}{.225\textwidth}
			\begin{tikzpicture}[scale=.54]

			\draw[step=1.5cm,gray,very thin] (-.5,.5) grid (6.3,8);
			\def\prop{.75}
			\def\shifthor{\prop*2}
%

    				%rectangle [040]
			 \draw[gray,thick] (\prop*2-\shifthor,\prop*6) --  (\prop*4-\shifthor,\prop*4);
 			 \draw[gray,thick] (\prop*2-\shifthor,\prop*6) -- (\prop*5-\shifthor,\prop*9);
  			 \draw[gray,thick] (\prop*4-\shifthor,\prop*4) -- (\prop*7-\shifthor,\prop*7);
 			 \draw[gray,thick]  (\prop*5-\shifthor,\prop*9) -- (\prop*7-\shifthor,\prop*7);

			\foreach \indeyc in {0,1,2}
			\foreach \indexc  in {2,...,5}
			\filldraw   					 (\prop*\indexc+\prop*\indeyc-\shifthor, \prop*4+\prop*\indexc-\prop*\indeyc)   	circle (.07);

			\draw[thick, -latex] (\prop*2-1.2*\shifthor, \prop*2) --  (\prop*9-\shifthor, \prop*2);
			\draw[thick, -latex] (\prop*2-\shifthor, \prop*2-.3) --  (\prop*2-\shifthor, \prop*11);
			\draw (\prop*2-\shifthor, \prop*2-.35) node[left] {\footnotesize 2};
			\draw (\prop*2-\shifthor, \prop*6) node[left] {\footnotesize 6};
			
			\draw (\prop*7.5-\shifthor, \prop*4.5) node[right] {\footnotesize [040]};
			
			\draw[blue,thick]  (\prop*3+\prop*0-\shifthor, \prop*6+\prop*1-\prop*1) ellipse (.22 and 1.2);

\end{tikzpicture}
\end{minipage}
\ \ 
  	\begin{minipage}{.225\textwidth}
			\begin{tikzpicture}[scale=.54]

			\draw[step=1.5cm,gray,very thin] (-.5,.5) grid (6.3,8);
			\def\prop{.75}
			\def\shifthor{\prop*2}
%
		
		%  %rectangle [000]
		% \draw[thick] (\prop*2-\shifthor,\prop*2) --  (\prop*9-\shifthor,\prop*9);
		
 		 %rectangle [020]
  		 \draw[gray, thick] (\prop*2-\shifthor,\prop*4) --  (\prop*3-\shifthor,\prop*3);
		 \draw[gray,thick] (\prop*3-\shifthor,\prop*3) -- (\prop*6-\shifthor,\prop*6);
 		 \draw[gray,thick] (\prop*2-\shifthor,\prop*4) --  (\prop*5-\shifthor,\prop*7);
 		 \draw[gray,thick]  (\prop*5-\shifthor,\prop*7) -- (\prop*6-\shifthor+.1,\prop*6-.1);

		\foreach \indeyc in {0,1}
		\foreach \indexc  in {2,...,5}
		\filldraw   					 (\prop*\indexc+\prop*\indeyc-\shifthor, \prop*2+\prop*\indexc-\prop*\indeyc)   	circle (.07);

			\draw[thick, -latex] (\prop*2-1.2*\shifthor, \prop*2) --  (\prop*9-\shifthor, \prop*2);
			\draw[thick, -latex] (\prop*2-\shifthor, \prop*2-.3) --  (\prop*2-\shifthor, \prop*11);
			\draw (\prop*2-\shifthor, \prop*2-.35) node[left] {\footnotesize 2};
			\draw (\prop*2-\shifthor, \prop*4) node[left] {\footnotesize 4};
			
			\draw (\prop*7.5-\shifthor, \prop*4.5) node[right] {\footnotesize [020]};
			
			\draw[blue,thick]  (\prop*3+\prop*0-\shifthor, \prop*5-\prop*1) ellipse (.22 and 1.2);

\end{tikzpicture}
\end{minipage}
\ \ 
  	\begin{minipage}{.225\textwidth}
			\begin{tikzpicture}[scale=.54]

			\draw[step=1.5cm,gray,very thin] (-.5,.5) grid (6.3,8);
			\def\prop{.75}
			\def\shifthor{\prop*2}
%

%		
  		%rectangle [000]
		 \draw[gray,thick] (\prop*2-\shifthor,\prop*2) --  (\prop*5-\shifthor,\prop*5);

		\foreach \indeyc in {0}
		\foreach \indexc  in {2,...,5}
		\filldraw   					 (\prop*\indexc+\prop*\indeyc-\shifthor, \prop*0+\prop*\indexc-\prop*\indeyc)   	circle (.07);

			\draw[thick, -latex] (\prop*2-1.2*\shifthor, \prop*2) --  (\prop*9-\shifthor, \prop*2);
			\draw[thick, -latex] (\prop*2-\shifthor, \prop*2-.3) --  (\prop*2-\shifthor, \prop*11);
			\draw (\prop*2-\shifthor, \prop*2-.35) node[left] {\footnotesize 2};
			\draw (\prop*2-\shifthor, \prop*2+.3) node[left] {\footnotesize 2};
			
			\draw (\prop*7.5-\shifthor, \prop*4.5) node[right] {\footnotesize [000]};
			
			\draw[blue,thick]  (\prop*3+\prop*0-\shifthor, \prop*3-\prop*1) ellipse (.22 and 1.2);

\end{tikzpicture}
\end{minipage}
\end{minipage}
\end{center}
This figure wants to show that a physical pair of operators with level splitting label $m^*=2$ exists
for any $[0b0]$ with $b\ge 2$ and $\tau\ge b+2a+6$, but when we go to $[000]$ or $[010]$ 
we find only one physical operator. 
For clarity, let us remark  that $\mathcal{P}^*(T,B,a,l)$ is defined in the domain \eqref{physical_domain_E} 
and here we are discussing what happens outside that domain, as it was the case for the rank reduction in twist, but this time in the $b$ direction.

Consider then the characteristic polynomial given in \eqref{db_dege_ch_poly}, and check what 
happens in the case of the figure. From right to left the significative cases are $[000]$ and $[020]$. 
The result is included in the general formula
\begin{gather}
[a0a]%\qquad{\rm even\ spins}
\qquad n=a+l+2\\[.2cm]
%
%\scalebox{1.1}{$
\begin{array}{c}
\mathcal{P}^*=\left( \tilde{\eta}+\frac{ \tau(\tau+2l+4)(8+4a+4\tau+2l\tau+\tau^2)}{16}\right)\left( \tilde{\eta} 
			+\frac{(\tau-2a-4)(2n+4+\tau)}{4}\big(T+\frac{3a-3n+an-n^2}{2n+5} \big)\!\right)%$}
\end{array}
			\notag
\end{gather}
Thus the characteristic polynomial factorises! and the $n$-dependent root coincides with the anomalous dimension
of the physical operator at $\tau=b+2a+6$ upon inspection.\footnote{This is the first value of the twist for 
an $m^*=2$ two-particle operator with $a+l=n-2$.} Similarly, 
\begin{gather}
[a1a]%\qquad{\rm even\ spins}
\qquad n=a+l+2\\[.2cm]
%
%\scalebox{1.1}{$
\begin{array}{c}
\mathcal{P}^*=\left( \tilde{\eta}+\frac{ (\tau-1)(\tau+1)(\tau+2l+3)(\tau+2l+5) }{16}\right)\left( \tilde{\eta} 
			+\frac{(\tau-2a-5)(2n+5+\tau)}{16}\big(T+\frac{12+2a+4l}{2n+5} \big)\!\right)%$}
			\end{array}
			\notag
\end{gather}
Notice the appearance of a fully factorised root, and recall now that it is $b=1$, rather than $b=0$, 
the first value of $b$ to lie outside the definition domain of the characteristic polynomial.
From $[020]$ upward we will find two physical roots with square root splitting, as shown in section \ref{sec_mstar2}. 

Consider also the characteristic polynomial given in \eqref{db_dege_ch_poly_odd}. Again we find
\begin{gather}
[a2a]%\qquad{\rm odd\ spins}
\qquad n=a+l+3\\[.2cm]
\!\!\!\!\!\!\!\!\!\!\!\!\!\scalebox{.9}{$
\mathcal{P}^*=\left( \tilde{\eta}+\frac{ (\tau-2)\tau(\tau+2)(\tau+2l+2)(\tau+2l+4)(\tau+2l+6)}{64}\right) \left( \tilde{\eta} 
								+\frac{(\tau-2a-6)(2n+4+\tau)}{4}\Big(T^2-\frac{22+2a+12n+an+n^2}{2n+5}+\frac{(n+3)(2-4a-a^2+5n+an)}{2n+5} \Big)\!\right)$}
								\notag
\end{gather}
Again the $n$-dependent root coincides with the physical anomalous dimensions of the operator 
at the minimum twist. Even more interestingly, the value $b=2$ is the first 
value of $b$ to lie outside the domain of definition of the characteristic polynomial, and again
we find a fully factorise root. 

The pattern of factorised non physical roots continues for $m^*=3,4$, in particular, 
\begin{align}
\mathcal{P}^*_{m,[aba]}\Big|_{b=2m-3}=& \Big( \tilde{\eta} + (\tfrac{\tau+1}{2}-m+1)_{2m-2} (\tfrac{\tau+1}{2}+l-m+3)_{2m-2}\Big)\Big( \ldots \Big)\qquad a+l\ {even} \notag\\
\\
\mathcal{P}^*_{m,[aba]}\Big|_{b=2m-2}=& \Big( \tilde{\eta} + (\tfrac{\tau}{2}-m+1)_{2m-1} (\tfrac{\tau}{2}+l-m+3)_{2m-1}\Big)\Big( \ldots \Big)\qquad \rule{.8cm}{0pt} a+l\ {odd}\notag
\end{align}

The transition in $b$ is thus different compared to the rank reduction in the twist $\tau$. 
It involves remarkable factorisations of the characteristic polynomial, but
brings some of the analytically continued operators to a non physical sheet.
Nevertheless, all physical roots are correctly captured, and we verified this statement 
for all characteristic polynomials attached.

Finally, consider combining the transition in $b$ and the reduction in $\tau$. Take first 
a ${\cal P}^*_{m}(T,B,a,l)$ and vary $b$ to a value such that $m^c$ analytically continued 
operators do not belong to the physical sheet, while $m-m^c$ instead remain. At this point
the characteristic polynomial factorises a polynomial of degree  $m^c$, which we discard, 
and an $m-m^c$ polynomial on the physical sheet remains. We can now reduce the 
latter in $\tau$, and ask whether the rank reduction works in the same way
as for the generic case. For all the examples we have, it works! 

We find all the properties of the characteristic polynomial quite fascinating.

%=============================================================
%=============================================================

\section{Conclusions}

%=============================================================
%=============================================================

In this paper we have taken another concrete step towards 
the determination of string theory amplitudes in curved space.
In particular, we gained access to the yet unknown 
Virasoro-Shapiro amplitude in $AdS_5\times S^5$
by bootstrapping  four-point correlators of single-particle operators $\langle \cO_{p_1} \cO_{p_2}\cO_{p_3}\cO_{p_4}\rangle$ 
in ${\cal N}=4$ SYM at genus zero, and arbitrary charges. 
To achieve the desired result we put together the understanding of the spectrum 
of two-particle operators in supergravity, classified in \cite{Aprile:2018efk},
the use of Mellin space techniques a la Penedones \cite{Penedones:2010ue}, 
the large $p$ limit \cite{Aprile:2020luw}, and previous results in \cite{Drummond:2019odu,Drummond:2020dwr}. 
Our key observation has been to relate polynomiality in Mellin space
with the spectral properties of the Virasoro Shapiro amplitude in $AdS_5\times S^5$ in the $\alpha'$ expansion.\footnote{Also, we expect that our discussion admits a 
generalisation to the study of the VS amplitude in $AdS_3\times S^3$,  
where hints of a hidden symmetry were observed in \cite{Rastelli:2019gtj,Giusto:2019pxc,Giusto:2020neo}.}

The inspiration to construct a crossing symmetric ansatz came from the large $p$ limit \cite{Aprile:2020luw},
which implies a precise relation between the Mellin amplitude and the flat space Virasoro-Shapiro amplitude in ten dimension, 
as well as a precise relation between the $AdS_5\times S^5$ Mellin variable and the Mandelstam momenta  in ten dimension.
Then we added in the information about the two-particle operators ${\cal K}_{(pq)}$ \cite{Aprile:2018efk}, 
and we assumed that the subset of visible operator  at each order in $\alpha'$, satisfies 
the inequality $l_{10}[{\cal K}_{(pq)}]\leq n$ where $n\in\mathbb{N}\ {\rm even}$. 
The l.h.s.~is the ten dimensional  spin of the two-particle operators
labelling the SUGRA anomalous dimensions, as recalled in section \ref{rew_sec_SUGRA}.
The value of $n$ on the r.h.s.~is controlled by  ${\bf s}^n+{\bf t}^n+{\bf u}^n$, which corresponds to the greatest spin contribution of the 
covariantised flat space amplitude at $(\alpha')^{n+3}$.
The bootstrap algorithm at this point simply becomes an `exclusion plot' on the initial ansatz, because
only certain operators can appear in the Virasoro-Shapiro amplitude at order $(\alpha')^{n+3}$. 
In section \ref{STRINGY_eige_sec} we formalised this statement as a set of \emph{rank constraints}.

After implementing  the {rank constraints} on an ansatz for ${\cal V}_n(\alpha')^{n+3}$, we showed that 
the CFT data at the edge $l_{10}=n$ is uniquely determined. On the other hand, CFT data of 
operators with $l_{10}<n$ is affected by the (bootstrap) 
ambiguity, for example those of shifting the amplitude  by the same as the amplitudes at previous orders. Thus novel
inputs are needed to finally fix the VS amplitude. In this sense, we also pointed out that the OPE can be used once more
to generate other predictions on top of the rank constraints. In particular, data extracted unambiguously at the edge 
of $(\alpha')^{k+3}$ with $k<n$, provides additional sharp information for the amplitudes at higher orders.
Another source of information comes from the results between the integrated correlator and the deformations of the partition functions 
computed by localisation \cite{Binder:2019jwn,Chester:2019pvm,Chester:2019,Chester:2020vyz}.
The current data can be used to fix the amplitudes at  $(\alpha')^{3,4,5}$, but was not enough at $(\alpha')^{6}$.
It would be interesting to understand whether more constraints can be derived in this context.  
 
 The integral transform $\tilde{\mathcal{V}}\rightarrow \mathcal{V}$ we introduced in \eqref{integral_tr_James} is perhaps 
 the most natural generalisation in $AdS_5\times S^5$  of the integral transform defining the flat space limit a la Penedones \cite{Penedones:2010ue}.
The way $\tilde{\mathcal{V}}$ simplifies is highly non trivial because it requires different strata of the amplitude 
to recombine together.  The simplicity of the final result appears to us quite remarkable, and we
think this might be the optimal way to understand the connection with other techniques. 
For example, it would be very interesting to bridge our results with the computation of the VS amplitude 
through vertex operator insertions \cite{Berkovits:2012ps, Azevedo:2014rva,Fleury:2021ieo}, and it would also be 
interesting to see how  octagon-like string configurations  at genus zero and strong coupling 
\cite{Coronado:2018ypq,Bargheer:2019exp,Belitsky:2019fan,Belitsky:2020qrm} overlap with the VS amplitude.

With the amplitudes at $(\alpha')^{5,7,9}$ at our disposal, we then studied the CFT data at the edge, and shown that it organises nicely as a splitting problem 
for the partial degeneracy of tree level supergravity,
with very interesting analyticity properties w.r.t.~twist, spin, and $su(4)$ indexes $[aba]$, 
both for the corrections to the anomalous dimensions and the three-point couplings.
The characteristic polynomial of the level splitting matrix is a new object in $\mathcal{N}=4$ SYM and it appears to be quite special.
As discussed and exemplified in section \ref{sec_mstar3}, two of the most intriguing properties
of the characteristic polynomial of the level splitting matrix are, the sequential splitting of the anomalous dimensions away from the flat space limit, 
and the rank reduction when the space of physical operators varies in number. 
We cannot avoid mentioning a possible connection 
with the quantum spectral curve \cite{Beisert:2010jr}, even though it is not obvious at this stage.

All together, the simplification achieved by the integral transform $\tilde{\mathcal{V}}\rightarrow \mathcal{V}$, the uniqueness of the CFT data at the edge,
and the intriguing properties of the level splitting problem, such as the symmetry $T\leftrightarrow B$ and $a\leftrightarrow l$,
strongly suggest that a preferred sub-amplitude $\tilde{\cal V}$, in the full and resummed VS amplitude, exists and encodes smartly all the nice features we see on the spectrum. 
It would be interesting to pin down precisely which properties in Mellin space select this sub-amplitude, beyond the $rank=1$ case. 
Even more interestingly, it would be great to build this quantity with a `top-down' approach, alternative to our bootstrap approach. 
For example, we expect that the symmetry $T\leftrightarrow B$ and $a\leftrightarrow l$ at the level of the spectrum should then
become a non trivial duality in Mellin space for this sub-amplitude. 
We hope to return on this problem in the near future.

\section*{Acknowledgements}
We thank Pedro Vieira for enlightening conversations, Theresa Abl, Paul Heslop and Arthur Lipstein for many discussions on related topics, and
Shai Chester for providing us with the latest localisation results. 
MS thanks Davide Bufalini and Sami Rawash for useful discussions.
FA is partially supported by the ERC-STG grant 637844 HBQFTNCER.
JD is supported in part by the ERC Consolidator grant 648630 IQFT.
HP acknowledges support from the ERC Starting Grant 679278 Emergent-BH.
MS is supported by a Mayflower studentship from the University of Southampton.

\appendix

%==========================================================================
%========================================================

\section{On the correlators $\langle {\cal O}_{p_1} {\cal O}_{p_2} {\cal O}_{p_3}  {\cal O}_{p_4}  \rangle$ at genus zero}\label{CONVENTIONS}

%=======================================================
%===========================================================================

A correlator of half-BPS operators is subject to the constraints from the partial non renormalisation theorem \cite{Eden:2000bk}, 
and can be splitted into free and dynamical part. This dynamical part can then be written by factoring out certain $\texttt{kinematics}$. 
The truly interacting part is given by the amplitude $\mathcal{A}$ so defined,
\be
\!\!\!\langle {\cal O}_{p_1}({\tt x}_1) {\cal O}_{p_2}({\tt x}_2)  {\cal O}_{p_3}({\tt x}_3)  {\cal O}_{p_4}({\tt x}_4)  \rangle\Big|_{\texttt{dynamical}} =
\underbrace{  \mathcal{P}[\{ g_{ij} \}]\  \prod_{i,j=1}^2(x_i-y_j) }_{\texttt{kinematics} } \ \mathcal{A}_{}(U,V,\tilde{U},\tilde{V}; \vec{p})
\ee
where the prefactor of superpropagators is \footnote{The 
propagator is defined by $g_{ij}=Y_i\cdot Y_j/X_i\cdot X_j$ where ${\tt x}=(X,Y)$ are embedding coordinates for a point in $AdS_5\times S^5$.
Then, the super cross-ratios are $g_{13}g_{24}/(g_{12}g_{34})=U/\tilde U$ and $g_{14}g_{23}/(g_{13}g_{24})=\tilde V/V$.} 
\begin{align}\label{prefactor_Pgij}
 \mathcal{P}[\{ g_{ij} \}]=g_{12}^{\frac{1}{2}( p_1+p_2-p_3-p_4) } g_{14}^{\frac{1}{2}( p_1+p_4-p_2-p_3) }  g_{24}^{\frac{1}{2}( p_2+p_4-p_1-p_3) } \big( g_{13} g_{24} \big)^{p_3} 
\end{align}
and 
$\mathcal{A}$ is function of the cross ratios $U,V$ and $\tilde U,\tilde V$ \footnote{$U,V$, are the cross ratios on spacetime, i.e.~at the boundary of $AdS_5$, then
$\tilde U, \tilde V$ are those for the internal sphere. Then $U=x_1x_2$, ${\tilde U}=y_1y_2$, and  $V=(1-x_1)(1-x_2)$, ${\tilde V}=(1-y_1)(1-y_2)$.} 
and the external charges.

The Mellin transform of the amplitude, generically denoted by $\mathcal{M}$, is defined by the integral transform
\begin{align}\label{app_amplitude_def}
 \mathcal{A}_{}=\oint\!ds dt \oint\!d\tilde{s} d\tilde{t}\ \ U^s V^t \tilde{U}^{\tilde s} \tilde{V}^{\tilde t}\times \Gamma_{\otimes} \times \mathcal{M}(s,t,\tilde s, \tilde t ; p_1p_2p_3p_4) 
\end{align}
where the kernel $\Gamma_{\otimes}$ is factorised onto $AdS_5\times S^5$, and reads
\be
\begin{array}{c}
\displaystyle
{\Gamma}_{\otimes}= \mathfrak{S} \
\frac{ \Gamma[-s]\Gamma[-t]\Gamma[-u]\Gamma[-s+c_s\,]\Gamma[-t+c_t]\Gamma[-u+c_u]}{ 
\Gamma[1+\tilde s] \Gamma[1+\tilde t]\Gamma[1+\tilde u] \Gamma[1+\tilde s + c_s]\Gamma[1+\tilde t + c_t] \Gamma[1+\tilde u + c_u] } %\\[.7cm]
\end{array} \label{GammaUnbalanced} 
\ee
where 
\ba
s+t+u=-p_3-2\qquad;\qquad \tilde s+\tilde t +\tilde u = p_3-2  \qquad;\qquad \mathfrak{S}=\pi^2 \frac{\ (-)^{\tilde t}(-)^{\tilde u} }{\sin(\pi \tilde t\, )\sin(\pi \tilde u)}\notag\\
\\
c_s=\tfrac{ p_1+p_2-p_3-p_4}{2}\quad;\quad 
c_t=\tfrac{p_1+p_4-p_2-p_3}{2} \quad;\quad 
c_u=\tfrac{ p_2+p_4-p_3-p_1}{2}\quad;\quad  
\Sigma=\tfrac{p_1+p_2+p_3+p_4}{2} \notag
\ea
The sign $\mathfrak{S}$ is conventional in the sense that the amplitude $\mathcal{A}$ is finite in $\tilde U$ and $\tilde V$, 
therefore we need to fix a fundamental domain of integration and flip two $\Gamma$ function from the denominator to the numerator in order to do a contour integration. 
For example, $\textit{T}=\{ \ \tilde s\ge max(0,-c_s)\ ;\  \tilde t,\tilde u\ge 0 \}$. We can also avoid the sign and instead of considering a sphere Mellin integral, simpy sum over $\tilde s,\tilde t$.
The contour integral in $s$ and $t$ instead is a standard 
Mellin Barnes contour, separating left from right poles both in the $s$ and $t$ (complex) planes

It is useful to recall, as in \cite{Aprile:2020luw}, that the Mellin variables might be subject to conventions.
To match conventions with other works, as for example \cite{Penedones:2010ue}, one might need to consider shifts and rescalings. 
%are \emph{not} physical Mandelstam invariants. 
In the flat space limit a la Penedones only rescalings are relevant. The flat space limit for the $AdS_5\times S^5$ amplitude becomes manifest by taking the limit of large $p_i$. 
Since the charges of the single-particle operators are external parameters, they can be used to tune the correlator to 
a saddle point. As shown in  \cite{Aprile:2020luw}, the Mellin amplitude at this saddle point is the flat space 10d amplitude in our bold font Mellin variables.
%and this limit acquires a nice physical interpretation through 
%saddle point analysis \cite{Aprile:2020luw}. 

%%=====================================================================================
%%=====================================================================================

\subsection{Ansatz at all orders: iterative scheme}\label{app_ansatz}

%%=====================================================================================
%%=====================================================================================

As motivated in section \ref{VS_section}, we can stratify the VS amplitude 
and accommodate each stratum in the ansatz
\begin{align}\label{inizio_con}
&
\mathcal{S}_{n,l}=\sum_{  0 \leq d_1+ d_2\leq \ell }^{} \underbrace{ K^{(n)}_{\ell;\, d_1d_2}(\tilde s,\tilde t,p_1p_2p_3p_4) } \, {\bf s}^{d_1}{\bf t}^{d_2} \\[-.2cm]
&
\rule{4cm}{0pt}K^{(n)}_{\ell;\, d_1d_2}=\sum_{ 0 \leq \delta_1+\delta_2\leq (n-\ell) } k^{(n)}_{\ell;\, d_1d_2, \delta_1\delta_2}(p_1p_2p_3p_4)\, \tilde{s}^{\delta_1} \tilde{t}^{\delta_2} \notag
\end{align}
at given order $n$ in the $(\alpha')^{n+3}$ expansion.

There are nonetheless two issues. Firstly \eqref{inizio_con} will contain the new 
stratum $\mathcal{M}_{n,\ell}$ we were looking for, but also pieces of the amplitude at
previous orders $<n$, which we have to  discard by hand. This is inevitable because of 
the inequalities in the sums, which allow to take into account powers of ${\bf u}^{}$ and 
$\tilde u^{}$ correctly,  but introduce much more freedom than the one really contained 
in a stratum.  Secondly, the crossing symmetric version of $\mathcal{S}_{n,\ell}$ 
is still written in the variables ${\bf s}, {\bf t}, \tilde s, \tilde t, p_{i=1,2,3,4}$ and
we want to make crossing symmetry manifest. Therefore, we rewrite 
it in terms of crossing invariant combinations built out of ${\bf s}, {\bf t}, {\bf u}$, $\tilde{\bf s}, \tilde{\bf t}, \tilde{\bf u}$ 
and $c_s, c_t, c_u,\Sigma$, in practise by making a second ansatz and checking that 
we can map free parameters with an invertible matrix.

To fix ideas consider how the case $n=3$ and $\ell=2$. After imposing crossing and 
rewriting the solution in terms of crossing invariant combinations, we find 
\begin{align}\label{example_explanation}
\mathcal{S}_{n=3,2}&=\overbrace{ \rule{0pt}{.4cm} \mathcal{H}_{3,(2,1)}+\mathcal{M}_{n=2,\ell=2}}^{\mathcal{M}_{n=3,2}}\ +\ \mathcal{S}_{n=2,\ell=1}\\[.2cm]
\mathcal{H}_{n=3,(2,1)}&=a_{}^{(1)}\left({\bf s}^2  \tilde{\bf s}+ {\bf t}^2 \tilde{\bf t}
+{\bf u}^2  \tilde{\bf u} \right)+   a_{}^{(2)} ({\bf s}^2 + {\bf t}^2 + {\bf u}^2)\Sigma \notag\\[.2cm]
\mathcal{M}_{n=2,2}&= a_{}^{(3)} ({\bf s}^2 + {\bf t}^2 + {\bf u}^2)\notag
% \\[.2cm]
%\mathcal{S}_{n=2,1}&=
\end{align}
In \eqref{example_explanation}, $\mathcal{M}_{n=3,2}$ is the stratum we are looking for, 
and it comes with two contributions, $\mathcal{M}_{n=2,2}$ and the polynomial 
$\mathcal{H}_{k, (\ell,k-\ell)}$, where
\begin{mdframed}
\begin{center}
$\mathcal{H}_{n, (\ell,n-\ell)}$ is defined to be a crossing symmetric polynomial
in all its variables, of degree $n$, such that only monomials of degree $\ell$ in 
${\bf s}$, ${\bf t}$ and ${\bf u}$ appear.
\end{center}
\end{mdframed}
Notice that $\mathcal{H}_{n, (\ell,n-\ell)}$ is \emph{homogeneous} and that its contribution 
is genuinely the new contribution in $\mathcal{M}_{n,\ell}$. 
In \eqref{example_explanation} in fact, $\mathcal{M}_{2,2}$ is known from $(\alpha')^5$.

Summarising
\begin{center}
\begin{tabular}{cc}
$\mathcal{S}_{n,\ell}$ & polynomial of \emph{max} degree $\ell$ in ${\bf s}$ and \emph{max} degree $n$ in the large $p$ limit\\
$\mathcal{M}_{n,\ell}$ & polynomial of \emph{fixed} degree $\ell$ in ${\bf s}$ and \emph{max} degree $n$ in the large $p$ limit\\
$\mathcal{H}_{n,(\ell,n-\ell)}$ & polynomial of \emph{fixed} degree $\ell$ in ${\bf s}$ and \emph{fixed} degree $n$ in the large $p$ limit 
\end{tabular}
\end{center}

The idea is the following: assume $\mathcal{S}_{n-1,\ell}$ (the crossing symmetric version of it) is 
known for $\ell=0,\ldots n-1$, then
\begin{align}
\mathcal{S}_{n,\ell}= \overbrace{ \rule{0pt}{.4cm} \mathcal{H}_{n,(\ell,n-\ell)}+ \mathcal{M}_{n-1,\ell} }^{\mathcal{M}_{n,\ell}}\ +\ \mathcal{S}_{n-1,\ell-1}
\end{align}
and as we anticipated with the example above, only $\mathcal{H}_{n,(\ell,n-\ell)}$ is new. 
Notice that $\mathcal{S}_{n,0}$ is not contaminated by previous orders, 
and always returns the corresponding stratum. 

The amplitudes $\mathcal{M}_{n,n}$ are known from covariantising the flat VS amplitude 
thus we don't have to construct them.  
The beginning of the recursion is peculiar due to the Mandelstam-type constraints on the Mellin variables,
thus~$\mathcal{H}_{n=1,(1,0)}=0$ and $\mathcal{M}_{1,1}=0$.
Then,
at $(\alpha')^5$ we find
\be
\mathcal{S}_{2,1}=\mathcal{H}_{2,(1,1)}+(\mathcal{M}_{1,1}=0)+\mathcal{S}_{1,0}
\qquad \rightarrow \qquad 
\mathcal{M}_{2,1}=a_{4,1}\left( {\bf s}\tilde{\bf s} + {\bf t}\tilde{\bf t}+ {\bf u}\tilde{\bf u}\right)\notag
\ee
and  $\mathcal{S}_{n=2,0}=\mathcal{M}_{2,0}=\mathcal{H}_{2,(0,2)}+\mathcal{M}_{1,0}$.
For the case of $(\alpha')^6$ all terms contribute in the recursion, %and we find
\begin{align}
\mathcal{S}_{3,2}= \overbrace{ \rule{0pt}{.4cm}  \mathcal{H}_{3,(2,1)}+\mathcal{M}_{2,2}}^{\mathcal{M}_{3,2}}+\mathcal{S}_{2,1}\qquad;\qquad
\mathcal{S}_{3,1}= \overbrace{ \rule{0pt}{.4cm} \mathcal{H}_{3,(1,2)}+\mathcal{M}_{2,1}}^{\mathcal{M}_{3,1}}+\mathcal{S}_{2,0}%\qquad;\qquad
%\mathcal{S}_{3,0}=\mathcal{N}_{3,(0,3)}+\mathcal{M}_{2,0}
\end{align}
and finally $\mathcal{S}_{3,0}=\mathcal{M}_{3,0}=\mathcal{H}_{3,(0,3)}+\mathcal{M}_{2,0}$.

Let us highlight some patterns  which we tested up to $(\alpha')^9$. %i.e.~$n\leq 6$. 
When we construct $\mathcal{H}_{n,(\ell,n-\ell)}$ we begin with 
${\bf s}^{\ell}\times \mathcal{P}(\tilde s, \tilde t, \tilde u, c_s,c_t,c_u,\Sigma)$ crossing symmetrised. 
The overall homogeneous scaling has to be $n$, therefore the polynomial $\mathcal{P}$ can have the structure
\begin{itemize}
\item 
monomials of the form $(\tilde{\bf s}^{d_1} c_s^{d_2}+\texttt{crossing})$ with $d_1+d_2=n-\ell$,
\item 
monomials of the form $ (\tilde{\bf s}^{d_1} c_s^{d_2}+\texttt{crossing})\,\times\, \mathcal{I}_{n-d_1-d_2}(\tilde{\bf s},\tilde{\bf t},\tilde{\bf u}, c_s,c_t, c_u, \Sigma)$ 
\end{itemize}
with $\mathcal{I}_{n-d_1-d_2}$, invariant under crossing.
Then, we can also have a structure like
\begin{itemize} 
\item
products of invariants under crossing of the form $\mathscr{I}_{d}({\bf s},{\bf t},{\bf u},\tilde{\bf s},\tilde{\bf t},\tilde{\bf u}, c_s,c_t,c_u )$.
\end{itemize}
Typically these invariants are found from the amplitudes at previous orders.

For $\mathcal{M}_{n,1}$ we cannot have products of invariants, because
$\mathscr{I}_{\ell=1}({\bf s},{\bf t},{\bf u})={\bf s}+{\bf t}+{\bf u}=-4$.  
This feature of $\mathcal{M}_{n,1}$ offers a starting point for the analysis of the various strata. 
For example, at $n=6$ the crossing invariant ansatz for $\mathcal{H}_{6,(1,5)}$ is the symmetrisation of 
\begin{align}
\begin{array}{cccl} 
{\bf s}^1 \otimes &\{ \tilde{\bf s}^5, \ldots ,  \tilde{\bf s}c_s^4\}  &\otimes & \{{  1} \}\\[.2cm]
{\bf s}^1 \otimes& \{ \tilde{\bf s}^4, \ldots,  c_s^4\}  &\otimes& \{ {\color{black} \Sigma}\}  \\ [.2cm]
{\bf s}^1 \otimes&\{ \tilde{\bf s}^3 , \ldots , \tilde{\bf s}c_s^2\} &\otimes& \{ {\color{black} \Sigma^2, (c_s^2 +c_t^2+c_u^2)}, (\tilde{\bf s}^2+\tilde{\bf t}^2+\tilde{\bf u}^2)\}  \\[.4cm]
{\bf s}^1 \otimes&\{ \tilde{\bf s}^2 , c_s^2\} &\otimes& \left\{ {\color{black} \Sigma^3, \Sigma (c_s^2 +c_t^2+c_u^2),c_sc_tc_u}, \,
(\tilde{\bf s}^3+\tilde{\bf t}^3+\tilde{\bf u}^3), (\tilde{\bf s} c_s^2+\tilde{\bf t} c_t^2+\tilde{\bf u} c_u^2)\right\}\\[.4cm]
{\bf s}^1 \otimes&\{ \tilde s, c_s\} &\otimes& \{  {\color{black} \Sigma^4 , \Sigma^2 (c_s^2+c_t^2+c_u^2),  (c_s^2+c_t^2+c_u^2)^2,(c_s^4+c_t^4+c_u^4), \Sigma\, c_s c_t c_u }\}
\end{array}\notag
\end{align}

The case $\mathcal{M}_{n,2}$ is the first case in which we can have an invariant in the 
boldfont variables, i.e. ${\bf s}^2+{\bf t}^2+{\bf u}^2$. For $\mathcal{M}_{n,\ell\ge 3}$ there is a similar story. 
Novelties in general come from the possibility of adding products of invariants. The basis up to $(\alpha')^9$ is
given in the ancillary file {\tt calHbasis}.

%%=====================================================================================
%%=====================================================================================

\subsection{OPE equations}\label{more_OPE}

%%=====================================================================================
%%=====================================================================================

Let us recall that the OPE at genus zero gives the following $\alpha'$-dependent constraints
\begin{align}
\label{again_OPE}
\mathbf{C}_{\vec{\tau}}(\alpha')  \mathbf{C}_{\vec{\tau}} ^T(\alpha') = \mathbf{L}_{\vec{\tau}}\qquad;\qquad 
\mathbf{C}_{\vec{\tau}}(\alpha')  \pmb{\eta}_{\vec{\tau}}(\alpha')  \mathbf{C}_{\vec{\tau}}^T(\alpha') = \mathbf{M}_{\vec{\tau}}(\alpha') \,,
\end{align}
where $\mathbf{M}_{\vec{\tau}}(\alpha')$ is the CPW of the $\log u$ discontinuity of the VS amplitude, while 
$\mathbf{L}_{\vec{\tau}}$ is the CPW from disconnected free theory, in the long sector. The $\alpha'$ 
expansion reads
\begin{align}
\pmb{\eta}_{} &= \ \pmb{\eta}^{(0)}_{} + \alpha'^{3}\pmb{ \eta}^{(3)}_{} \ + \alpha'^{5} \pmb{\eta}^{(5)}_{} + \ldots\, ,\notag  \\[.2cm]
\mathbf{C} &= \mathbf{C}^{(0)} + \alpha'^{3} \mathbf{C}^{(3)} + \alpha'^{5} \mathbf{C}^{(5)} + \ldots\,.
\end{align}
Inserting this in the OPE  we will find a tower of relations, of which the first one 
obviously coincides with the supergravity eigenvalue problem.
At order $(\alpha')^{n+3}$ we find 
\begin{align}\label{constr_3pt}
\Big( \mathbf{C}^{(n+3)}\mathbf{C}^{(0)T}\ +\ \mathbf{C}^{(0)}\mathbf{C}^{(n+3)T}\Big)\ + \sum_{\substack{ k_1+k_2=n+3 \\k_1 \neq n+3}}\ \mathbf{C}^{(k_1)}\mathbf{C}^{(k_2)T} =0
\end{align}
\be\label{constr_3pt_poi}
\!\!\!\!
\Big(\mathbf{C}^{(0)}{\pmb \eta}^{(n+3)}\mathbf{C}^{(0)T}+ \mathbf{C}^{(n+3)}{\pmb \eta}^{(0)}\mathbf{C}^{(0)T}\ +\ \mathbf{C}^{(0)}{\pmb \eta}^{(0)}\mathbf{C}^{(n+3)T}\Big)\ 
+ \sum_{\substack{ k_1+k_2+k_3=n+3 \\k_2 \neq n+3}}\ \mathbf{C}^{(k_1)}{\pmb \eta}^{(k_2)}\mathbf{C}^{(k_3)T} ={\bf M}^{(n+3)}
\ee
where we isolated the first term to emphasize 
that  $\mathbf{C}^{(n+3)}$ is new at this order, while the other matrices in the sum 
already featured at previous orders (when existing). Actually the sum is over distinct permutations.

We will now rewrite the two equations in \eqref{constr_3pt}-\eqref{constr_3pt_poi} by going to the eigenvector basis $\mathbf{c}_{\vec{\tau}}^{(0)} =\mathbf{L}^{-\frac{1}{2}}_{\vec{\tau}} \mathbf{C}^{(0)}_{\vec{\tau}}$, and using
the resolution of the identity
\be
\mathbf{C}^{(0)T}\, \mathbf{L}^{-1} \, \mathbf{C}^{(0)}= \mathbf{1}
\ee 
to split matrix products of three point functions and anomalous dimensions corresponding to different orders. 
To do so 
%
%where we have defined
%\begin{align}
%\mathbf{c}_{\vec{\tau}}^{(0)} =\mathbf{L}^{-\frac{1}{2}}_{\vec{\tau}} \mathbf{C}^{(0)}_{\vec{\tau}} \qquad;\qquad
% \mathbf{N}_{\vec{\tau}}^{(0)}= \mathbf{L}^{-\frac{1}{2}}_{\vec{\tau}}\mathbf{M}_{\vec{\tau}}^{(0)}\mathbf{L}^{-\frac{1}{2}}_{\vec{\tau}}
%\end{align}
%
it is convenient to introduce the matrix 
\be
\mathbf{D}^{(k)}_{}= \mathbf{L}^{-\frac{1}{2}}\left(  \mathbf{C}^{(k)}\mathbf{C}^{(0)T}\right) \mathbf{L}^{-\frac{1}{2}}=
\mathbf{L}^{-\frac{1}{2}}\, \mathbf{C}^{(k)}\, \mathbf{c}^{(0)T}
\ee
and rewrite both as 
\be
\Big( \mathbf{D}^{(n+3)}_{}+ \mathbf{D}^{(n+3)T} \Big)+ \sum_{\substack{ k_1+k_2=n+3 \\ k_1 \neq n+3}} \mathbf{D}^{(k_1)}_{-}\mathbf{D}^{(k_2)T}_{-}  =0
\ee
\be
\!\!\!\!\Big(\mathbf{c}^{(0)}{\pmb \eta}^{(n+3)}\mathbf{c}^{(0)T}+ \mathbf{D}^{(n+3)}\mathbf{N}^{(0)}\ +\ \mathbf{N}^{(0)} \mathbf{D}^{(n+3)T}\Big)\  
+ \sum_{\substack{ k_1+k_2+k_3=n+3 \\k_2 \neq n+3}} \mathbf{D}^{(k_1)} \Big[ \mathbf{c}^{(0)}{\pmb \eta}^{(k_2)}\mathbf{c}^{(0)T}\Big] \mathbf{D}^{(k_3)T}   = {\bf N}^{(n+3)}\notag
\ee
where $\mathbf{N}^{(0)}$ is by construction symmetric.

The matrix ${\bf D}^{(n+3)}$ has a block structure depicted below,
\be
\begin{tikzpicture}[scale=.9]
\def\step{.9};

%\filldraw[red!20] (0,-1*\step) -- (0,-3*\step) -- (2*\step,-3*\step) -- (2*\step,-2*\step) --(1*\step,-2*\step) -- (1*\step,-1*\step)  --  cycle;
%\filldraw[red!20] (1*\step,0) -- (3*\step,0) -- (3*\step,-2*\step)   -- (2*\step,-2*\step)  -- (2*\step,-1*\step) -- (1*\step,-1*\step) -- cycle;

\filldraw[red!20] (0,0) rectangle (2*\step,-2*\step);
\filldraw[green!20] (2*\step,0) rectangle (3*\step,-2*\step);
\filldraw[green!20] (0,-2*\step) rectangle (2*\step,-3*\step);

\draw[step=\step,gray] (0,0) grid (3,-3);
\draw[gray] (0,-3*\step) -- (0,-6*\step);
\draw[gray] (3*\step,0) -- (6*\step,0);
\draw[gray] (3*\step,-3*\step) -- (6*\step,-3*\step);
\draw[gray] (3*\step,-3*\step) -- (3*\step,-6*\step);

%\draw (.5*\step,-.5*\step) node[scale=1.5] {$0$};
\draw (1.5*\step,-1.5*\step) node[rotate=-45,scale=1.8] {$\ldots$};
\draw (2.5*\step,-2.5*\step) node[scale=1.5] {$0$};

\draw (1.5*\step,-4.5*\step) node[scale=4] {$0$};
\draw (4.5*\step,-4.5*\step) node[scale=4] {$0$};
\draw (4.5*\step,-1.5*\step) node[scale=4] {$0$};

\draw  (.5*\step,+.5*\step) node[scale=1] {$\mathbb{V}_{\vec{\tau},1}$};
\draw  (1.5*\step,+.5*\step) node[scale=1] {$\ldots$};
\draw  (2.5*\step,+.5*\step) node[scale=1] {$\mathbb{V}_{\vec{\tau},m^*}$};
\draw  (4.3*\step,+.5*\step) node[scale=1] {$\mathbb{V}_{\vec{\tau},m^*+1}\ \ldots$};
%\draw  (4.5*\step,+.5*\step) node[scale=1] {$\ldots$};

\draw  (+6.*\step,-.5*\step) node[scale=1] {$\mathbb{V}_{\vec{\tau},1}$};
\draw  (+6.*\step,-1.5*\step) node[scale=1] {$\vdots$};
\draw  (+6.2*\step,-2.5*\step) node[scale=1] {$\mathbb{V}_{\vec{\tau},m^*}$};
\draw  (+6.4*\step,-3.5*\step) node[scale=1] {$\mathbb{V}_{\vec{\tau},m^*+1}$};
\draw  (+6.*\step,-4.5*\step) node[scale=1] {$\vdots$};

\draw  (-2*\step,-3*\step) node[scale=1] {$\left(\mathbf{D}^{(n+3)}_{\vec{\tau}}\right)_{mm'}=$};

%space
\draw  (+8.2*\step,-3.5*\step) node[scale=1] {$\rule{.8cm}{0pt}$};

\end{tikzpicture}
\ee

The symmetric part of ${\bf D}$, contained in the red block, is fully determined by previous orders,
\be
\Big( \mathbf{D}^{(n+3)}_{}+ \mathbf{D}^{(n+3)T} \Big)=-\sum_{\substack{ k_1+k_2=n+3 \\ k_1 \neq n+3}} \mathbf{D}^{(k_1)}_{-}\mathbf{D}^{(k_2)T}_{-} 
 \ee 
 
The anomalous dimensions ${\pmb \eta}^{(n+3)}$ and the anti-symmetric part of ${\bf D}^{(n+3)}$ 
are determined by the other equation, therefore by ${\bf N}^{(n+3)}$ on the r.h.s. and $\sum_{} \mathbf{D}^{(k_1)} 
\Big[ \mathbf{c}^{(0)}{\pmb \eta}^{(k_2)}\mathbf{c}^{(0)T}\Big] \mathbf{D}^{(k_3)T}$.

%We can be more specific.
%
%{\color{red} TO BE CONTINUED}

%=====================================================================================
\subsection{Superblock decomposition}\label{superb_deco_app}
%=====================================================================================

To understand the spectral properties of the Virasoro Shapiro amplitude in $AdS_5\times S^5$, 
we have to compute its superblocks decomposition.
Considering the $\Gamma$ functions that define the correlator in position space, 
\be
\begin{array}{c}
\displaystyle
{\Gamma}_{\otimes}= \mathfrak{S} \
\frac{ \Gamma[-s]\Gamma[-t]\Gamma[-u]\Gamma[-s+c_s\,]\Gamma[-t+c_t]\Gamma[-u+c_u]}{ 
\Gamma[1+\tilde s] \Gamma[1+\tilde t]\Gamma[1+\tilde u] \Gamma[1+\tilde s + c_s]\Gamma[1+\tilde t + c_t] \Gamma[1+\tilde u + c_u] } %\\[.7cm]
\end{array} 
\ee
we will first integrate the Mellin amplitude in $\tilde s$ and $\tilde t$, thus obtaining an explicit polynomial 
in the sphere cross ratios ${\tilde U},{\tilde V}$, and then we will rewrite the $s$ and $t$ integral as sum over 
$\overline{D}$ functions. In this form we series expand and match the expansion 
with well understood formulae for the superblocks  \cite{Doobary:2015gia}. 

We will be interested in the superblock decomposition of the $\log(u)$ discontinuity of the amplitude.
Only long superblocks contribute. These are the simplest and have a factorised form 
into conformal and internal blocks. Below we describe them in details, for convenience of the 
reader.\footnote{There is a small difference compared with the long block $\mathbb{L}$ written as \cite{Aprile:2017xsp}, 
and it has to do with the prefactor of superpropagators that we used to define $\mathcal{A}$ in this paper. }
%In our conventions $p_{43}\ge p_{21}$. 

%=====================================================================================
\paragraph{Long (2,2) superblocks.}~\\[-.2cm]
%=====================================================================================

The main formulae are given in \cite{Doobary:2015gia}.  We first introduce the bosonic block
\be
%g_{\kappa}(z;\gamma,\alpha,\beta)=z^{\kappa} ~_2F_1\big[\,^{1+\kappa+\alpha ,1+\kappa+\beta}_{\ \ \ 2+2\kappa+\gamma}\big]\qquad;\qquad
B_{\underline{\lambda}}(\underline{z};\gamma,p_{12},p_{43})=\frac{1}{U^{1+\frac{\gamma}{2}-\ell} } 
\frac{\det\Big[ z_i^{1+\frac{\gamma}{2}+\lambda_j-j}
~_2F_1\big[\,^{1+\frac{\gamma}{2}+\lambda_j-j-\frac{p_{12}}{2},\,1+\frac{\gamma}{2}+\lambda_j-j-\frac{p_{43}}{2}  }_{\rule{1.5cm}{0pt}2+\gamma+2\lambda_j-2}\big]\,\Big] }{ \det\Big[ z_i^{\ell-j} \Big] }
\ee
with $\ell$ being equal to the number of variables.  
Then the long superconformal block is
\be\label{Long_notation}
\mathbb{L}_{\underline{\lambda},\gamma}={\tt prefa}_{\gamma,\vec{p}}\ \prod_{i,j} \frac{(x_i-y_j)}{x_i y_j} \ 
B_{[\lambda_1,\lambda_2]}(x_1,x_2;\gamma,p_{12},p_{43}) B_{[\lambda'_1,\lambda'_2]}(y_1,y_2;-\gamma,-p_{12},-p_{43})
\ee
where
\be
{\tt prefa}_{\gamma,\vec{p}}=
 g_{12}^{\frac{p_1+p_2}{2} } \!g_{34}^{\frac{p_3+p_4}{2} } \left[ \frac{ g_{14} }{g_{24} }\right]^{\frac{p_1-p_2}{2}}\!\left[ \frac{ g_{14} }{g_{13} }\right]^{\frac{p_4-p_3}{2}}
\!\!\!\bigg( \frac{U}{\tilde U}\bigg)^{\!\!\frac{\gamma}{2}}\,.
\ee
The Young diagrams are given in terms of quantum numbers as follows,
\be
1+\tfrac{\gamma}{2}+\lambda_i-i=\left\{\begin{array}{cc} 2+\frac{\tau}{2}+l  & ;\ i=1 \\[.2cm] 1+\frac{\tau}{2} &;\  i=2 \end{array}\right.\quad;\quad
1-\tfrac{\gamma}{2}+\lambda'_j-j=\left\{\begin{array}{cc} -\frac{b}{2}  & ;\ j=1 \\[.2cm] -1-\frac{b}{2}-a &;\  j=2 \end{array}\right.
\ee
These values fix what goes into the entries of the$~_2F_1$ functions.  Note the symmetries (up to a sign), 
\be
\begin{array}{clll}
(I)\qquad& -\frac{b}{2}\leftrightarrow  2+\frac{\tau}{2}+l  &\quad;\quad a\leftrightarrow l &\quad;\quad y_i\leftrightarrow x_i \\[.2cm]
(II)\qquad&-\frac{b}{2}\leftrightarrow 1+\frac{\tau}{2}      &\quad;\quad a\leftrightarrow -l-2    &\quad;\quad y_1 \leftrightarrow x_2  \quad  y_2 \leftrightarrow x_1
\end{array}
\ee

In free theory $\gamma$ counts the propagator lines in a diagram going from $(p_1p_2)$ to $(p_3p_4)$, and the superconformal block can be performed diagram by diagram. 
However, a long superblock does not actually depend on $\gamma$: 
The explicit $\gamma$ factor in ${\tt prefa}_{\gamma,\vec{p}}$ cancels with corresponding prefactors in the product $B\times B$ in \eqref{Long_notation}.
Note at this point that the arguments of the$~_2F_1$ depend on $\tau-p_{12}$ and $\tau-p_{43}$, therefore $\tau\ge max(p_{43},p_{12})$. On the other hand, we can always
arrange the charges of our correlators $\langle p_1p_2p_3p_4\rangle$, without loss of generality, such that $p_{43}\ge p_{21}\ge 0$, 
and therefore fix conventions in such a way that $\tau\ge p_{43}$. Diagrammatically, 
we are then comparing any diagram to the one in which $\gamma=\frac{p_{43}}{2}$ lines go from $(p_1p_2)$ to $(p_3p_4)$.
We understand in this way that a  convenient way to arrange the long superblock is
\be\label{final_long}
\mathbb{L}_{p_{12},p_{43}}={\tt pref}_{\gamma=\frac{p_{43}}{2},\,\vec{p}} \times    {\tt Blc}_{\tau,l}(x_1,x_2)\,  {\tt Hrm}_{[aba]} (y_1,y_2)
\ee
with the functions   ${\tt Blc}_{\tau,l}(x_1,x_2)$ and  ${\tt Hrm}_{[aba]} (y_1,y_2)$ arranged from $B\times B$ in \eqref{Long_notation}. These are explicitly written in the next sections. 
For the superconformal block decomposition in the long sector we only need to remember that 
\be
 \mathcal{P}[\{ g_{ij} \}]{\rm\ in\ \eqref{prefactor_Pgij} }\ =\  {\tt pref}_{\gamma=\frac{p_{43}}{2},\,\vec{p}} \left( \frac{U}{\tilde U}\right)^{\!p_3}
\ee
in order to match \eqref{final_long}.

%=====================================================================================
\paragraph{Decomposition on $S^5$.}~\\[-.2cm]
%=====================================================================================

The integration over $\tilde s$ and $\tilde t$ will admit the following decomposition in $su(4)$ harmonics,
\be
\frac{1}{ \tilde{U}^{p_3} } \oint d\tilde{s} d\tilde{t}\ \tilde{U}^s \tilde{V}^t  \Big[ f(s,t,\tilde s,\tilde t) \Big]  = 
\sum_{[aba]} \ {\tt h}_{[aba]}(s,t)\  {\tt Hrm}_{[aba]}  
\ee
where the precise $f(s,t,\tilde s,\tilde t)$ comes from ${\Gamma}_{\otimes}\times \mathcal{V}_n(s,t,\tilde s,\tilde t)$, and 
\begin{gather}\label{hrm_decomp}
{\tt Hrm}_{[aba]}= (-)^{1+a}\frac{ \tilde{U}^{\frac{-2-b+p_{43} }{2}}}{y_1-y_2} \times \Bigg[ y_1^{-a-1} \mathscr{F}^+_{-1-\frac{b}{2}-a }(y_1) \mathscr{F}^+_{-\frac{b}{2}}(y_2) - (1\leftrightarrow 2)\Bigg]\\[.2cm]
 \mathscr{F}^{+}_{s}(x)=\!\!~_2F_1[s+\tfrac{p_{12}}{2},s+\tfrac{p_{43}}{2},2s](x)\qquad;\qquad \tilde{U}=y_1y_2\qquad;\qquad \tilde{V}=(1-y_1)(1-y_2)\notag
\end{gather}
For given charges $p_1p_2p_3p_4$, the value of $[aba]$ are organised as, 
\be\label{set_aba_p1p2p2p3p4}
\begin{array}{ccl}
a=0 					& \qquad  &\ \ \quad b_{min}\leq b \leq b_{max}  \\[.2cm]
a=1 					& \qquad &\quad b_{min}\leq b \leq b_{max}-2  \\[.2cm]
%a=2 					& \qquad &\quad b_{min}\leq b \leq b_{max}-4  \\[.2cm]
\vdots				& \qquad &\quad\ldots \\[.2cm]
a= \kappa_{\vec{p}}		 & \qquad &b_{min}\leq b \leq b_{min}
\end{array}
\ee
where $b_{min}= p_{43}$ and $b_{max}=min(p_1+p_2,p_3+p_4)-4$.

Using the notation ${\tt Hrm}_{[aba]}[ f(s,t,\tilde s,\tilde t)]$ to refer to the coefficient ${\tt h}_{[aba]}$ of $f$ in \eqref{hrm_decomp},
a useful formula to know is 
\be
{\tt Hrm}_{[aba]}\left[ \frac{  \mathfrak{S} \times (\Sigma-1-a)_{a} \times (-\tilde{u})^{d_1} \tilde{t}^{d_2}  }{ 
\Gamma[1+\tilde s] \Gamma[1+\tilde t]\Gamma[1+\tilde u] \Gamma[1+\tilde s + c_s]\Gamma[1+\tilde t + c_t] \Gamma[1+\tilde u + c_u] }\right]_{d_1+d_2=a}
= 
 {\tt Y}_{[aba],\vec{p}} \notag 
\ee
\be
\label{formulaBMichele}
\ee
\be
\!\!\!\!\!\!\!\!\!\!\!\!\!\!\!\!\!{\tt Y}_{[aba],\vec{p}}=
\frac{(\Sigma-2)! b!(b+1)!(2+a+b)  }{  \Gamma[ \pm \frac{p_1-p_2}{2} + \frac{b+2}{2}]  
\Gamma[\frac{p_1+p_2}{2} + \frac{b+2}{2}]   \Gamma[\frac{p_1+p_2}{2} - \frac{b+2a+2}{2}]  
\Gamma[ \pm \frac{p_3-p_4}{2} + \frac{b+2}{2}] \Gamma[\frac{p_3+p_4}{2} + \frac{b+2}{2} ] \Gamma[\frac{p_3+p_4}{2} - \frac{b+2a+2}{2} ] }\notag
\ee
which was spotted in \cite{Drummond:2020dwr}. Notice that ${\tt Y}$ vanishes for 
$a$ and $b$ not in the set \eqref{set_aba_p1p2p2p3p4}.

By construction, the coefficients ${\tt h}_{[aba]}(s,t)$, at given order $(\alpha')^{n+3}$, 
are at most polynomials in $s$ and $t$ of degree $n$. Upon measuring the max spin per $su(4)$ channel 
(measured by the max power of $t$, with the constraint on $u$ implemented), we find the pattern
\begin{align}
\begin{array}{c|cccccc}
	      & [0b0] & [1b1] & [2b2]  & [3b3] & [4b4] & \ldots \\
	      \hline\\
(\alpha')^3 &  l \leq 0 	& - 	& - 	& -  	& -  & \ldots \\[.2cm] \\
(\alpha')^5 & l  \leq 2	&  l \leq 1 	&  l \leq 0 	& - 	& -  & \\[.2cm] 
(\alpha')^6 & l  \leq 2	&  l \leq 1 	&  l \leq 0 	& - 	& -  & \\[.2cm] \\
%
%{\color{red} \alpha^6 }&  3 	&  2 	&  1 	&  0 	& -& \ldots  \\
(\alpha')^7 &  l\leq 4 	&  l\leq 3 	&  l\leq 2 	&  l\leq 1 	&  l\leq 0 \\
\vdots
\end{array}
\end{align}
When the degree would go negative we find a truncation.

%=====================================================================================
\paragraph{Decomposition on $AdS^5$}~\\[-.2cm]
%=====================================================================================

We then integrate $ \Gamma_{AdS_5}\times {\tt h}_{[aba]}(s,t)$, in $s$ and $t$, 
and perform the conformal block decomposition according to the formula, 
\be\label{block_deco_app_g}
 {U}^{p_3}  \oint d{s} d{t}\ {U}^s {V}^t\  \Big[ f(s,t) \Big]  = 
\sum_{\tau,l} \ {\tt b}_{\tau,l}\  {\tt Blc}_{\tau,l}   
\ee
where
\begin{gather}\label{Blc_def}
{\tt Blc}_{\tau,l}= 
(-)^l  \frac{ U^{\frac{\tau-p_{43}}{2} } }{x_1-x_2}\Bigg[ x_1^{l+1} \mathscr{F}^-_{2+\frac{\tau}{2}+l}(x_1) \mathscr{F}^-_{1+\frac{\tau}{2}}(x_2) - (1\leftrightarrow 2)\Bigg] \\
 \mathscr{F}^{-}_{s}(x)=\!\!~_2F_1[s-\tfrac{p_{12}}{2},s-\tfrac{p_{43}}{2},2s](x)\qquad;\qquad U=x_1x_2\qquad;\qquad V=(1-x_1)(1-x_2)\notag
\end{gather}
Together ${\tt Blc}$ and ${\tt Hrm}$ are the building blocks of the long superblock.%\footnote{Notice the symmetry 
%$b\leftrightarrow -\tau-2$ and $a\leftrightarrow -l-2$, with $p_i\leftrightarrow -p_i$ and $y_i\leftrightarrow x_i$}

Notice at this point that upon integrating $\Gamma_{AdS_5}\times 1$ we obtain $\overline{D}_{p_4+2,p_3+2,p_2+2,p_1+2}$.
Therefore, when we integrate the VS amplitude, what we can do is to rearrange   $\Gamma_{AdS_5}\times {\tt h}_{[aba]}(s,t)$
 as a polynomial in $U\partial_U$ and $V\partial_V$ acting on $\overline{D}_{p_4+2,p_3+2,p_2+2,p_1+2}$. 
In sum, at any given order $(\alpha')^{n+3}$, we understand what is the space of functions we will work with. 
For example, at order $(\alpha')^5$ we would find at most a second order polynomial in $s$ and $t$, therefore
\be
\texttt{Span}\left[  
\begin{array}{lll}
					  \overline{D}_{p_4+2,p_3+2,p_2+2,p_1+2},  \\
					  \overline{D}_{p_4+2,p_3+2,p_2+3,p_1+3},    & \overline{D}_{p_4+3,p_3+2,p_2+2,p_1+3} \\
					  \overline{D}_{p_4+2,p_3+2,p_2+4,p_1+4},    &\overline{D}_{p_4+3,p_3+2,p_2+3,p_1+4},  & \overline{D}_{p_4+4,p_3+2,p_2+2,p_1+4}  
\end{array}\right]									
\ee

A useful intermediate step for the next section is to consider the decomposition in conformal blocks of various
$\overline{D}$ functions, for example those contributing to maximal spin. We will use the notation ${\tt Blc}_{\tau,l}[f]$ to 
mean the coefficients ${\tt b}_{\tau,l}$ in the block decomposition of $f$ in \eqref{block_deco_app_g}.

Define for later convenience the following factorials 
\begin{align}
&
\!\!\!\!\!
{\tt F}_{\vec{p}}(\tau,l) = (-)^{\frac{p_1+p_2-p_3-p_4}{2}} \frac{ \left(\frac{ 2+2l+\tau+p_4-p_3}{2}\right)!\left(\frac{ 2+2l+\tau+p_1-p_2}{2}\right)! }{(3+2l+\tau)! } 
\frac{ \left(\frac{\tau+p_1+p_2+2}{2}\right)! \left(\frac{\tau+p_3+p_4+2}{2}\right)! }{ (\tau+1)!} 
\end{align}

%%%%%%%%%%%%%%%%%%%%%%%%%%%%%%%%%%%%%%%%%%%%%%%%%%%%%%%%%%%%%%
\paragraph{Decomposition of $\,\overline{D}_{p_4+2,p_3+2,p_2+2+k,p_1+2+k}$}~\\[-.2cm]
%%%%%%%%%%%%%%%%%%%%%%%%%%%%%%%%%%%%%%%%%%%%%%%%%%%%%%%%%%%%%%

This class of $\overline{D}$ functions only contributes at $l=0$, since it only comes from $U\partial_U$ derivatives on $\overline{D}_{p_4+2,p_3+2,p_2+2,p_1+2}$. We find
\begin{align}\label{cpw_dbaralpha3}
&
\texttt{Blc}_{\tau,l=0}\Big[ \overline{D}_{p_4+2,p_3+2,p_2+2+k,p_1+2+k} \Big]=  
\\[.2cm]
&
{\tt F}_{\vec{p}}(\tau,0) \times \frac{ (-)^{1+k}}{(\Sigma+1+k)!} \ 
 \prod_{t=+\frac{1}{2}p_{43}-1}^{\frac{1}{2}(p_3+p_4)-1} \!\! \left(\tfrac{\tau}{2}-t\right)
 \prod_{t=+\frac{1}{2}p_{12}-1}^{\frac{1}{2}(p_1+p_2)-1+k} \!\! \left(\tfrac{\tau}{2}-t\right) 
 \ \prod_{t=1}^{k} \left( \tfrac{\tau}{2}+\tfrac{p_1+p_2}{2} +1+t\right) \notag
\end{align}

%%%%%%%%%%%%%%%%%%%%%%%%%%%%%%%%%%%%%%%%%%%%%%%%%%%%%%%%%%%%%%
\paragraph{Decomposition of $\,\overline{D}_{p_4+2+k,p_3+2,p_2+2,p_1+2+k}$}~\\[-.2cm]
%%%%%%%%%%%%%%%%%%%%%%%%%%%%%%%%%%%%%%%%%%%%%%%%%%%%%%%%%%%%%%

This class of $\overline{D}$ functions has a more complicated CPW expansion, whose spin support depends on $k$. 
The case $k=0$ provides the starting point, and it is included above. We have found

The series $l=k$
\begin{align}
\label{Dbar_deco1}
&
\texttt{Blc}_{\tau,l=k}\Big[\overline{D}_{p_4+2+k,p_3+2,p_2+2,p_1+2+k} \Big]=  
\\[.2cm]
&
 {\tt F}_{\vec{p}}(\tau,k)  \ \tfrac{ (\frac{\tau+p_1+p_2}{2}+2)_{k}(\frac{\tau+p_3+p_4}{2}+2)_{k} }{(\tau+2)_k (\tau+k+3)_k }\times
\frac{ (-)^{1+k}}{(\Sigma+1+k)!} \ 
 \prod_{t=+\frac{1}{2}p_{43}-1-k}^{\frac{1}{2}(p_3+p_4)-1} \!\! \left(\tfrac{\tau}{2}-t\right)
 \prod_{t=+\frac{1}{2}p_{12}-1-k}^{\frac{1}{2}(p_1+p_2)-1} \!\! \left(\tfrac{\tau}{2}-t\right) 
\notag
\end{align}
Notice that $(\tau+2)_k (\tau+k+3)_k$ concatenates and becomes $=(\tau+2)_{2k+1}/(\tau+k+2)$.
 
The series $l=k-1$
\begin{align}
\label{Dbar_deco2}
&
\texttt{Blc}_{\tau,l=k-1}\Big[ \overline{D}_{p_4+2+k,p_3+2,p_2+2,p_1+2+k} \Big]=   \\
&
{\tt F}_{\vec{p}}(\tau,k-1)\tfrac{ (\frac{\tau+p_1+p_2}{2}+2)_{k-1}(\frac{\tau+p_3+p_4}{2}+2)_{k-1} }{(\tau+2)_{k-1} (\tau+k+2)_{k-1} }\times
 \frac{ (-)^{1+k}}{(\Sigma+1+k)!} 
 \prod_{t=+\frac{1}{2}p_{43}-k}^{\frac{1}{2}(p_3+p_4)-1} \!\! \left(\tfrac{\tau}{2}-t\right)
 \prod_{t=-\frac{1}{2}p_{21}-k}^{\frac{1}{2}(p_1+p_2)-1} \!\! \left(\tfrac{\tau}{2}-t\right)   \notag\\
&\rule{1.8cm}{0pt}
\times\tfrac{k}{8}
\tfrac{ (p_3^2-p_4^2-(2k+2) \tau-\tau^2)(p_1^2-p_2^2+(2k+2) \tau+\tau^2)- 4 (p_1+k+1)(p_4+k+1)\tau(\tau+(2k+2))}{ \tau(\tau+2k+2)}
 \notag
\end{align}
~\\

The series $l\leq k-2$ is already complicated. The case $l=0$ of $\overline{D}_{p_4+4,p_3+2,p_2+2,p_1+4}(U,V)$ has the degree eight polynomial, which is quite cumbersome.

%%%%%%%%%%%%%%%%%%%%%%%%%%%%%%%%%%%%%%%%%%%%%%%%%%%%%%%%%%%%%%
\paragraph{Decomposition of $\,\overline{D}_{p_4+2+k,p_3+2,p_2+3,p_1+3+k}$}~\\[-.2cm]
%%%%%%%%%%%%%%%%%%%%%%%%%%%%%%%%%%%%%%%%%%%%%%%%%%%%%%%%%%%%%%

This case corresponds to a $U\partial_U$ derivative 
on the previous case $\overline{D}_{p_4+2+k,p_3+2,p_2+2,p_1+2+k}$.

The series $l=k$
\begin{align}
\label{Dbar_deco3}
&
\texttt{Blc}_{\tau,l=k}\Big[ \overline{D}_{p_4+2+k,p_3+2,p_2+3,p_1+3+k} \Big]=  
\\[.2cm]
&
{\tt F}_{\vec{p}}(\tau,k)  \ \tfrac{ (\frac{\tau+p_1+p_2}{2}+2)_{k+1}(\frac{\tau+p_3+p_4}{2}+2)_{k} }{(\tau+2)_k (\tau+k+3)_k } \times
\frac{ (-)^{k}}{(\Sigma+2+k)!} \ 
 \prod_{t=+\frac{1}{2}p_{43}-1-k}^{\frac{1}{2}(p_3+p_4)-1} \!\! \left(\tfrac{\tau}{2}-t\right)
 \prod_{t=+\frac{1}{2}p_{12}-1-k}^{\frac{1}{2}(p_1+p_2)} \!\! \left(\tfrac{\tau}{2}-t\right) 
\notag
\end{align}

%%==============================================================================
%%==============================================================================

\subsection{Details on the $m^*=1$ anomalous dimensions}\label{rank1_app}

%%==============================================================================
%%==============================================================================

The study of STRINGY corrections to anomalous dimension at $m^*=1$ for $a+l=n,n-1$ 
can be done independently of the bootstrap program 
for the VS amplitude, as we anticipated in section \ref{sec_rank_equal1_anadim}.  
The $m^*=1$ operators are those labelled by the left most corner in the rectangle $R_{\vec{\tau}}$, 
\be
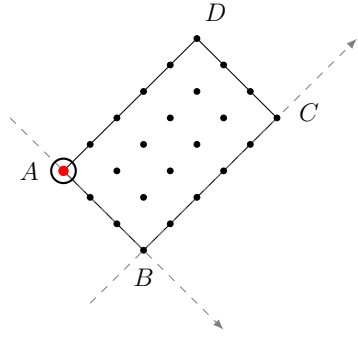
\begin{tikzpicture}[scale=.54]
%
%\draw[step=2cm,gray,very thin] (-4,2) grid (8,8);
%
\def\prop{.65}
\def\shifthor{\prop*2}
\def\ptuno{(\prop*2-\shifthor,\prop*8)}
\def\ptdue{(\prop*5-\shifthor,\prop*5)}
\def\pttree{(\prop*7-\shifthor,\prop*13)}
\def\ptquattro{(\prop*10-\shifthor,\prop*10)}
%
%axis horizontal
\draw[-latex, line width=.6pt]		(\prop*1   -\shifthor-4,         \prop*12          -0.5*\shifthor)    --  (\prop*1  -\shifthor-2.5  ,   \prop*12-      0.5*\shifthor) ;
\node[scale=.8] (oxxy) at 			(\prop*1   -\shifthor-2.5,  \prop*13.7     -0.5*\shifthor)  {};
\node[scale=.9] [below of=oxxy] {$p$};
%
%axis vertical
\draw[-latex, line width=.6pt] 		(\prop*1   -\shifthor-4,     \prop*12       -0.5*\shifthor)     --  (\prop*1   -\shifthor-4,        \prop*15-      0.5*\shifthor);
\node[scale=.8] (oxyy) at 			(\prop*1   -\shifthor-2,   \prop*15.2   -0.5*\shifthor) {};
\node[scale=.9] [left of= oxyy] {$q$};
%
%rectangle
\draw[] 								\ptuno -- \ptdue;
\draw[black]							\ptuno --\pttree;
\draw[black]							\ptdue --\ptquattro;
\draw[]								\pttree--\ptquattro;
\draw[-latex,gray, dashed]					(\prop*0-\shifthor,\prop*10) --(\prop*8-\shifthor,\prop*2);
\draw[-latex,gray, dashed]					(\prop*3-\shifthor,\prop*3) --(\prop*13-\shifthor,\prop*13);
%		
%dots
%
\foreach \indeyc in {0,1,2,3}
\foreach \indexc  in {2,...,7}
\filldraw   					 (\prop*\indexc+\prop*\indeyc-\shifthor, \prop*6+\prop*\indexc-\prop*\indeyc)   	circle (.07);
\filldraw[red,thick]  (\prop*2+\prop*0-\shifthor, \prop*6+\prop*2-\prop*0) circle (.1);
\draw[black,thick]  (\prop*2+\prop*0-\shifthor, \prop*6+\prop*2-\prop*0) circle (.3);
%
%letters
%
\node[scale=.8] (puntouno) at (\prop*4-1.5*\shifthor,\prop*8) {};
\node[scale=.8]  [left of=puntouno] {$A$};   
\node[scale=.8] (puntodue) at (\prop*5-\shifthor,\prop*5.5+.5) {};
\node[scale=.8] [below of=puntodue]  {$B$}; 
\node[scale=.8] (puntoquattro) at (\prop*11.2-\shifthor,\prop*12.5) {};
\node[scale=.8] [below of=puntoquattro] {$C$};
\node[scale=.8] (puntotre) at (\prop*7.7-\shifthor,\prop*11.7) {};
\node[scale=.8] [above of=puntotre] {$D$}; 
\node at  (8.5,\prop*13) {\phantom{space}};
%legend
%\node[scale=.84] (legend) at (11,5) {$\begin{array}{l}  
%													\displaystyle A=(2,8); \\[.1cm]
%													\displaystyle B=(5,5); \\[.1cm]
%													\displaystyle C=(12,12); \\[.1cm]
%													\displaystyle D=(9,15); \\[.1 cm] \end{array}$  };
\end{tikzpicture}
\notag
\ee
and the crucial observation is that 
the anomalous dimension we are looking for is given by the formula
\begin{align}\label{component_rank1_appe}
\eta^*_{\vec{\tau}}=
\frac{1}{\left(\mathbf{M}^{(n+3)}_{\vec{\tau}}\right)}_{p_1p_2,p_3p_4  }\!\!\!\!\times \ \ \sum_{r,s}\  
\frac{ \left(\mathbf{M}^{(n+3)}_{\vec{\tau}}\right)_{p_1p_2,rs} \left( \mathbf{M}^{(n+3)}_{\vec{\tau}} \right)_{rs,p_3p_4}  }{ \left(\mathbf{L}_{\vec{\tau}}\right)_{rs,rs} } 
\end{align} 
where ${\bf L}$ is given by the long superblock decomposition of disconnected free theory, 
which we repeat here below, since it will be important in the computations.

%==================================================================
\subsubsection*{Disconnected free theory}
%==================================================================

\begin{gather}
\!\! \mathbf{L}_{\vec{\tau}}=\frac{1+\delta_{pq}}{pq} \frac{(a+1)(a+b+2)\Gamma[2+b]\Gamma[4+2a+b](l+1)(l+\tau+2)}{ 
\Gamma[ \pm \frac{p-q}{2}+\frac{2+b}{2} ] \Gamma[ \pm \frac{p-q}{2}+\frac{4+2a+b}{2} ] \Gamma[ \frac{p+q}{2}\pm\frac{1\pm 1+b}{2} ] \Gamma[  \frac{p+q}{2}\pm\frac{1\pm1+2+2a+b}{2} ] } 
\times \Pi_{\frac{\tau}{2}}\Pi_{\frac{\tau}{2}+l+1} \notag\\[.2cm]
\Pi_s=\frac{ \left( s-\frac{q-p}{2}\right)! \left( s+\frac{q-p}{2}\right)! }{(2s)!} \frac{\left(s+\frac{p+q}{2}\right)!}{ \left( s-\frac{p+q}{2}\right)!} 
{\color{black} \frac{1}{ (s\pm\frac{1\pm1+2+2a+b}{2})(s\pm\frac{1\pm1+b}{2})}}
\label{disco_FT_formula}
\end{gather}
Notice the common factor.
This reproduces the numerator of the tree level anomalous dimensions

\begin{align}
\delta^{(8)}_{\tau,l,[aba]}=(\tfrac{\tau}{2}\pm\tfrac{1\pm1+2+2a+b}{2})(\tfrac{\tau}{2}\pm\tfrac{1\pm1+b}{2})(\tfrac{\tau}{2}+l+1\pm\tfrac{1\pm1+2+2a+b}{2})(\tfrac{\tau}{2}+l+1\pm\tfrac{1\pm1+b}{2})\notag\\
\end{align}

%==============================================================
\subsection*{Warming up: spin zero with pencil and paper}
%==============================================================

To start with consider some examples from the VS amplitude. 
The amplitude $\mathcal{V}_{n,\vec{p}}$ projected onto the $su(4)$ channels $[nbn]$ ($n$ even)
is given by ${\tt Y}_{[nbn],\vec{p}}$ introduced in \eqref{formulaBMichele} multiplying a constant in $s$ and $t$, which therefore can only contribute to spin zero. 
\footnote{We repeat it here for convenience.
\be
\!\!\!\!\!\!\!
{\tt Y}_{[aba],\vec{p}}=\frac{(\Sigma-2)! b!(b+1)!(2+a+b)  }{  \Gamma[ \pm \frac{p_1-p_2}{2} + \frac{b+2}{2}]  
\Gamma[\frac{p_1+p_2}{2} + \frac{b+2}{2}]   \Gamma[\frac{p_1+p_2}{2} - \frac{b+2a+2}{2}]  \Gamma[ \pm \frac{p_3-p_4}{2} + \frac{b+2}{2}] \Gamma[\frac{p_3+p_4}{2} + \frac{b+2}{2} ] \Gamma[\frac{p_3+p_4}{2} - \frac{b+2a+2}{2} ] }\notag
\ee
}
The first case study is obviously %$a=n$ find
\begin{align}
{\tt Hrm}_{[0b0]}\big[ \mathcal{V}_{0}\big]= 2\zeta_3 (\Sigma-1)_{3} {\tt Y}_{[0b0]} \qquad;\qquad \alpha'^3
\end{align}
where the factor of $ 2 (\Sigma-1)_{3}$ is the value of the Mellin amplitude. This comes from \cite{Drummond:2019odu}. We will claim that 
\begin{align}\label{inizio_rankn}
\ \ {\tt Hrm}_{[nbn]}\big[  \mathcal{V}_{n}\big]= 2 \zeta_{n+3} (\Sigma-1)_{3} {\tt Y}_{[nbn]}  \qquad;\qquad (\alpha')^{n+3}
\end{align}
This would be the amplitude coming from $\tilde{t}^n$ and there is already something interesting about the $(\Sigma-1)_{3}$ term, 
that we want to anticipate. This pochhammer 
is obvious in the case $(\alpha')^3$ but not for $(\alpha')^{(n+3)}$ because of the $(\Sigma-1-n)_n\tilde{t}^n$ in \eqref{formulaBMichele}. 
In practise the amplitude $\mathcal{V}_n\Big|_{{\tilde t}^n}$ is such that $(\Sigma-1-n)_n$ 
cancels out. 

The above cancelation must take place because, as we will motivate better in the next few paragraphs, 
%from the factorisation properties in \eqref{component_rank1_appe}, 
~the anomalous dimension from \eqref{component_rank1_appe} cannot depend on the external charges $p_{i=1,2,3,4}$.
Notice then that there is a nice fine tuning among all the strata in the amplitude going on. For example, if we look at the term $\tilde{t}^2$ in the amplitude at $(\alpha')^5$,
we find
\begin{align}
\mathcal{V}_2\Bigg|_{\tilde{t}^2}&=\zeta_{5}\,(\Sigma-1)_3\times \{+40,-20,+2\}.\{1,\Sigma+2,(\Sigma+2)(\Sigma+3)\}\notag\\
&=\zeta_5\,(\Sigma-1)_3\times 2(\Sigma-3)(\Sigma-2)
\end{align}
By using \eqref{formulaBMichele} the result will now coincide with \eqref{inizio_rankn} for $n=2$.

The superblock decomposition of \eqref{inizio_rankn} thus leads to
\begin{align}
\mathbf{M}_{\tau,[nbn],l=0}=2 \zeta_{n+3}(\Sigma-1)_{3}\, {\tt Y}_{[nbn]}\times \texttt{Blc}_{\tau,l=0}\Big[ \overline{D}_{p_4+2,p_3+2,p_2+2,p_1+2} \Big]
\end{align}
With the help of \eqref{Dbar_deco1} we obtain,\\[-.1cm] 
\begin{align}
&
\!\!\!\!\!\!\!\mathbf{M}_{\tau,[aba],l=0}= 2\zeta_{n+3}\ (-)^{\frac{p_1+p_2-p_3-p_4}{2}+1}  \ \frac{ b!(b+1)! (2+a+b) }{ (\tau+3)!   (\tau+1)!}\times
{\color{black} \prod_{t=+\frac{1}{2}p_{43}-1}^{\frac{1}{2}(p_3+p_4)-1} \!\! \left(\tfrac{\tau}{2}-t\right)
 \prod_{t=-\frac{1}{2}p_{21}-1}^{\frac{1}{2}(p_1+p_2)-1} \!\! \left(\tfrac{\tau}{2}-t\right) }  \bigg[   \notag\\
&
\!\!\!\!\!\!\!\!\frac{ \left(\frac{\tau+p_1\pm p_2+2}{2}\right)! }{
\Gamma[ \pm \frac{p_1-p_2}{2} + \frac{b+2}{2}]  \Gamma[\frac{p_1+p_2}{2} + \frac{b+2}{2}]   \Gamma[\frac{p_1+p_2}{2} - \frac{b+2a+2}{2}]} 
\frac{  \left(\frac{\tau+p_4\pm p_3+2}{2}\right)! }{ 
\Gamma[ \pm \frac{p_3-p_4}{2} + \frac{b+2}{2}] \Gamma[\frac{p_3+p_4}{2} + \frac{b+2}{2} ] \Gamma[\frac{p_3+p_4}{2} - \frac{b+2a+2}{2} ]  } \bigg] \notag \\
\label{spin0alpha3boot}
\end{align}
where $a\rightarrow n$.
The simplification to notice in $\mathbf{M}_{\tau,[aba],l=0}$
is the one between $(\Sigma-1)_{3}(\Sigma-2)!$ in the numerator, with the latter from 
${\tt Y}_{[nbn]}$, and $(\Sigma+1)!$ in the denominator of $\texttt{Blc}_{\tau,l=0}$. We notice this 
simplification because it involves objects depending on external charges, 
which individually would not simplify in the r.h.s. of
\begin{align}
\!\!\!\!\!\!\!\!\!\!\!\eta_{\vec{\tau}}
=\frac{1}{\left(\mathbf{M}_{\vec{\tau}}\right)}_{p_1p_2,p_3p_4  }\!\!\!\!\times \ \ \sum_{r,s}\  
\frac{ \left(\mathbf{M}_{\vec{\tau}}\right)_{p_1p_2,rs} \left( \mathbf{M}_{\vec{\tau}} \right)_{rs,p_3p_4}  }{ \left(\mathbf{L}_{\vec{\tau}}\right)_{rs,rs} }
\end{align}
Thanks to the simplification explained above
we will now show that $\mathbf{M}_{\tau,[aba],l=0}$ in \eqref{spin0alpha3boot} is a self consistent CFT data.

The net result is more precisely
\be\label{ultimate_rank1}
\!\!\!\!\!\!\!\eta_{\tau,[nbn],l=0}= \frac{ 2\zeta_{n+3} \times \delta^{(8)}_{\tau,0,[aba]}    }{ (a+1)(b+1)_{2a+3} (\tau+1)_3 } 
\sum_{r,s}\,\frac{ rs  }{(1+\delta_{rs}) }\left((\tfrac{r\pm s}{2} )^2-(\tfrac{\tau}{2}+1)^2 \right)  \prod_{m=1}^{1+a} \left((\tfrac{r\pm s}{2} )^2- (\tfrac{b}{2}+m)^2 \right)
\ee
The sum above depends on $a$ both explicitly in the product $\prod_{m=1}^{a+1}$, 
but also implicitly as the sum over $(r,s)$ is the sum 
over the rectangle $R_{\vec{\tau}}$ which labels the exchanged two-particle operators.

%==============================================================
\subsection*{General case $a+l=n$ even}
%==============================================================

Here we guess that the CFT data is 
\begin{align}\label{Mabaln}
 \mathbf{M}_{\tau,[aba],l=n-a}={2}{\color{black}\,\texttt{Bin}[\substack{n\\ l}]}\zeta_{n+3}\, (\Sigma-1)_{3+l}\, {\tt Y}_{[aba]}\times \texttt{Blc}_{\tau,l}\Big[\overline{D}_{p_4+2+l,p_3+2,p_2+2,p_1+2+l} \Big]
\end{align}
and we will show that is solves 
\begin{align}
\!\!\!\!\!\!\!\!\!\!\!\eta_{\vec{\tau}}
=\frac{1}{\left(\mathbf{M}_{\vec{\tau}}\right)}_{p_1p_2,p_3p_4  }\!\!\!\!\times \ \ \sum_{r,s}\  
\frac{ \left(\mathbf{M}_{\vec{\tau}}\right)_{p_1p_2,rs} \left( \mathbf{M}_{\vec{\tau}} \right)_{rs,p_3p_4}  }{ \left(\mathbf{L}_{\vec{\tau}}\right)_{rs,rs} }
\end{align}
as in the previous section.

The formula for $\mathbf{M}_{\tau,[aba],l=n-a}$ would only take into account the contribution 
from $\tilde{t}^{a} t^{l}$ in the amplitude, and again the amplitude has to be such that 
only $(\Sigma-1)_{3+l}$ appears and the term $(\Sigma-1-a)_a$ cancels out in \eqref{formulaBMichele}. 
Let us consider again the amplitude $(\alpha')^5$ as an example, then
\begin{align}
\mathcal{V}_{2}\Bigg|_{\tilde{t}t}&= \zeta_5 (\Sigma-1)_4\times \{ 0,-10,+2\}.\{ (\Sigma-2)^{-1},  1, (\Sigma+3)\} \notag\\
						&=\zeta_5(\Sigma-1)_4\times 2(\Sigma-2)
\end{align}
and upon using \eqref{formulaBMichele} it will lead to \eqref{Mabaln}.

The binomial factor $\texttt{Bin}$ will be justified a posteriori by insisting that the large twist 
limit of the anomalous dimensions does not depend on $[aba]$, thus it is the same as for 
$l=n$ and $l=0$. In this two cases in fact we know that the overall normalisation coming from 
$\mathcal{M}^{flat}_{n,n}$ of \eqref{Mabaln} is precisely a factor of $2$. This is a nice fun fact about the flat VS amplitude.
The rest is a straightforward generalisation of the case $l=0$ we discussed in the previous section.

We will now repeat the derivation of the anomalous dimensions as in the previous section, with 
minor modifications due to the spin. Let us begin by making explicit, 
\begin{align}
&
\!\!\!\!\!\!\!\!\!\mathbf{M}_{[aba],l}={2}{\color{black}\,\texttt{Bin}[\substack{n\\ l}]}\zeta_{n+3}\times\,\\
&
\!(-)^{\frac{p_1+p_2-p_3-p_4}{2}+l+1}  \ \frac{  b!(b+1)! (2+a+b)(2+l+\tau) }{ (\tau+2l+3)!   (\tau+2l+2)!}\times
{\color{black} \prod_{t=+\frac{1}{2}p_{43}-l-1}^{\frac{1}{2}(p_3+p_4)-1} \!\! \left(\tfrac{\tau}{2}-t\right)
 \prod_{t=-\frac{1}{2}p_{21}-l-1}^{\frac{1}{2}(p_1+p_2)-1} \!\! \left(\tfrac{\tau}{2}-t\right) }  \bigg[   \notag\\
&
\!\!\!\!\!\frac{ \left(\frac{\tau+2l+p_1\pm p_2+2}{2}\right)! }{
\Gamma[ \pm \frac{p_1-p_2}{2} + \frac{b+2}{2}]  \Gamma[\frac{p_1+p_2}{2} + \frac{b+2}{2}]   \Gamma[\frac{p_1+p_2}{2} - \frac{b+2a+2}{2}]} 
\frac{  \left(\frac{\tau+2l+p_4\pm p_3+2}{2}\right)! }{ 
\Gamma[ \pm \frac{p_3-p_4}{2} + \frac{b+2}{2}] \Gamma[\frac{p_3+p_4}{2} + \frac{b+2}{2} ] \Gamma[\frac{p_3+p_4}{2} - \frac{b+2a+2}{2} ]  } \bigg] \notag \\
\label{prima_formulazza_M}
\end{align}
where $l\rightarrow n-a$.
Then, from the relation in \eqref{component_rank1} we obtain
\begin{align}
\!\!\!\!\!\!\!\!\eta^*_{\vec{\tau}}\Big|_{a+l=n} =&\  {2}{\color{black}\,\texttt{Bin}[\substack{n\\ l}]}\zeta_{n+3} \times  \frac{ (-)^{a+l+2} \ \delta^{(8)}_{\tau,l,[aba]}}{ (a+1)(l+1)(b+1)_{2a+3} (\tau+1)_{2l+3}  }\\
&
\ \sum_{r,s}\,
\bigg[\ \frac{ rs}{ (1+\delta_{rs}) }\prod_{m=1}^{1+l} \left( (\tfrac{\tau}{2}+m)^2 -(\tfrac{r\pm s}{2} )^2 \right) \prod_{m=1}^{1+a} \left( (\tfrac{b}{2}+m)^2 -(\tfrac{r\pm s}{2} )^2 \right)\bigg] \notag
\end{align}
%\end{mdframed}
Finally, notice that $\texttt{Bin}[\substack{n\\ l}]\frac{1}{(a+1)(l+1)}=\frac{n!}{(a+1)!(l+1)!}$, therefore 
we reproduce the formula given in \eqref{intro_thesums} for $a+l=n$ even.

%=======================================================
\subsection*{General case $a+l=n-1$ odd }
%=======================================================

The relevant part of the VS amplitude here would have three terms of the form, 
\begin{align}
{\tt Hrm}_{[aba]}\big[ \mathcal{V}_n\big]_{a+l=n-1}=t^l( \tilde k_{0} + \tilde k_{s} s + \tilde k_{t} t ) \zeta_{n+3}
\end{align}
We will determine
$\tilde k_{0}$, $ \tilde k_{s}$ and $\tilde k_{t}$ from the self-consistency of the $rank=1$ bootstrap. 
Let us extract some piece of data we certainly know of, by defining 
\begin{align}
\tilde k_{0,s,t}=2(\Sigma-1)_{l+4}\, {\tt Y}_{[aba]}\Big|_{a+l=n-1}  k_{0,s,t}
\end{align} 
The factor  $(\Sigma-1)_{l+4}$ is the one associated to the Penedones transform for $t^{l}s$ and $t^{l+1}$. 
These two terms comes from the flat space amplitude therefore we do not expect extra $p_1p_2p_3p_4$ dependence. 
For $\tilde k_0$ the Penedones transform would give $(\Sigma-1)_{l+3}$, since this is not a top term. However, 
we also expect to find non trivial dependence on the external charges. 

From the amplitude we go to 
\begin{align}
&
\!\!\!\!\!\mathbf{M}_{\tau,[aba],a+l=n-1} =\,2 \zeta_{n+3}\ B_{[aba]} 
\Big[ 
%\rule{2cm}{0pt}  
- k_{s}  (\Sigma-1)_{l+4} \texttt{Blc}_{\tau,l}\big[  \overline{D}_{p_4+2+l,p_3+2,p_2+3,p_1+3+l}  \big]  \notag\\
&
\rule{6cm}{0pt} 
- k_{t} (\Sigma-1)_{l+4} \texttt{Blc}_{\tau,l}\big[ \overline{D}_{p_4+3+l,p_3+2,p_2+3,p_1+3+l}  \big]  \notag\\
  &
  \rule{1.7cm}{0pt} 
 + (k_{0}+ k_{s} c_s + k_{t} (l+1)(  c_t + \tfrac{l}{2} ) ) (\Sigma-1)_{l+4}\texttt{Blc}_{\tau,l}\big[ \overline{D}_{p_4+2+l,p_3+2,p_2+2,p_1+2+l}\big] \Big]  \label{matMalnmenos1}
\end{align}
The last term is the contribution from $U\partial_U$ in $t^{l}s$ and $V\partial_V$  in $t^{l+1}$, which add to the contribution from $t^l$ in the amplitude. 
As we anticipated above, the presence of $c_s$ and $c_t$ in this combination also implies that $k_0$ has a non trivial dependence on the external charges.
Of course the amplitude generates more $\overline{D}$ but we are interested in the spin $l$ given by $a+l=n-1$, and the ones above are the only relevant ones.

The various $\overline{D}$ entering \eqref{matMalnmenos1} have a common term in their decomposition, 
\begin{align}
&
\!\!\!\!\!\!\!\!\!\texttt{Comm}_{\tau,[aba],l}= 2\zeta_{n+3}\times \\  %(-)^l
&
\!(-)^{\frac{p_1+p_2-p_3-p_4}{2}+l+1}  \ \frac{  b!(b+1)! (2+a+b)(2+l+\tau) }{ (\tau+2l+3)!   (\tau+2l+2)!}\times
{\color{black} \prod_{t=+\frac{1}{2}p_{43}-l-1}^{\frac{1}{2}(p_3+p_4)-1} \!\! \left(\tfrac{\tau}{2}-t\right)
 \prod_{t=-\frac{1}{2}p_{21}-l-1}^{\frac{1}{2}(p_1+p_2)-1} \!\! \left(\tfrac{\tau}{2}-t\right) }  \bigg[   \notag\\
&
\!\!\!\!\!\frac{ \left(\frac{\tau+2l+p_1\pm p_2+2}{2}\right)! }{
\Gamma[ \pm \frac{p_1-p_2}{2} + \frac{b+2}{2}]  \Gamma[\frac{p_1+p_2}{2} + \frac{b+2}{2}]   \Gamma[\frac{p_1+p_2}{2} - \frac{b+2a+2}{2}]} 
\frac{  \left(\frac{\tau+2l+p_4\pm p_3+2}{2}\right)! }{ 
\Gamma[ \pm \frac{p_3-p_4}{2} + \frac{b+2}{2}] \Gamma[\frac{p_3+p_4}{2} + \frac{b+2}{2} ] \Gamma[\frac{p_3+p_4}{2} - \frac{b+2a+2}{2} ]  } \bigg] \notag \\
\end{align}
Notice that $\texttt{Comm}_{\tau,[aba],a+l=n-1}$ coincides with formula \eqref{prima_formulazza_M} which we used to compute $\mathbf{M}_{\tau,[aba],a+l=n}$. 
Therefore $\texttt{Comm}_{\tau,[aba],a+l=n-1}$ dependend on the external charges $p_1p_2$ and $p_3p_4$, as 
in the previous case $a+l=n$. However, now we find more stuff
\begin{align}
&
\!\!\!\!\!\mathbf{M}_{\tau,[aba],a+l=n-1} =\texttt{Comm}_{\tau,[aba],l} \underbrace{ 
\Big[ - k_{s}  \texttt{cpw}_s - k_{t}  \texttt{cpw}_t  + \Big(k_{0}+ k_{s} c_s + k_{t} (l+1)(  c_t + \tfrac{l}{2}) \Big) (\Sigma+l+2)\Big] }_{\displaystyle \texttt{poly}_{k_s,k_t,k_0}(\tau,l,\vec{p})}
\end{align}\\[-.6cm]
i.e.~a non trivial polynomial shows up. Upon defining $T\equiv\tau(4+2l+\tau)$ we have
\begin{align}
&
\texttt{cpw}_s= 
\tfrac{1}{4} \Big[ (p_1+p_2)(4+2l+p_1+p_2)-T\Big]
\notag\\
&
\texttt{cpw}_t=\tfrac{1+l}{8}\, \Big[ 4(2+l+p_1)(2+l+p_4) +(p_1^2-p_2^2-p_3^2+p_4^2)+ \frac{(p_1^2-p_2^2)(p_4^2-p_3^2)}{T}+ T \Big]\notag
\end{align}

There should be a restriction on $\mathbf{M}_{\tau,[aba],a+l=n-1}$ because of the large twist behaviour, 
and in fact it should not grow faster than $\texttt{Comm}_{\tau,[aba],l}$, since the latter gives 
the growth of $\mathbf{M}_{\tau,[aba],a+l=n}$.  This implies the term linear in $T$ has to cancel. 
We thus find the constraint
\begin{align}
k_s=\frac{l+1}{2} k_t
\end{align}
Computing the anomalous dimension as in \eqref{component_rank1_appe}, we find 
\begin{align}
 \rule{1cm}{0pt} \frac{ \texttt{poly}_{\frac{1}{2}(l+1)k_t,k_t,k_0}(\tau,l,p_1,p_2,r,s)\ \texttt{poly}_{\frac{1}{2}(l+1)k_t,k_t,k_0}(\tau,l,r,s,p_3,p_4) }{  \texttt{poly}_{\frac{1}{2}(l+1)k_t,k_t,k_0}(\tau,l,p_1,p_2,p_3,p_4) }
\end{align}
The only way the dependence on the external charges cancels out in the above ratio is if the individual 
$\texttt{poly}$ has itself a factorised dependence on $p_1p_2$ and $p_3p_4$. 
The structure of such a polynomial is $\texttt{poly}= -k_t\frac{(l+1)}{8} \frac{(p_1^2-p_2^2)(p_4^2-p_3^2)}{T} 
+ \ldots + k_0 (\Sigma+l+2)$. Therefore the $1/T$ term automatically 
satisfies our requirement. What is natural to do is to solve for $k_0$ so that also the constant term in 
$T$ is proportional to $k_t\frac{(l+1)}{8}(p_1^2-p_2^2)(p_4^2-p_3^2)$. This equation gives
\begin{align}\label{constr_a0}
k_0= -\frac{k_t(l+1)}{8(\Sigma+l+2)} \Big( K (p_1^2-p_2^2)(p_4^2-p_3^2)  +( c_u^2-c_t^2+2(c_s+2c_t-\Sigma-4)(\Sigma+l+2))\Big)
\end{align}
with $K$ a yet unknown constant. 
As a result  $ \texttt{poly}$ can be put in the following form,
\begin{align}
 \texttt{poly}_{\frac{1}{2}(l+1)k_t,k_t,k_0=\eqref{constr_a0}}=-k_t \frac{(l+1)}{8}(p_1^2-p_2^2)(p_4^2-p_3^2)\left( \frac{1}{T}+K\right)
\end{align}

At this point $K$ can only be function of $[aba]$, since our reasoning has led us 
to conjecture what is the dependence on $l$ and $p_1p_2p_3p_4$.  
From the amplitude at $(\alpha')^{5,7}$ we have enough data to overconstrain 
our conjecture, and in fact we find the consistent solution $K=-1/(b(b+2a+4))$. 
We can now put together formula \eqref{component_rank1} and find the anomalous 
dimensions as a sum over exchanged two-particle operators. A posteriori, the value of 
$k_t$ is fixed by imposing that the large twist limit of the anomalous dimensions is the same for any $[aba]$ (or spin), 
\begin{align}
k_t= \texttt{Bin}[\substack{n\\ l+1}]
\end{align}

In sum, our result is
\begin{gather}
%&
\!\!\!\!\!\!\!\!\!\eta^*_{\vec{\tau}}\Big|_{a+l=n-1}=\  {2}{\color{black}\,\texttt{Bin}[\substack{n\\ l+1}]}\zeta_{n+3} \times  \frac{ (-)^{a+l+2} \ \delta^{(8)}_{\tau,l,[aba]}}{ (a+1)(b+1)_{2a+3} (\tau+1)_{2l+3}  }
 \left[ \frac{1}{b(b+2a+2)}-\frac{1}{T}\right]\times\\
%
%&
\rule{1.5cm}{0pt}\times\sum_{r,s}\,
\bigg[\ \frac{ rs(r-s)^2(r+s)^2}{ 8(1+\delta_{rs}) }\prod_{m=1}^{1+l} \left( (\tfrac{\tau}{2}+m)^2 -(\tfrac{r\pm s}{2} )^2 \right) \prod_{m=1}^{1+a} \left( (\tfrac{b}{2}+m)^2 -(\tfrac{r\pm s}{2} )^2 \right)\bigg] \notag 
\end{gather}
Notice that $\texttt{Bin}[\substack{n\\ l+1}]\frac{1}{(a+1)}= \frac{n!}{(a+1)!(l+1)!}$ 
where $a+l=n-1$ odd. Finally we can absorb the sign $(-)^{a+l}$ 
by reversing the polynomial.

%=======================================================
\subsection{Spin structures of the flat VS}\label{more_10dspin}
%=======================================================

The value of the 10d spin of a monomial in ${s,t,u}$ contributing to the $\alpha'$ 
expansion of the VS amplitude is counted its power in $t$, with the constraint on ${u}$ implemented. 
From the exponential form of VS, parametrise the possible contributions as 
\begin{equation}\label{contri_flat_space_michele}
\text{VS}\Bigg|_{{\zeta_{n_1}\cdots\zeta_{n_r}}}\!\!\propto\ \frac{\sigma_{n_1}\cdots\sigma_{n_r}}{stu}\,(\alpha')^{\sum_{i=1}^r n_i}\qquad;\qquad
\sigma_n \equiv s^n+t^n+u^n 
\end{equation}
and
recall that $\sigma_n$ decomposes as 
\begin{equation}\label{formula_green}
\sigma_{n}\ \propto \sum_{2p+3q=n} \frac{(p+q-1)!}{p!q!}\left( \frac{\sigma_2}{2}\right)^p \left(\frac{\sigma_3}{3}\right)^q; 
\end{equation}
Notice that both $\sigma_2$ and $\sigma_3$ have spin $2$ therefore $\sigma_{n}$ has even spin by default, 
and therefore any term contributing to the VS amplitude in the $\alpha'$ expansion.

We can be more precise and find a formula to compute the 10d spin of \eqref{contri_flat_space_michele}. 
A term $(\sigma_2)^p(\sigma_3)^q$ in \eqref{formula_green} counts $2p+2q=n-q$. We have a sum in 
\eqref{formula_green} and therefore various possible $(\sigma_2)^p(\sigma_3)^q$, i.e.~monomials in ${s,t,u}$. 
Consider first the the contribution of maximum 10d spin, which is obtained for the minimal $q$ in the sum. 
The value of $n$ is odd by default, it means we take $q=1$ and $p=(n-3)/2$ to solve $2p+3q=n$, thus the spin is $n-1$. 
This reasoning after all is quite obvious. 
Putting all together,
\be
l_{10}\ \ {\tt of\ the\ }\ {\zeta_{n_1}\cdots\zeta_{n_r}}\ {\tt contribution}
\leq \ -2+\sum_{i=1}^{r} (n_i -1)
\ee
Now consider the case in which $\sigma_n$ decompose in various $(\sigma_2)^p(\sigma_3)^q$ terms. 
The spin of each term is still counted by $2p+2q=n-q$, thus the value of 
$q$ parametrises the various terms. To fix ideas consider
\be
\zeta_9 \times \frac{ s^9 + t^9+u^9}{stu}= \zeta_9 \times\Big( s^6 +t^6 +u^6 - \#( st u)^2\Big)
\ee
Then we find both $s^6 +t^6 +u^6$ with 10d spin $6$ and $( st u)^2$ with 10d spin $4$. 
The latter is the same contribution as $\zeta_3^3$. In order to generalise our previous formula we need, 
together with the information about the $\zeta_{n_i}$, also the values of $q_i$ we are looking at. 
Thus,
\be
l_{10}\ \ {\tt of\ the\ }\ {\zeta_{n_1}\cdots\zeta_{n_r}}\Big|_{\{q_1,\ldots q_r\}}\ {\tt contribution}=
 -2+\sum_{i=1}^{r} (n_i  -q_i) 
\ee
for the possible values of $\sum q_i$.

%======================================================================
\section{On the 3pt couplings $\langle \cO_{p}\cO_{q} {\cal K}_{(rs)}\rangle$} \label{singlet_eigenvectors}
%======================================================================

Let us illustrate some general features of ${\bf c}^{(0)}_{\vec{\tau}}$.  
A column of this matrix is a vector of $\mu(t-1)$ components with the following block structure 
\be
\begin{tikzpicture}
\def\step{.6}

\draw[thick,gray] (.25*\step,.3*\step)-- (.25*\step-.25,.3*\step)  -- (.25*\step-.25,-6.6*\step)  --  (.25*\step,-6.6*\step);
\draw[thick,gray] (1.75*\step,.3*\step)-- (1.75*\step+.25,.3*\step)  -- (1.75*\step+.25,-6.6*\step)  --  (1.75*\step,-6.6*\step);

\filldraw[green!20,draw=black] (.4*\step,0) rectangle (1.6*\step,-1.5*\step);
\filldraw[green!20,draw=black] (.4*\step,-1.6*\step) rectangle (1.6*\step,-1.5*\step-1.6*\step);

\draw[] (1*\step,-3.7*\step) node {$\vdots$};

\filldraw[green!20,draw=black] (.4*\step,-3*1.6*\step) rectangle (1.6*\step,-1.5*\step-3*1.6*\step);

\draw[] (1*\step,-.75*\step) node {${\cal T}_1$};
\draw[] (1*\step,-2.25*\step) node {${\cal T}_2$};
\draw[] (1*\step,-5.5*\step) node {${\cal T}_{\mu}$};

\draw[] (9.2*\step,-3.25*\step) node {${\cal T}_{\beta,\vec{\tau}}={\tt Table}\Big[\,\ldots\, ,\{i,1,t-1\}\Big]$};

\draw[] (-2.8*\step,-3.25*\step) node {${\tt cln}\Big({\bf c}^{(0)}_{\vec{\tau}}\Big)=$};

\end{tikzpicture}
\ee
The number of blocks is always fixed by the parameter $\mu$ in \eqref{multiplicity}, 
but the length of each block grows with the twist.  At the same time also the size of the matrix changes 
as we vary the twist, and it changes accordingly to the dimension of the rectangle $R_{\vec{\tau}}$.
The first column corresponds to the left-most corner $A$ of the rectangle, i.e.~the most negative SUGRA anomalous dimension. 
The last column corresponds to the right-most corner of the rectangle, therefore the less negative SUGRA anomalous dimension. 
Both these columns are uniquely determined by the SUGRA eigenvalue problem, being the corresponding anomalous 
dimension non degenerate. In between the SUGRA eigenvalue problem is degenerate, but for the special 
case at the minimum twist $\tau=b+2a+4$, where the rectangle collapses to a line with $-45^\circ$ orientation.

%==================================================================
\subsection*{All singlet eigenvectors of ${\bf N}^{(0)}$ for $a+l$ even}%\label{singlet_eigenvectors}
%==================================================================

The anomalous dimensions we bootstrapped in the previous section is directly computed from the VS amplitude as
\be
\eta^*_{\vec{\tau}}=  \mathbb{V}^T_{\vec{\tau},m^*=1}\, {\bf N}^{(n+3)}_{\vec{\tau}}\, \mathbb{V}_{\vec{\tau},m^*=1}
\ee
where $\mathbb{V}_{\vec{\tau},1}$ is the (only) most negative eigenvector 
of the SUGRA eigenvalue problem ${\bf N}^{(0)}_{\vec{\tau}}$, i.e. the one corresponding 
to the left most corner in $R_{\vec{\tau}}$ and given by $\mathbf{c}^{(0)}_{\vec{\tau}}\big|_{cln=1}$.

For $a+l$ even this is
\begin{align}
&
\!\!\!\!\!\mathbf{c}^{(0)}_{\vec{\tau}}\Big|_{cln=1}= \\
&
\!\!\!\mathcal{N}_{\vec{\tau}}\  
\bigoplus_{\beta=1}^{\mu}  \sqrt{ 
\frac{ (\beta)_{a+1}(b+2-\beta)_{a+1}}{(1+\delta_{\mu=\beta,b\in 2\mathbb{N}})\ \ } 
\frac{   ( \frac{\tau-b}{2}+2+l )_{\beta-1}  ( \frac{\tau+b}{2}+2-\beta)_{\beta-1}  }{ 
	    ( \frac{\tau-b}{2}+1)_{\beta-1}   ( \frac{\tau+b}{2}+3+l-\beta  )_{\beta-1}  }
 }\ \widetilde{\mathcal{T}}_{\beta,\vec{\tau}}\notag
\end{align}
where the multi-index  $(\beta,i)$ corresponds to the decomposition  
that splits the vector into various blocks, as in the figure above. 
Each block is given by
\begin{align}
&
\!\!\!\!\!\!\!\!\!\!\!\!\!\!\!\!\!\!\!\!\!\widetilde{\mathcal{T}}_{\beta,\vec{\tau}}=
\texttt{Table}\Big[ (-)^{i+1}\sqrt{  \chi_{\beta,i,[aba]} { (\tfrac{\tau+b}{2}+a+2+i)_{t-i-1} (l+2)_{t-1-i} (\tfrac{\tau+b}{2}+a+l+4)_{i-1}(t-i)_{i-1} }  }\, ,\, \{ i,1,t-1\} \Big] %\Gamma[t+l+1-i] \Gamma[ t+l+b+2a+3+i]
\notag\\
&
\rule{2.6cm}{0pt}\chi_{\beta,i,[aba]}= (i)_{a+1}(i+\beta+a)(i+b+a+2-\beta)(i+b+a+2)_{a+1} 
%\frac{ t+b-\beta+2)_{\beta-1} }{ t+1,\beta-1} 
\end{align}
where $t=(\tau-b)/2-a$, and the remaining data is
Finally, the normalisation is
\begin{align}
&
\!\!\!\!\!\mathcal{N}_{l,\tau,[aba]}=\\
&
\sqrt{ \frac{2^{2a+5-\tau} }{(\frac{\tau+b}{2}+l+2)_{1-\mu}(l+a +\frac{9}{2})_{t-3+\mu}}   %  (l+\tfrac{\tau}{2}+\frac{1-(-1)^b}{4}+2)_{\mu-1} 
\frac{\Gamma[b+1]}{\Gamma[b+2a+4]\Gamma[a+2]  \Gamma[t+a+1]}  }
\notag
\end{align}
For $b$ even,
\begin{align}
\mathbf{c}^{(0)}_{cln={\rm last}}=\big[ \mathfrak{S}\big]\cdot \mathbf{c}^{(0)}_{cln=1}\Big|_{l\rightarrow -l-\tau-3}
\end{align}
where $\mathfrak{S}$ is a diagonal matrix that changes the signs as follows: $\mathbf{c}^{(0)}_{cln=1}$ is 
such that the signs are equals on the blocks, i.e. they do not depend on the $\beta$ index, but within each block 
the signs are alternating, i.e.~they depend as $(-)^{i+1}$; instead $\mathbf{c}^{(0)}_{cln={\rm last}}$ is such that the 
signs don't depend on the index $i$ and alternate on the block index $\beta$.

%==================================================================
\subsection*{Examples at $m^*=2$}%\label{more_3pt_app}
%==================================================================

In the main text, when discussing the bootstrap of the $(\alpha')^{6,7}$ amplitudes 
we referred to the 3pt couplings of $m^*=2$ two-particle operators in the $su(4)$ channel $[0b0]$ with $a+l=0$, 
which are unmixed uniquely by the VS amplitude at $(\alpha')^5$. 
We collect here other explicit examples, in addition to the one given in \eqref{formulas_040_3pt}. 

We shall refer to the general parametrisation given in \eqref{formulazza_3pt}, which we repeat here for convenience
\begin{align}
&
{\cal T}^{\pm}_{\beta,\vec{\tau}}=\sqrt{ {\cal N}_{\vec{\tau},\beta}\times
\frac{ (\frac{\tau+b}{2}-\beta+2)_{\beta-2}  (\frac{\tau-b}{2}+l+3)_{\beta-2} }{  (\frac{\tau-b}{2}+\mu-2)_{\beta+2-\mu}  (\frac{\tau+b}{2}+l+3-\beta)^{}_{\beta+2-\mu}} 
 }\times \widetilde{\cal T}_{\beta,\vec{\tau}}^{\pm} 
\\[.3cm]
& 
\!\!{\cal N}_{\vec{\tau},\beta}=\frac{1}{(\tau-1)_{2l+7} } \frac{(\frac{\tau+b}{2}+a+l+5)_{-a-\mu} }{ (\frac{\tau-b}{2}-a-2)_{+a+\mu}}
\left\{ \begin{array}{lc} 
(\frac{\tau+b}{2}) (\frac{\tau-b}{2}+l+2) ; 			&				 \beta=1\\[.2cm]
1 										& 				  2\leq \beta\leq \mu-1 \\[.2cm]
(\frac{\tau-b}{2}+\mu-1)(\frac{\tau+b}{2}-\mu+l+3) ;	 &				 \beta=\mu
  \end{array}\right. \notag 
\end{align}
where
\be\label{formula_calT_app}
\!\!\!\!\!\!\!\!\!\!\!\widetilde{\cal T}_{\beta,\vec{\tau}}^{\pm}= {\tt Table}\Big[ \sigma_{\beta,i} \left[(\tfrac{\tau+b}{2}+a+i+2)_{l+1}(\tfrac{\tau-b}{2}-i-a)_{l+1}\!
\Bigg(\widetilde{\cal P}_{\beta,1}(T,I)\pm  \frac{ \widetilde{\cal P}_{\beta,2}(T,I) }{ \sqrt{ \gamma_{2,1}^2-4 \gamma_{2,0}} }\Bigg)^{}\right]^{\!\!\frac{1}{2}}\!\!\!,\,\{i,1,t-1\} \Big]
\ee
with ${\cal P}_{\beta}$ polynomials in the variables $T$ and $I\equiv i(i+b+2a+2)$.

A number of polynomials ${\cal P}_{\beta,1}$ and ${\cal P}_{\beta,2}$ are attached in the ancillary files.
The notation is 
\be
{\cal P}_{\beta,1}\rightarrow {\tt savingP1}[a,b,a,l][\beta]\qquad;\qquad {\cal P}_{\beta,2}\rightarrow {\tt savingP2}[a,b,a,l][\beta].
\ee 
Let us write here the first of each, for orientation. This is the case $[020]$ spin $l=0$.  
\begin{align}
&\rule{.8cm}{0pt}
\widetilde{\cal P}_{\beta,1}\!=\!\!\left[ \begin{array}{c}
\scalebox{.8}{
$132(-8400 - 4200I - 525I^2 + 3180T + 
      1290IT + 123I^2T - 553T^2 - 126IT^2 + 35T^3)$ }\\
\scalebox{.8}{
  $88(-1875 - 1350I - 243I^2+ 1100T + 
      792*IT + 135I^2 T- 392T^2 - 126IT^2 + 35T^3)$} 
   \end{array}\right]\notag\\
   \\
%\label{formulas_040_3pt} \\
 &
\!\!\!\!\!\!\!\!\!\widetilde{\cal P}_{\beta,2}\!=\!\!\left[\!\!\!\!\begin{array}{c}
\scalebox{.8}{
$528 (-5775 (4 + I)^2 + 
   6 (5930 + I (2465 + 247 I)) T - (15103 + 5772 I + 
      519 I^2) T^2 + 14 (197 + 45 I) T^3 - 175 T^4)$ }\\
\scalebox{.8}{
  $-352 (-33 (25 + 9 I)^2 + (25675 + 18 I (859 + 129 I)) T + 
   9 (372 + I (274 + 51 I)) ccT^2 - 63 (31 + 10 I) T^3 + 
   175 T^4)$} 
   \end{array}\!\!\!\right]\notag
\end{align}

Only the signs $\sigma^2=1$ are not explicit in \eqref{formula_calT_app}. These can be determined 
either by direct computation of the eigenvalue problems, as we have actually done, or alternatively by 
imposing orthogonality w.r.t.~the first (singlet) eigenvector given explicitly in appendix \ref{singlet_eigenvectors}. 
We verified that upon fixing $ \sigma_{1,1}=1$, for convention, there is a unique assignment of signs such 
that orthogonality holds, and it obviously agrees with the assignment from the actual solution of
the eigenvalue problems.

%%%%%%%%%%%%%%%%%%%
\bibliographystyle{apsrev4-1}
%\bibliography{refs}

\end{document}